\documentclass[prb,twocolumn,showpacs,superscriptaddress]{revtex4}

\usepackage{amsmath,amssymb,graphicx}
\usepackage{psfrag}

\newcommand{\bs}[1]{\boldsymbol{#1}}
\newcommand{\comm}[2]{\left[#1,#2\right]}

\newcommand{\ket}[1]{\left|#1\right\rangle}
\newcommand{\bra}[1]{\left\langle#1\right|}
\newcommand{\sgn}[1]{\text{sgn}(#1)\,}
\newcommand{\ii}{\text{i}}
\newcommand{\up}{\uparrow}
\newcommand{\dw}{\downarrow}

\allowdisplaybreaks[4]

\begin{document}
\title{Dynamical spin-spin correlation functions in the Kondo model out of
equilibrium}  
\author{Dirk Schuricht}
\author{Herbert Schoeller}
\affiliation{Institut f\"ur Theoretische Physik A, RWTH Aachen, 
  52056 Aachen, Germany}
\affiliation{JARA-Fundamentals of Future Information Technology}
\date{26 August 2009}

\begin{abstract}
  We calculate the dynamical spin-spin correlation functions of a Kondo dot
  coupled to two noninteracting leads held at different chemical potentials.
  To this end we generalize a recently developed real-time renormalization
  group method in frequency space (RTRG-FS~\cite{Schoeller09}) to allow the
  calculation of dynamical correlation functions of arbitrary dot operators in
  systems describing spin and/or orbital fluctuations. The resulting two-loop
  RG equations are analytically solved in the weak-coupling regime. This
  implies that the method can be applied provided either the voltage $V$
  through the dot or the external magnetic field $h_0$ are sufficiently large,
  $\max\{V,h_0\}\gg T_K$, where the Kondo temperature $T_K$ is the scale where
  the system enters the strong-coupling regime. Explicitly, we calculate the
  longitudinal and transverse spin-spin correlation and response functions as
  well as the resulting fluctuation-dissipation ratios.  The correlation
  functions in real-frequency space can be calculated in Matsubara space
  without the need of any analytical continuation.  We obtain analytic results
  for the line-shape, the small- and large-frequency limits and several other
  features like the height and width of the peak in the transverse
  susceptibility at $\Omega\approx\tilde{h}$, where $\tilde{h}$ denotes the
  renormalized magnetic field. Furthermore, we discuss how the developed
  method can be generalized to calculate dynamical correlation functions of
  other operators involving reservoir degrees of freedom as well.
\end{abstract}
\pacs{05.10.Cc, 73.63.Kv, 75.40.Gb}
\maketitle

\section{Introduction}
The single-impurity Kondo model~\cite{Kondo64} is unquestionably one of the
most important models studied in condensed matter physics over the past
decades.  The investigation of its equilibrium properties has caused the
development of important theoretical tools such as renormalization group
methods or the Bethe Ansatz for impurity
systems~\cite{Wilson75,TsvelikWiegmann83,Andrei-83,Hewson93}. More than 20
years ago it was also realized~\cite{GlazmanRaikh88,NgLee88,Kawabata91} that
the Kondo model can be used to describe transport experiments through quantum
dots. The developments in the ability to engineer devices on the nanoscale has
led to the experimental realization of Kondo physics in such
systems~\cite{Goldhaber-98,Cronenwett-98,Schmid-98,Goldhaber-98prl,Simmel-99,Wiel-00,Nygard-00}.
One particular advantage of these quantum dots is the almost full control over
system parameters like temperature, bias and gate voltages, magnetic field,
and exchange couplings.  These possibilities have triggered a great interest
in the theoretical study of quantum dots out of equilibrium. A wide range of
theoretical methods has been applied in the past, including non-equilibrium
perturbation
theory~\cite{Rosch-01,ParcolletHooley02,Mao-03,ShnirmanMakhlin03,Rosch-03prl,
  Paaske-04prb1,Paaske-04prb2,Rosch-05,DoyonAndrei06,Chung-09}, the
flow-equation
method~\cite{Kehrein05,Kehreinbook,FritschKehrein09ap,FritschKehrein09},
Coulomb gas representations~\cite{MitraMillis07,Segal-07}, real-time and
functional renormalization group
methods~\cite{Schoeller00,SchoellerKoenig00,KeilSchoeller01,
  Jakobs-07,Korb-07,Schoeller09,SchoellerReininghaus09}, a non-equilibrium
extension of the numerical renormalization group (NRG)
method~\cite{Anders08prl,Anders08}, and time-dependent
density matrix renormalization group (DMRG) techniques~\cite{BohrSchmitteckert07,Boulat-08,Kirino-08,Dias-08,Heidrich-Meisner-09}.
These studies established the importance of relaxation and decoherence effects
for the understanding of non-equilibrium physics. From an experimental point
of view various measurable quantities like the steady-state current, the
magnetization, and the static susceptibility have been calculated.

Beside the application of these techniques there were also attempts to employ
the known integrability of certain impurity models in equilibrium, notably the
Anderson impurity model and the interacting resonant level model, to
investigate their non-equilibrium properties. Konik et
al.~\cite{Konik-01,Konik-02} calculated the differential conductance in the
Anderson impurity model by combining the well-known scattering states of the
equilibrium system~\cite{TsvelikWiegmann83} with a Landauer-B\"uttiker
formalism. In this work the chemical potentials in the leads were coupled to
dressed excitations rather than free electrons and the calculation was
restricted to a subset of the excitations. In contrast, a different approach
was recently put forward by Metha and Andrei~\cite{MehtaAndrei06} to treat the
interacting resonant level model.  They constructed a new set of scattering
states of Bethe-Ansatz form which share the quantum numbers of free electrons
in the incoming channel, hence allowing the application of a finite voltage in
the usual manner.  However, questions concerning the existence of these
scattering states and issues related to the used regularization scheme of the
theory remain open.  Nevertheless, the interacting resonant level model has
become one of the benchmark systems in the study of non-equilibrium
physics~\cite{Borda-07,Doyon07,Golub07,NishinoHatano07,SchillerAndrei07,BoulatSaleur08,Boulat-08,Borda-08,Nishino-09}.

Despite the large number of studies of impurity models out of equilibrium only
few results are known for the dynamical correlation functions. The spin
dynamics of a non-equilibrium quantum dot has been studied by using a Majorana
fermion representation~\cite{Mao-03,ShnirmanMakhlin03}, which yields the
qualitative low-frequency properties of the correlation functions.  The
transverse susceptibility in a Kondo model was studied by Paaske et
al.~\cite{Paaske-04prb2} using non-equilibrium perturbation theory together
with a pseudo-fermion representation of the Kondo spin. They showed that the
Fourier transform of the transverse susceptibility possesses a peak if its
frequency equals the value of the applied magnetic field, $\Omega\approx h_0$,
and that the width of this peak is given by the transverse spin relaxation
rate $\tilde{\Gamma}_2$. Their derivation was, however, restricted to either
the regime $h_0\ll\tilde{\Gamma}_2$ or
$\max\{|\Omega-h_0|,\tilde{\Gamma}_2\}\ll\max\{h_0,V\}$, where $V$ denotes the
applied voltage. Very recently, Fritsch and
Kehrein~\cite{FritschKehrein09ap,FritschKehrein09} applied the flow-equation
method to study the longitudinal correlation function as well as the
magnetization and T-matrix in a Kondo model in and out of equilibrium. The
numerical solution of the two-loop scaling equations allowed them to study the
correlation function for all combinations of the parameters voltage,
temperature, and magnetic field, provided the weak-coupling condition (i.e.
the presence of a large enough infrared cutoff) was satisfied. In particular,
these numerical solutions were used to compare the effects of an applied
voltage and a finite temperature, revealing qualitative differences such as
the appearance of Kondo splitting in the non-equilibrium situation. In
general, however, this method cannot provide analytic expressions for the line
shape.

In this article we will generalize the real-time renormalization group method
in frequency space~\cite{Schoeller09} to allow the calculation of dynamical
correlation functions of arbitrary dot operators in systems describing spin
and/or orbital fluctuations. In this setting the quantum dot is coupled to
non-interacting leads which are held at different chemical potentials. The
derived two-loop RG equations can be solved analytically in the weak-coupling
regime. Explicitly, we calculate the longitudinal and transverse spin-spin
correlation and response functions in a two-lead Kondo model in a magnetic
field $h_0$ up to order $J_c^2$. Here $J_c$ denotes the effective coupling at
the energy scale $\Lambda_c=\max\{V,h_0\}$ where the flow of the coupling
constant is cut-off.  In order to satisfy the weak-coupling condition $J_c\ll
1$ either the applied voltage or the magnetic field have to be sufficiently
large compared to the Kondo temperature, $\Lambda_c\gg T_K$, where $T_K$ is
the scale where the system enters the strong-coupling regime.  We note that
the applied formalism does not rely on a fermionic representation of the Kondo
spin but rather deals with its matrix representation in Liouville space
directly.  The longitudinal response function possesses a peak at the spin
relaxation rate $\tilde{\Gamma}_1$, which gets suppressed in a finite magnetic
field.  Interestingly, in the case of a strong magnetic field, $V<\tilde{h}$
where $\tilde{h}=(1-J_c+\ldots)h_0$ denotes the renormalized magnetic field
(see Eq.~\eqref{eq:tildeh} for the precise value), the longitudinal
correlation and response function show ``kink-like'' structures at the
frequencies $\Omega=\tilde{h},\tilde{h}\pm V$, which were also observed using
the flow-equation method~\cite{FritschKehrein09}.  Here we additionally
provide the line shape close to these ``kinks'' and show that the real part of
the response functions shows characteristic logarithmic features at
$\Omega=\tilde{h},\tilde{h}\pm V$.  Furthermore we study the longitudinal and
transverse fluctuation-dissipation ratios. As expected these ratios show a
revival of the fluctuation-dissipation theorem~\cite{CallenWelton51,Landau5}
provided the applied voltage is small compared to the magnetic field or the
considered oscillation frequency.

This article is organized as follows. In the next two sections we will define
the general set up we want to study and define the used notations. This will
include the Kondo model, the notion of Liouville operators as well as the
definition of the symmetrized correlation function and susceptibility. In
Sec.~\ref{sec:PE} we will then derive perturbative expansions for the kernels
needed to calculate these correlation functions. This will be performed in
Liouville space; the expansion is done in powers of the exchange coupling
between the dot and the reservoirs. These perturbative expansions can be
applied to any model describing spin and/or orbital fluctuations as well as
correlation functions of arbitrary operators. In Sec.~\ref{sec:genRG} we will
use these results to derive the RG equations for the kernels of pure dot
operators. In the following section we will further specialize to the two-lead
Kondo model, where we use the explicit expressions of the Liouville operator
and the coupling between the dot and the reservoirs to derive analytic results
for the effective kernels appearing in the correlation functions.  Finally,
these expressions for the kernels are used in Secs.~\ref{sec:CF}
and~\ref{sec:TCF} to calculate the longitudinal and transverse correlation and
response functions.

\section{Kondo model}\label{sec:Kondo}
The real-time renormalization group in frequency space was applied in
Ref.~\onlinecite{SchoellerReininghaus09} to calculate various quantities
including the spin relaxation and dephasing rates, the renormalized magnetic
field, the magnetization and the current in the anisotropic Kondo model in a
finite magnetic field out of equilibrium.  In this reference all notations
which we will use in the following were originally set up.  In order to
increase the readability of the present manuscript we will briefly recall the
basic formulas and notations.

We consider a quantum dot with fixed charge in the Coulomb blockade regime
coupled to external reservoirs. As shown in detail in
Ref.~\onlinecite{Korb-07}, a standard Schrieffer-Wolff transformation leads to
a Hamiltonian of the form
\begin{equation}
\label{H_total}
H\,=\,H_{res}\,+\,H_S\,+\,V\,=\,H_0\,+\,V,
\end{equation}
where $H_{res}$ is the reservoir part, $H_S$ characterizes the isolated
quantum dot, and $V$ describes the coupling between reservoirs and quantum
dot. They are given explicitly by
\begin{eqnarray}
\label{H_res}
H_{res}\,&=&\,\sum_{\nu\equiv \alpha\sigma\dots}
\int d\omega\,(\omega+\mu_\alpha)\,a_{+\nu}(\omega)a_{-\nu}(\omega),
\label{H_S}\\
H_S\,&=&\,\sum_s\,E_s\,|s\rangle\langle s|,\\
\nonumber
V\,&=&\,{1\over 2}\,\sum_{\eta\eta'}\sum_{\nu\nu'}\int d\omega \int d\omega'
\,g_{\eta\nu,\eta'\nu'}(\omega,\omega')\\*
&&\hspace{20mm}\,
\times 
:a_{\eta\nu}(\omega)\,a_{\eta'\nu'}(\omega'):.\label{coupling}
\end{eqnarray}
Here, $a_{\eta\nu}$ are the fermionic creation ($\eta=+$) and annihilation
($\eta=-$) operators in the reservoirs and $\nu$ is an index characterizing
all quantum numbers of the reservoir states, which contains the reservoir
index $\alpha\equiv L,R\equiv \pm$ and the spin quantum number $\sigma\equiv
\uparrow,\downarrow\equiv\pm$.  We measure the energy $\omega$ of the
reservoir states relative to the chemical potential $\mu_\alpha$ of reservoir
$\alpha$. The eigenstates and eigenenergies of the isolated quantum dot are
denoted by $\ket{s}$ and $E_s$. The interaction $V$ is quadratic in the
reservoir field operators, which arises from second order processes of one
electron hopping on and off the quantum dot coherently. This keeps the charge
fixed and allows only spin/orbital fluctuations. The coupling vertex
$g_{\eta\nu,\eta'\nu'}(\omega,\omega')$ is an arbitrary operator acting on the
dot states.  It is written in its most general form, depending on the quantum
numbers and energies of the reservoir states in an arbitrary way.  As
explained in Ref.~\onlinecite{Schoeller09}, the RG approach can be set up in
its most convenient form if one assumes that the frequency dependence of the
initial vertices is rather weak and varies on the scale of the band width $D$
of the reservoirs. For the model we have in mind, the isotropic
spin-$\tfrac12$ Kondo model in a magnetic field, this is certainly satisfied.
Therefore, we will assume this in the following and introduce below [see
Eq.~(\ref{contraction})] a convenient cutoff function into the free reservoir
Green's functions.

To achieve a more compact notation for all indices, we write $1\equiv
\eta\nu\omega$ and sum (integrate) implicitly over all indices 
(frequencies). The interaction is then written in the compact form
\begin{equation}
\label{spin_orbital}
V\,=\,{1\over 2}\,g_{11'}\,:a_1 a_{1'}:.
\end{equation}
$:\dots:$ denotes normal-ordering of the reservoir field operators, meaning
that no contraction is allowed between reservoir field operators within the
normal-ordering.  Within the normal-ordering of Eq.~(\ref{spin_orbital}), the
field operators can be arranged in an arbitrary way (up to a fermionic sign),
therefore the coupling vertex can always be chosen such that antisymmetry
holds:
\begin{equation}
\label{antisymmetry}
g_{11'}\,=\,-\,g_{1'1}.
\end{equation}
Furthermore, due to the hermiticity of $V$, the vertex has the property
\begin{equation}
\label{hermiticity}
g_{11'}^\dagger\,=\,g_{\bar{1}'\bar{1}},
\end{equation}
where $\bar{1}\equiv-\eta,\nu,\omega$.

\begin{figure}[t]
  \includegraphics[scale=0.18]{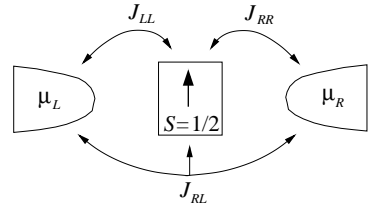}
  \caption{Isotropic spin-1/2 Kondo model coupled via exchange
    to two reservoirs. $J_{LL}$ and $J_{RR}$ involve exchange between the
    electron spins of the left/right reservoir and the local spin,
    $J_{RL}=J_{LR}$ transfers an electron from one reservoir to the other
    during the exchange process. We assume that the Kondo model was derived
    from an Anderson impurity model via a Schrieffer-Wolff transformation,
    which implies the relation $J_{LL}\,J_{RR}=J_{RL}^2$.}
\label{fig:Kondomodel}
\end{figure}
The specific model we want to study is the isotropic Kondo model in an
external magnetic field $h_0>0$ (see Fig.~\ref{fig:Kondomodel}). In this case
the above relations read explicitly
\begin{eqnarray}
\label{H_S_kondo}
\hspace{-0.5cm}
H_S&=&h_0\,S^z,\\
\label{g_kondo}
\hspace{-0.5cm}
g_{11'}&=&{1\over 2}
\left\{
\begin{array}{ll}
(J_{\alpha\alpha'})_0
S^i\sigma^i_{\sigma\sigma'}
& \text{for $\eta=-\eta'=+$} \\[2mm]
-(J_{\alpha'\alpha})_0
S^i\sigma^i_{\sigma'\sigma}
& \text{for $\eta=-\eta'=-$}
\end{array}
\right.\,,
\end{eqnarray}
where $i\in\{x,y,z\}$, $S^i$ is the $i$-component of the spin-$\tfrac12$
operator of the quantum dot, $\sigma^i$ is a Pauli matrix, and
$(J_{\alpha\alpha'})_0$ are the initial exchange couplings. We will be
interested in the antiferromagnetic model here, i.e., we assume
$(J_{\alpha\alpha'})_0>0$ initially. If one derives the Kondo model via a
Schrieffer-Wolff transformation from an Anderson impurity model (see, e.g.,
Ref.~\onlinecite{Korb-07}), one further finds
\begin{equation}
\label{J_form}
(J_{\alpha\alpha'})_0\,=
\,2\sqrt{x_\alpha x_{\alpha'}}\,J_0,\quad
\sum_\alpha\,x_\alpha\,=\,1.
\end{equation}
Although the general formalism and many of the following formulas are also
valid for an arbitrary number of reservoirs, we will consider the case of two
reservoirs only with chemical potentials given by
\begin{equation}
\mu_L={V\over 2},\quad \mu_R=-{V\over 2},
\label{eq:voltagedef}
\end{equation}
where $V$ is the applied voltage which we assume to be positive, $V>0$.

A contraction is defined with respect to a grand-canonical distribution of the
reservoirs, given by
\begin{equation}
\label{contraction}
{a_1\,a_{1'}
  \begin{picture}(-20,11) 
    \put(-22,8){\line(0,1){3}}
    \put(-22,11){\line(1,0){12}}
    \put(-10,8){\line(0,1){3}}
  \end{picture}
  \begin{picture}(20,11) 
  \end{picture}
}
\,\equiv\,
\langle a_1 a_{1'}\rangle_{\rho_{res}}
\,=\,\delta_{1\bar{1}'}\,\rho(\omega)\,f_\alpha(\eta\omega).
\end{equation}
$f_\alpha(\omega)=(e^{\omega/T_\alpha}+1)^{-1}=1-f_\alpha(-\omega)$ is the
Fermi distribution function corresponding to temperature $T_\alpha$ (note that
the chemical potential does not enter this formula since $\omega$ is measured
relative to $\mu_\alpha$). Furthermore,
$\delta_{11'}\equiv\delta_{\eta\eta'}\delta_{\nu\nu'}\delta(\omega-\omega')$
is the $\delta$-function in compact notation.  Furthermore, we have introduced
the cutoff by the band width $D$ into the reservoir contraction via the
density of states
\begin{equation}
\label{band_cutoff}
\rho(\omega)\,=\,{D^2\over D^2 + \omega^2}.
\end{equation}

In order to calculate the dynamical spin-spin correlation functions we have to
know the time evolution of the density matrix $\rho(t)$.  Formally, this
follows from the solution of the von Neumann equation
\begin{eqnarray}
\rho(t)&=&e^{-\ii H(t-t_0)}\,\rho(t_0)\,e^{\ii H(t-t_0)}\nonumber\\
&=&e^{-\ii L(t-t_0)}\,\rho(t_0),
\label{rhoformalL}
\end{eqnarray}
where
\begin{equation}
L\,=\,\comm{H}{.}_-,
\end{equation}
is the Liouvillian acting on usual operators in Hilbert space via the
commutator.  The form \eqref{H_total} of the Hamiltonian yields a similar
decomposition of the Liouvillian,
\begin{equation}
  L=L_{res}+L_S^{(0)}+L_V,
\end{equation}
with $L_{res}=\comm{H_{res}}{.}_-$, $L_S^{(0)}=\comm{H_S}{.}_-$, and
$L_V=\comm{H_V}{.}_-$. We would like to note that the concept of Liouville
space and superoperators have been used in various contexts, for example in
quantum statistical mechanics~\cite{FickSauermann,Lawrie94}.

Initially, we assume that the density matrix is a product of an arbitrary dot
part $\rho_S(t_0)$ and a grandcanonical distribution $\rho_{res}$ for the
reservoirs,
\begin{equation}
  \rho(t_0)=\rho_S(t_0)\,\rho_{res}.
  \label{eq:rhoinitial}
\end{equation}
Furthermore, we introduce the Laplace transform
\begin{equation}
  \tilde{\rho}(z)=\int_{t_0}^\infty dt\,e^{\ii z(t-t_0)}\,\rho(t)
  ={\ii\over z-L}\,\rho(t_0),
\label{laplace_rho}
\end{equation}
where we will frequently use the notation $z=E+\ii\omega$. The stationary
density matrix is defined as
\begin{equation}
  \rho^{st}=\lim_{t\rightarrow\infty} \rho(t)
  =\lim_{t_0\rightarrow-\infty} \rho(t),
  \label{eq:stationaryrhodef}  
\end{equation}
which is understood in the sense
$\mbox{Tr}(O\rho^{st})=\lim_{t\rightarrow\infty}\mbox{Tr}(O\rho(t))$ for any
local operator $O$, and can be calculated using 
\begin{equation}
  \rho^{st}=-\ii\lim_{z\rightarrow\ii 0+}\,z\,\tilde{\rho}(z)
  =\lim_{z\rightarrow\ii 0+}\frac{z}{z-L}\rho(t_0).
  \label{eq:stationaryrho}
\end{equation}
The existence of a stationary state was proven in
Ref.~\onlinecite{DoyonAndrei06} using non-equilibrium pertubation theory to
all orders as well as in Ref.~\onlinecite{Schoeller09} using the RTRG-FS,
which in particular clarified the generation of the relaxation and dephasing
rates under the RG flow. The reduced density matrix of the dot is obtained by
tracing out the reservoir degrees of freedom
\begin{equation}
  \tilde{\rho}_S(z)=\mbox{Tr}_{res}\,\tilde{\rho}(z)
  ={\ii\over z-L_S^{eff}(z)}\,\rho_S(t_0),
\end{equation}
where $L_S^{eff}(z)$ denotes the effective Liouvillian of the quantum dot
formally defined in \eqref{L_eff} below. The stationary reduced density matrix
can then be obtained similar to \eqref{eq:stationaryrho},
\begin{equation}
  \rho_S^{st}=\lim_{t\rightarrow\infty} \rho_S(t)=
  \lim_{z\rightarrow \ii 0+}\frac{z}{z-L_S^{eff}(z)}\,\rho_S(t_0).
  \label{eq:pstdef}
\end{equation}

\section{Correlation functions}
The quantities of interest in this article are the two-point correlation
function of two operators $A$ and $B=A^\dagger$ as well as their dynamical
susceptibility with respect to the steady state,
\begin{eqnarray}
  S_{AB}(t)\!&=&\!\frac{1}{2}\Bigl\langle
  \comm{A(t)_\mathrm{H}\!-\!\bigl\langle A\bigr\rangle_{st}}
  {B(0)_\mathrm{H}\!-\!\bigl\langle B\bigr\rangle_{st}}_+\Bigr\rangle_{st},
  \qquad  \label{eq:defPhi}\\
  \chi_{AB}(t)\!&=&\!\ii\,\Theta(t)\Bigl\langle
  \comm{A(t)_\mathrm{H}}{B(0)_\mathrm{H}}_-\Bigr\rangle_{st}.
  \label{eq:defchi}
\end{eqnarray}
where
\begin{equation}
  \bigl\langle O\bigr\rangle_{st}=\lim_{t_0\rightarrow-\infty}
  \mbox{Tr}\Bigl(O\,e^{\ii L t_0}\,\rho(t_0)\Bigr)
  =\lim_{t_0\rightarrow-\infty}
  \mbox{Tr}\Bigl(O\,\rho(0)\Bigr).
\end{equation}
Here the trace is taken over the dot states as well as the reservoir degrees
of freedom, $\mbox{Tr}=\mbox{Tr}_S\,\mbox{Tr}_{res}$. The time-evolution of
the operators in the Heisenberg picture is given by
\begin{equation}
  A(t)_\mathrm{H}=e^{\ii H t}\,A\,e^{-\ii H t}=e^{\ii L t}\,A.
  \label{eq:timedependenceA}
\end{equation}
Instead of calculating \eqref{eq:defPhi} and \eqref{eq:defchi} in real time
we will study their respective Fourier transforms
\begin{eqnarray}
  S_{AB}(\Omega)&=&
  \int_{-\infty}^\infty dt\,e^{\ii\Omega t}\,S_{AB}(t),
  \label{eq:PhiFT}\\
  \chi_{AB}(\Omega)&=&
  \int_{-\infty}^\infty dt\,e^{\ii\Omega t}\,\chi_{AB}(t),
  \label{eq:chiFT}
\end{eqnarray}
where $\Omega=\Omega\pm\ii\delta$ for $t>0$ ($t<0$). The susceptibility admits
the standard decomposition $\chi_{AB}(\Omega)=\chi_{AB}'(\Omega)+
\ii\,\chi_{AB}''(\Omega)$.

In order to calculate $S_{AB}(\Omega)$ and $\chi_{AB}(\Omega)$ we introduce
the auxiliary correlation functions 
\begin{equation}
  C^\pm_{AB}(\Omega)=
  \int_{-\infty}^0 dt\,e^{-\ii\Omega t}\,\Bigl\langle
  \comm{A(0)_\mathrm{H}}{B(t)_\mathrm{H}}_\pm\Bigr\rangle_{st}
  \label{eq:defbarCOmega}
\end{equation}
with $\Omega=\Omega+\ii\delta$. Its relations to the correlation functions are
given by (see App.~\ref{sec:Crel})
\begin{eqnarray}
  S_{AB}(\Omega)&=&Re\,C^+_{AB}(\Omega)
  -2\pi\,\bigl\langle A\bigr\rangle_{st}\bigl\langle B\bigr\rangle_{st}\,
  \delta(\Omega),\quad\label{eq:rel1}\\
  \chi_{AB}(\Omega)&=&\ii\,C^-_{AB}(\Omega).\label{eq:rel2}
\end{eqnarray}
The static susceptibility is related to the dynamical susceptibility via
\begin{equation}
  \chi_{AB}=\frac{\partial M}{\partial h_0}=
  -\lim_{\Omega\rightarrow 0}\,\chi_{AB}'(\Omega),
  \label{eq:staticchi}
\end{equation}
where $M=\langle S^z\rangle_{st}$ denotes the magnetization.

Some general properties of the correlation functions
can be obtained by considering their spectral representations. Let
$\{\ket{n}\}$ be a complete set of basis states of the full Hamiltonian $H$,
i.e.  $H\ket{n}=E_n\ket{n}$.  Furthermore, the stationary density matrix
$\rho^{st}$ satisfies $\comm{H}{\rho^{st}}_-=0$, thus the basis states
$\ket{n}$ can be chosen such that
\begin{equation}
  \bra{n}\rho^{st}\ket{m}=\rho_n^{st}\,\delta_{nm}.
\end{equation}
Using this one easily verifies the spectral representations
\begin{eqnarray}
  C^\pm_{AB}(\Omega)&=&\sum_{mn}(\rho_n^{st}\pm\rho_m^{st})\,
  \bra{n}A\ket{m}\,\bra{m}B\ket{n}\nonumber\\*
  & &\!\!\!\!\!\!\!\!\!\!\!\!\!\!\!\!\!\!\!\!
  \times\left(\pi\delta(\Omega+E_n-E_m)+
    \ii\frac{\text{P}}{\Omega+E_n-E_m}\right).\quad
  \label{eq:spectralbarC}
\end{eqnarray}
These relations imply $S_{AA^\dagger}(\Omega)\ge 0$ as well as 
$S_{BA}(\Omega)=S_{AB}(-\Omega)$, $\chi_{BA}'(\Omega)=\chi_{AB}'(-\Omega)$,
and $\chi_{BA}''(\Omega)=-\chi_{AB}''(-\Omega)$.
In equilibrium the matrix elements of the density matrix are given by
$\rho_n=e^{-E_n/T}/Z$ with the partition sum $Z$, which implies the well-known
fluctuation-dissipation theorem~\cite{CallenWelton51,Landau5}
\begin{equation}
  \chi_{AA^\dagger}''(\Omega)=\tanh\frac{\Omega}{2T}\,
  S_{AA^\dagger}(\Omega)
  \label{eq:FDT}
\end{equation}
as well as $\Omega\,\chi_{AA^\dagger}''(\Omega)\ge 0$.

\section{Perturbative expansion for the correlation functions}\label{sec:PE}
In this section we derive a perturbative expansion in Liouville space for the
auxiliary correlation functions $C^\pm_{AB}(\Omega)$, which will serve
as the starting point for the derivation of the RG equations below. A similar
perturbative expansion for the effective Liouvillian of the quantum dot
$L_S^{eff}$ has been derived in
Refs.~\onlinecite{Schoeller09,SchoellerReininghaus09}. We will generalize
these results to $C^\pm_{AB}(\Omega)$ while closely following the
presentation of Ref.~\onlinecite{SchoellerReininghaus09}.

As starting point to set up the formalism we assume that the operators $A$ and
$B$ admit a representation similar to \eqref{spin_orbital}, 
\begin{equation}
  A=\frac{1}{m!}\,a_{1\ldots m}\,:a_1\ldots a_m:,\quad
  B=\frac{1}{n!}\,b_{1\ldots n}\,:a_1\ldots a_n:,
  \label{eq:ABdef}
\end{equation}
where we recall the short-hand notation $1\equiv \eta\nu\omega$ and sum
(integrate) implicitly over all indices (frequencies). We further assume the
operators $A$ and $B$ to be bosonic which implies $m$ and $n$ to be even.
Eventually we will be concerned with the correlation functions of the spin
operators on the dot, i.e. $A,B=S^+,S^-,S^z$. In this case the operators do
not couple dot and reservoir degrees of freedom and hence only the terms with
$m=n=0$ are non-vanishing. However, we will keep the general forms
\eqref{eq:ABdef} throughout this section, which for example include the case
of current operators~\cite{Schoeller09,SchoellerReininghaus09} where $m=n=2$.

In order to set up the perturbative expansions in Liouville space we define
the operators
\begin{equation}
  L_A=\frac{\ii}{2}\comm{A}{.}_+,\quad
  L_B^\pm=\ii\comm{B}{.}_\pm.
\end{equation}
Then using \eqref{eq:timedependenceA} together with \eqref{rhoformalL} we
obtain after some algebra
\begin{equation}
  \begin{split}
  C^\pm_{AB}(\Omega)=&(-\ii)^2\lim_{t_0\rightarrow -\infty}\,
  \int_{-\infty}^0 \,dt\,e^{-\ii\Omega t}\\
  &\times\mbox{Tr}\left(L_A\,e^{\ii Lt}\,L_B^\pm\,e^{-\ii L(t-t_0)}\,
    \rho(t_0)\right).
  \end{split}
  \label{eq:CAB1}
\end{equation}
In the next step we use \eqref{eq:stationaryrhodef} and
\eqref{eq:stationaryrho} and furthermore perform the Laplace transform
$t\rightarrow\Omega$ in \eqref{eq:CAB1} to obtain
\begin{equation}
  C^\pm_{AB}(\Omega)=-\ii\lim_{\xi\rightarrow\ii 0+}\,
  \mbox{Tr}\left(L_A\,\frac{1}{\Omega-L}\,L_B^\pm\,\frac{\xi}{\xi-L}\,
    \rho(t_0)\right).
  \label{eq:CAB2}
\end{equation}
Here we have used $\Omega=\Omega+\ii\delta$ to ensure convergence of the
integral. The limit $\xi\rightarrow \ii 0+$ has to be taken before
$\delta\rightarrow 0$ in order to reach the stationary state.

The next step is to expand the expression \eqref{eq:CAB2} in the interacting
part $L_V$ of the Liouvillian and to integrate out the reservoir part. This
procedure was outlined for the reduced density matrix $\tilde{\rho}_S(z)$ in
detail in Ref.~\onlinecite{Schoeller09}; we will generalize this to the case
of the auxiliary correlation functions \eqref{eq:CAB2} here. First, using
$L=L_0+L_V$ with $L_0=L_{res}+L_S^{(0)}$, \eqref{eq:CAB2} can be formally
expanded in $L_V$,
\begin{equation}
  \begin{split}
  C^\pm_{AB}(\Omega)=&-\ii\!\lim_{\xi\rightarrow\ii 0+}\!
  \sum_{k,l=0}^\infty\!\mbox{Tr}\Biggl[L_A\,\frac{1}{\Omega-L_0}
    \left(L_V\,\frac{1}{\Omega-L_0}\right)^k\\*
    &\times L_B^\pm\,\frac{\xi}{\xi-L_0}\,
    \left(L_V\,\frac{1}{\xi-L_0}\right)^l\,\rho(t_0)\Biggr].
  \label{eq:CAB3}
  \end{split}
\end{equation}
Second, in order to integrate out the reservoir degrees of freedom we write
$L_V$ in the form
\begin{equation}
\label{coupling_product}
L_V\,=\,{1\over 2}\,p'\,G^{pp'}_{11'}\,
:J^{p}_1 J^{p'}_{1'}:,
\end{equation}
where we implicitly sum (integrate) over $1=\eta\nu\omega$ as well as
$p,p'=\pm$.  $J^p_1$ is a quantum field superoperator in Liouville space for
the reservoirs, defined by ($C$ is an arbitrary reservoir operator)
\begin{equation}
\label{liouville_field_operators}
J_1^p\,C \,=\,
\left\{
\begin{array}{cl}
a_1\,C\, &\mbox{for }p=+ \\
C\,a_1\, &\mbox{for }p=-
\end{array}
\right..
\end{equation}
Here $p=\pm$ serves as an auxiliary index which is similar to the Keldysh
index indicating whether the field operator is acting on the upper or the
lower part of the Keldysh contour. $G^{pp'}_{11'}$ is a superoperator acting
in Liouville space of the quantum dot, and is defined by ($C$ is an arbitrary
operator on the quantum dot)
\begin{equation}
\label{G_vertex_liouville}
G^{pp'}_{11'}\,C\,=\,
\delta_{pp'}\,
\left\{
\begin{array}{cl}
g_{11'}\,C\, &\mbox{for }p=+ \\
-C \,g_{11'}\, &\mbox{for }p=-
\end{array}
\right..
\end{equation}
In the same way we define 
\begin{eqnarray}
  L_A&=&\frac{1}{m!}\,\sigma^{p_1\ldots p_m}\,
  \mathcal{A}^{p_1\ldots p_m}_{1\ldots m}\,:J_1^{p_1}\ldots J_m^{p_m}:,
  \label{eq:LAdef}\\
  L_B^\pm&=&\frac{1}{n!}\,\sigma^{p_1\ldots p_n}\,
  \mathcal{(B_\pm)}^{p_1\ldots p_n}_{1\ldots n}\,:J_1^{p_1}\ldots J_n^{p_n}:,
  \label{eq:LBdef}
\end{eqnarray}
where the dot superoperators $\mathcal{A}^{p_1\ldots p_m}_{1\ldots m}$ and
$\mathcal{(B_\pm)}^{p_1\ldots p_n}_{1\ldots n}$ act on arbitrary dot operators
$C$ as 
\begin{eqnarray}
\mathcal{A}^{p_1\ldots p_m}_{1\ldots m}\,C&=&
\frac{\ii}{2}\,\delta_{p_1p_2}\ldots\delta_{p_1p_m}\nonumber\\*
& &\quad\times\left\{
\begin{array}{ll}
a_{1\ldots m}\,C,& p_1=+ \\
C\,a_{1\ldots m},& p_1=-
\end{array}
\right.,\\
\mathcal{(B_\pm)}^{p_1\ldots p_n}_{1\ldots n}\,C&=&\ii\,
\delta_{p_1p_2}\ldots\delta_{p_1p_n}\nonumber\\*
& &\quad\times\left\{
\begin{array}{ll}
b_{1\ldots n}\,C,& p_1=+ \\
\pm C\,b_{1\ldots n},& p_1=-
\end{array}
\right..
\end{eqnarray}
For $m=0$ or $n=0$ we define
$\mathcal{A}C=\frac{\ii}{2}\comm{a}{C}_+$ and
$\mathcal{B_\pm}C=\ii\comm{b}{C}_\pm$, respectively.  As we
consider only bosonic operators $A$ and $B$ which change the number of
fermions by an even integer, the sign-superoperator is given by
\begin{equation}
  \sigma^{p_1\ldots p_n}=p_2\,p_4\cdots p_n.
\end{equation}
This operator was introduced to compensate additional signs due to
interchanges of fermionic reservoir field operators as explained in detail in
Ref.~\onlinecite{Schoeller09}.

\begin{figure}[t]
  \psfrag{Avertex}{$\mathcal{A}_{12}$}
  \psfrag{Bvertex}{$\mathcal{B}_{\pm,78}$}
  \includegraphics[scale=0.28]{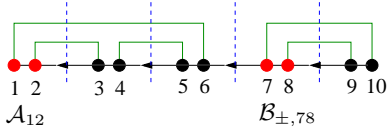}
  \caption{(color online) Example of a diagram contributing to
    the auxiliary correlation function $C^\pm_{AB}(\Omega)$. The time
    direction is to the left. Each vertex $G$ is represented by two adjacent
    black dots indicating the two reservoir field operators associated with
    each vertex.  The vertices $\mathcal{A}$ and $\mathcal{B}$ are represented
    in the same way by red dots. For this example we have chosen $m=n=2$ in
    \eqref{eq:ABdef} as it would be the case for a current-current correlation
    function. The black horizontal lines connecting the vertices denote the
    free time propagation of the quantum system, leading to the resolvents
    ${1\over E+X_i-L_S^{(0)}}$ in Laplace space. The green lines are the
    reservoir contractions arising from the application of Wick's theorem. The
    vertical blue lines between the vertices are auxiliary lines to determine
    the energy argument $X_i$ of the resolvents.}
\label{fig:diagram_example}
\end{figure}

\begin{figure}[t]
  \psfrag{Avertex}{$\mathcal{A}_{12}$}
  \psfrag{Bvertex}{$\mathcal{B}_{\pm,78}$}
  \includegraphics[scale=0.28]{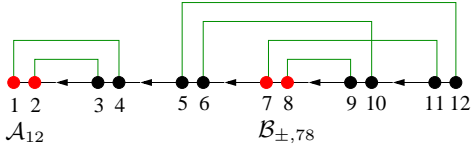}
  \caption{(color online) Example of a diagram contributing to 
    \eqref{eq:barCABterm1}.}
\label{fig:block1example}
\end{figure}

\begin{figure}[t]
  \psfrag{Avertex}{$\mathcal{A}_{12}$}
  \psfrag{Bvertex}{$\mathcal{B}_{\pm,78}$}
  \includegraphics[scale=0.28]{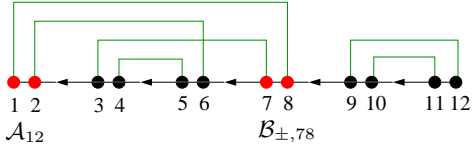}
  \caption{(color online) Example of a diagram contributing to 
    \eqref{eq:barCABterm2}.}
\label{fig:block2example}
\end{figure}

Inserting the representations (\ref{coupling_product}) as well as
\eqref{eq:LAdef} and \eqref{eq:LBdef} into \eqref{eq:CAB3} and shifting all
reservoir field superoperators $J_1^p$ to the right using
\begin{equation}
  J_1^p\,L_{res}=(L_{res}-x_1)\,J_1^p,
\end{equation}
where we have introduced the short-hand notation
$x_i=\eta_i(\omega_i+\mu_{\alpha_i})$, one can show\cite{Schoeller09} that
each term of perturbation theory can be written as a product of a dot part and
an average over a sequence of field superoperators of the reservoirs with
respect to $\rho_{res}$.  Evaluating the latter with the help of Wick's
theorem, one can represent each term of the Wick decomposition by a diagram
(see Fig.~\ref{fig:diagram_example} for an example) describing a certain
process contributing to the auxiliary correlation function
$C^\pm_{AB}(\Omega)$.  Each process consists of a sequence of
interaction vertices $G_{11'}^{pp'}$ between the dot and the reservoirs, and a
free time propagation of the dot in between (leading to resolvents in Laplace
space). Since the reservoirs have been integrated out, the vertices are
connected by reservoir contractions (the green lines in
Fig.~\ref{fig:diagram_example}).  This means that the various diagrams
represent terms for the effective time evolution of the dot in the presence of
dissipative reservoirs.
Each diagram for the auxiliary correlation function has the form
\begin{eqnarray}
\nonumber
C^\pm_{AB}(\Omega)&\,\rightarrow\,&
-{\ii \over S}\, 
(-1)^{N_p}\,\left(\prod\gamma\right)
\lim_{\xi\rightarrow\ii 0+}\xi\\*
\nonumber && \hspace{-1.5cm}
\mathcal{A}\,{1\over \Omega+X_1-L_S^{(0)}}\,G\,{1\over \Omega+X_2-L_S^{(0)}}\,
\,G\,\dots\,\mathcal{B}_\pm\\*
\label{value_diagram}
&& \hspace{-1.5cm}
\ldots \,G\,
{1\over \xi+X_r-L_S^{(0)}}\,G\,{1\over \xi-L_S^{(0)}}\,\rho_S(t_0),
\end{eqnarray}
where $L_S^{(0)}=\comm{H_S}{.}_-$, $G\equiv G_{ij}^{p_i p_j}$ indicates an
interaction vertex, and $\gamma\equiv\gamma^{p_i p_j}_{ij}$ is a contraction
between the reservoir field superoperators, defined by
\begin{eqnarray}
\nonumber
\gamma_{11'}^{pp'}\,&=&\,{J_1^p\,J_{1'}^{p'}
  \begin{picture}(-20,11) 
    \put(-22,8){\line(0,1){3}}
    \put(-22,11){\line(1,0){12}} 
    \put(-10,8){\line(0,1){3}}
  \end{picture}
  \begin{picture}(20,11) 
  \end{picture}
}
\,=\,p'\,\mbox{Tr}_{res}\,J_1^p J_{1'}^{p'}\,\rho_{res}\\
\label{liouville_contraction}
\,&=&\,\delta_{1\bar{1}'}\,\rho(\omega)\,p'\,f_\alpha(\eta p' \omega).
\end{eqnarray}
We stress that only the initial reduced density matrix of the dot
$\rho_S(t_0)$ defined in \eqref{eq:rhoinitial} appears in
\eqref{value_diagram} as we have already performed the trace over the
reservoir degrees of freedom (and hence $\rho_{res}$) to obtain the reservoir
contractions $\gamma$. To factorize the Wick decomposition, a fermionic sign
has to be assigned to each permutation of reservoir field superoperators,
indicated by the sign factor $(-1)^{N_p}$ in (\ref{value_diagram}).  For each
pair of vertices connected by two reservoir lines, a combinatorial factor
${1\over 2}$ occurs, leading to the prefactor ${1\over S}$ in
(\ref{value_diagram}). The value of the frequencies $X_i$ in the resolvents
between the interaction vertices is determined by the sum over all variables
$x=\eta(\omega+\mu_\alpha)$ of those indices belonging to the reservoir lines
which are crossed by a vertical line at the position of the resolvent (the
blue lines in Fig.~\ref{fig:diagram_example}). Thereby, the index of the left
vertex has to be taken of the corresponding reservoir line.  For example, the
diagram shown in Fig.~\ref{fig:diagram_example} is given by (the obvious
dependence on the Keldysh indices has been omitted for simplicity, i.e.
$\gamma_{ij}\equiv\gamma_{ij}^{p_i p_j}$ and $G_{ij}\equiv G_{ij}^{p_i p_j}$)
\begin{equation}
\begin{split}
&-\ii\lim_{\xi\rightarrow\ii 0+}\xi
\Bigl(\gamma_{16}\gamma_{23}\gamma_{45}\,
\mathcal{A}_{12}\,\Pi_{12}(\Omega)\,G_{34}\,\Pi_{14}(\Omega)\,
G_{56}\Bigr)\\*
&\quad\times{1\over \Omega-L_S^{(0)}}
\left({1\over 2}
\gamma_{7,10}\gamma_{89}\,
\mathcal{B}_{\pm,78}\,\Pi_{78}(\xi)\,G_{9,10}\right)\\*
&\quad\times{1\over \xi-L_S^{(0)}}\rho_S(t_0),
\label{diagram_example}
\end{split}
\end{equation}
where the resolvents are defined by
\begin{equation}
\Pi_{1\dots n}(z)={1\over z_{1\dots n}+\bar{\omega}_{1\dots n}-L_S^{(0)}},
\end{equation}
with
\begin{equation}
\label{E_definition}
z_{1\dots n}=z+\sum_{i=1}^n\,\bar{\mu}_i,
\end{equation}
as well as
\begin{equation}
\bar{\omega}_{1\dots n}=\sum_{i=1}^n\,\bar{\omega}_i,\quad
\bar{\mu}_i=\eta_i\,\mu_{\alpha_i},\quad
\bar{\omega}_i\,=\,\eta_i\,\omega_i.
\label{mu_omega_bar} 
\end{equation}
As can be seen from the example (\ref{diagram_example}), each diagram consists
of a sequence of irreducible blocks (where a vertical line always cuts at
least one reservoir line) and free resolvents $1/(\Omega-L_S^{(0)})$ or
$1/(\xi-L_S^{(0)})$ in between.  Now there are two possibilities: (i) The
vertices $\mathcal{A}$ and $\mathcal{B}$ do not belong to the same block (see
Fig.~\ref{fig:block1example} for an example). (ii) The vertices $\mathcal{A}$
and $\mathcal{B}$ belong to the same block (see Fig.~\ref{fig:block2example}
for an example).
In the first case (i) one can formally resum those terms between the vertices
$\mathcal{A}$ and $\mathcal{B}$ which are not connected to them similar to
Dyson equations with the result
\begin{equation}
\label{reduced_dm}
\frac{1}{\Omega-L_S^{eff}(\Omega)},
\end{equation}
where 
\begin{equation}
\label{L_eff}
L_S^{eff}(\Omega)\,=\,L_S^{(0)}\,+\,\Sigma(\Omega).
\end{equation}
Here the kernel $\Sigma(\Omega)$ contains the sum over all irreducible
diagrams,
\begin{eqnarray}
\nonumber
\Sigma(\Omega)
\,&\rightarrow&\, 
{1\over S} \, (-1)^{N_p} \, \left(\prod\gamma\right)_{irr}\\*
\label{value_kernel}
&& \!\!\!\!\!\!\!\!\!\!\!\!\!\!\!\!\!\!\!\!\!\!\!\!\!\!\!
\times G\,{1\over \Omega+X_1-L_S^{(0)}}\,G\,\dots \,G\,
{1\over \Omega+X_r-L_S^{(0)}}\,G,
\end{eqnarray} 
where the subindex $irr$ indicates that only irreducible diagrams are allowed
where any vertical line between the vertices cuts through at least one
reservoir contraction. We further introduce irreducible blocks
$\Sigma_A(\Omega)$ as well as $\Sigma_B^\pm(\Omega,\xi)$ which are given as the
sum over all irreducible diagrams containing the vertices $\mathcal{A}$ and
$\mathcal{B}_\pm$,
\begin{eqnarray}
\Sigma_A(\Omega)
\,&\rightarrow&\, 
{1\over S} \, (-1)^{N_p} \, \left(\prod\gamma\right)_{irr}\nonumber\\*
\label{value_kernelA}
&& \!\!\!\!\!\!\!\!\!\!\!\!\!\!\!\!\!\!\!\!\!\!\!\!\!\!\!\!\!\!
\times \mathcal{A}\,{1\over \Omega+X_1-L_S^{(0)}}\,G\,\dots \,G\,
{1\over \Omega+X_r-L_S^{(0)}}\,G,\\
\Sigma_B^\pm(\Omega,\xi)
\,&\rightarrow&\, 
{1\over S} \, (-1)^{N_p} \, \left(\prod\gamma\right)_{irr}\nonumber\\*
\label{value_kernelB}
&& \!\!\!\!\!\!\!\!\!\!\!\!\!\!\!\!\!\!\!\!\!\!\!\!\!\!\!\!\!\!
\times G\,{1\over \Omega+X_1-L_S^{(0)}}\,G\,\dots \,G\,
{1\over \Omega+X_r-L_S^{(0)}}\,\mathcal{B}_\pm\nonumber\\*
&& \!\!\!\!\!\!\!\!\!\!\!\!\!\!\!\!\!\!\!\!\!\!\!\!\!\!\!\!\!\!
\times {1\over \xi+X_{r+1}-L_S^{(0)}}\,G\,\dots \,G\,
{1\over \xi+X_{s}-L_S^{(0)}}\,G.
\end{eqnarray} 
Obviously, in the case of spin operators, $A,B\in\{S^+,S^-,S^z\}$, the vertex
$\mathcal{A}$ has no external legs and hence $\Sigma_A(\Omega)=\mathcal{A}$.
In contrast, although the vertex $\mathcal{B}_\pm$ does not possess external
legs either, there exist irreducible diagrams containing $\mathcal{B}_\pm$ and
at least one vertex $G$ to the left and one to the right of $\mathcal{B}_\pm$.
If we now proceed by resumming the irreducible blocks right to (and not
connected to) the vertex $\mathcal{B}_\pm$ similar to \eqref{reduced_dm} and
perform the limit $\xi\rightarrow\ii 0+$ using \eqref{eq:pstdef}, we deduce
that all terms of type (i) contribute to
\begin{equation}
  -\ii\,\mbox{Tr}_S
  \Biggl[\Sigma_A(\Omega)\,\frac{1}{\Omega-L_S^{eff}(\Omega)}\,
  \Sigma_B^\pm(\Omega,\ii 0+)\,\rho_S^{st}\Biggr].
  \label{eq:barCABterm1}
\end{equation}

In the second case (ii) we introduce a kernel similar to \eqref{value_kernelA}
and \eqref{value_kernelB} which contains all irreducible diagrams containing
both vertices $\mathcal{A}$ and $\mathcal{B}$,
\begin{eqnarray}
\Sigma_{AB}^\pm(\Omega,\xi)
\,&\rightarrow&\, 
{1\over S} \, (-1)^{N_p} \, \left(\prod\gamma\right)_{irr}\nonumber\\*
\label{value_kernelAB}
&& \!\!\!\!\!\!\!\!\!\!\!\!\!\!\!\!\!\!\!\!\!\!\!\!\!\!\!\!\!\!\!\!\!\!\!
\times \mathcal{A}\,{1\over \Omega+X_1-L_S^{(0)}}\,G\,\dots \,G\,
{1\over \Omega+X_r-L_S^{(0)}}\,\mathcal{B}_\pm\nonumber\\*
&& \!\!\!\!\!\!\!\!\!\!\!\!\!\!\!\!\!\!\!\!\!\!\!\!\!\!\!\!\!\!\!\!\!\!\!
\times {1\over \xi+X_{r+1}-L_S^{(0)}}\,G\,\dots \,G\,
{1\over \xi+X_{s}-L_S^{(0)}}\,G.
\end{eqnarray} 
In the case of spin operators, $A,B\in\{S^+,S^-,S^z\}$, there exist no
irreducible diagrams connecting $\mathcal{A}$ and $\mathcal{B}_\pm$, hence
$\Sigma_{AB}^\pm(\Omega,\xi)=0$ in this case.  Now using again
\eqref{eq:pstdef} for the sum of the irreducible blocks right to (and not
connected to) the vertex $\mathcal{B}_\pm$ we deduce that all terms of type
(ii) contribute to
\begin{equation}
  -\ii\,\mbox{Tr}_S\Bigl[\Sigma_{AB}^\pm(\Omega,\ii 0+)\,\rho_S^{st}\Bigr].
  \label{eq:barCABterm2}
\end{equation}
Hence, taking together (i) and (ii) we finally arrive at the main result of
this section,
\begin{eqnarray}
C_{AB}^\pm(\Omega)\!\!&=&\!\!
  -\ii\,\mbox{Tr}_S
  \Biggl[\Sigma_A(\Omega)\,\frac{1}{\Omega-L_S^{eff}(\Omega)}\,
  \Sigma_B^\pm(\Omega,\ii 0+)\,\rho_S^{st}\Biggr]\nonumber\\*
  & &\!\!
  -\ii\,\mbox{Tr}_S\Bigl[\Sigma_{AB}^\pm(\Omega,\ii 0+)\,\rho_S^{st}\Bigr],
  \label{eq:barCresult}
\end{eqnarray}
where the kernels are defined by \eqref{value_kernel}, \eqref{value_kernelA},
\eqref{value_kernelB}, and \eqref{value_kernelAB}, respectively.

In addition we note that the diagrammatic series can be partially resummed by
taking all closed sub-diagrams between two fixed vertices together which
contain only contractions connecting vertices between the two fixed ones. This
has the effect that the resolvents in (\ref{value_kernel}),
\eqref{value_kernelA}, \eqref{value_kernelB}, and \eqref{value_kernelAB}, are
replaced by
\begin{equation}
\label{resolvent_replacement}
{1\over \Omega+X_i-L_S^{(0)}}\rightarrow
{1\over \Omega+X_i-L_S^{eff}(\Omega+X_i)}
\end{equation}
(and similar for $\Omega\rightarrow\xi$), i.e., the full effective Liouville
operator occurs in the denominator. In this formulation the number of diagrams
is reduced, i.e. diagrams containing closed sub-diagrams between two vertices
are no longer allowed.

\section{Generic RG equations}\label{sec:genRG}
In this section we will set up the generic RG equations for the kernels
$\Sigma_A(\Omega)$, $\Sigma_B^\pm(\Omega,\xi)$, and
$\Sigma_{AB}^\pm(\Omega,\xi)$, for spin operators on the dot in a model with
spin/orbital fluctuations.  Hence we will assume the form
\eqref{coupling_product} for the coupling between the reservoirs and the
quantum dot, but will keep the vertex $G_{11'}^{pp'}$ arbitrary at this stage.
The derivation will require some relations between the initial Liouvillian
$L_S^{(0)}$, the vertex $\mathcal{B}_\pm$ and the effective dot Liouvillian
$L_S^{eff}$ which can be explicitly checked for the Kondo model to be studied
in the next section but have to be assumed here. These relations are
\eqref{eq:defkappa}, \eqref{eq:assumption1}, \eqref{eq:assumption2}, and
\eqref{eq:commapp}.  The generic RG equations for the vertex $G$ and the
effective Liouvillian $L_S^{eff}$ have been derived and solved in
Ref.~\onlinecite{SchoellerReininghaus09}, we will quote these results without
derivation when they are needed.

Furthermore, we will restrict ourselves to the calculation of the dynamical
spin-spin correlations only, i.e. we will assume $A,B\in\{S^+,S^-,S^z\}$ in
what follows. This implies in particular, that the initial values of the
vertices $\mathcal{A}$ and $\mathcal{B}_\pm$ defined in \eqref{eq:LAdef} and
\eqref{eq:LBdef} do not possess any external lines, i.e. $m=n=0$.
Explicitly, 
\begin{eqnarray}
  \mathcal{A}&=&\frac{\ii}{2}\comm{A}{.}_+,\label{eq:Avertexinitial}\\
  \mathcal{B}_\pm&=&\ii\comm{B}{.}_\pm.\label{eq:Bvertexinitial}
\end{eqnarray}
We will see below that this form of the vertex $\mathcal{A}$ is conserved
under the RG flow. In contrast, a new effective $B$-type vertex
$\mathcal{B}_{\pm,11'}$ with two external lines will be generated.  The fact
that the initial vertex $\mathcal{A}$ has no external lines directly implies
the final results for the kernels \eqref{value_kernelA} and
\eqref{value_kernelAB}, namely
\begin{eqnarray}
  \Sigma_A(\Omega)&=&\mathcal{A},\label{eq:resultSigmaA}\\
  \Sigma_{AB}^\pm(\Omega,\xi)&=&0.
\end{eqnarray}
Hence, in the following we have to consider the kernel \eqref{value_kernelB}
only. We note that the results of this section remain valid for any pure dot
operators $A$ and $B$, i.e. any operators \eqref{eq:ABdef} with $m=n=0$.

The RG procedure is divided into two steps. In the first step we will
integrate out the symmetric part of the reservoir contractions
$\gamma_{11'}^{pp'}$. The reason for this is as follows: The effective
dot Liouvillian $L_S^{eff}(z)$ can be diagonalized as
\begin{equation}
  L_S^{eff}(z)=\sum_i \lambda_i(z)\,P_i(z),
  \label{eq:Ldiagonalization}
\end{equation}
where $\lambda_i(z)$ and $P_i(z)$ denote the eigenvalues and corresponding
projectors, respectively. This diagonalization implies for the resolvents
\begin{equation}
  {1\over z-L_S^{eff}(z)}=\sum_i \frac{1}{z-\lambda_i(z)}\,P_i(z).
  \label{eq:resprojectors}
\end{equation}
Now there exists a zero eigenvalue $\lambda_0(z)=0$ whose eigenstate for
$z\rightarrow \ii 0+$ corresponds to the stationary state. The appearance of
this zero eigenvalue can lead to infrared divergencies of the frequency
integrations in the perturbative expansions for the vertex $G$ and the
effective Liouvillian $L_S^{eff}$ as is elaborated on in detail in
Ref.~\onlinecite{Schoeller09}.  However, after the discrete RG step we can
trivially sum over the Keldysh indices $p$ and $p'$ by introducing
\begin{equation}
  \bar{G}_{11'}=\sum_p G_{11'}^{pp},\quad
  \tilde{G}_{11'}=\sum_p p\,G_{11'}^{pp}.
  \label{eq:barGdef}
\end{equation}
In the resulting RG equations only the symmetric vertex $\bar{G}$ will
appear. This vertex has the important property
\begin{equation}
  P_0(z)\,\bar{G}_{11'}=0,
  \label{eq:proj0G}
\end{equation}
which is independent of the model specifics. Hence, after the discrete RG step
the zero eigenvalue can no longer appear in any resolvent standing left to
$\bar{G}$ (i.e. in no resolvent except the one standing left to the vertex
$\mathcal{B}_\pm$). This resolves the problem of infrared divergent internal
frequency integrations, as the remaining eigenvalues $\lambda_i(z)$ have a
strictly negative imaginary part (see \eqref{eq:lambda1} and
\eqref{eq:lambdapm}). We would like to refer to Ref.~\onlinecite{Schoeller09}
for a general discussion of this topic.

In the second step we introduce a cut-off $\Lambda$ into the
reservoir contractions via the Fermi function. We then integrate out the
reservoirs by sending $\Lambda\rightarrow 0$, which results in a description
of the system in terms of effective dot quantities like $L_S^{eff}$.
This second, continuous, RG step is further divided into two substeps, first
we integrate out the reservoir degrees of freedom down to an energy scale
$\Lambda_c$, and second we complete the flow from $\Lambda_c$ down to
$\Lambda=0$.

\subsection{Discrete RG step}
\begin{figure}[t]
  \psfrag{Bvertex}{$\mathcal{B}_\pm$}
  \includegraphics[scale=0.28]{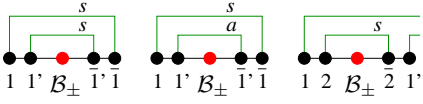}
  \caption{(color online) The lowest order diagrams for the kernel
    $\Sigma_{B}^{\pm,a}(\Omega,\xi)$ (left and middle diagram) and the
    effective vertex $\mathcal{B}_{\pm,11'}^a(\Omega,\xi)$ (right diagram) when
    the symmetric part of the contraction is integrated out.  $s$ ($a$)
    denotes the symmetric (antisymmetric) contraction $\gamma^s$
    ($\gamma^a$).}
\label{fig:discrete_LG}
\end{figure}

In the first discrete RG step we integrate out the symmetric part ${1\over
  2}(f_\alpha(\omega)+f_\alpha(-\omega))={1\over 2}$ of the Fermi function in
the contraction (\ref{liouville_contraction}).  The discrete RG step for the
kernel (\ref{value_kernel}) and the vertex $G$ has been performed in
Ref.~\onlinecite{SchoellerReininghaus09}. Here we will derive the analog
results for the kernel \eqref{value_kernelB} and the vertex
$\mathcal{B}_{\pm,11'}$. This is achieved by decomposing the contraction
(\ref{liouville_contraction}) according to
\begin{eqnarray}
\label{contraction_decomposition}
\gamma_{11'}^{pp'}\,&=&\delta_{1\bar{1}'}\,p'\,\gamma^s_1
\,+\,\delta_{1\bar{1}'}\,\gamma^a_1,\\*
\label{sym_antisym_contraction}
\gamma^s_1\,&=&\,{1\over 2}\,\rho(\bar{\omega}),\quad 
\gamma^a_1\,=\,\rho(\bar{\omega})\,\left[f_\alpha(\bar{\omega})-
  {1\over 2}\right],
\end{eqnarray}
with $\bar{\omega}\,\equiv\,\eta\,\omega$.  Using this decomposition in
\eqref{value_kernelB}, one finds that each diagram decomposes into a series of
blocks which are irreducible with respect to the symmetric part $\gamma^s$
(i.e., any vertical line hits at least one symmetric contraction) and
connected to each other by antisymmetric contractions $\gamma^a$. The blocks
which are irreducible with respect to $\gamma^s$ can be formally resummed into
an effective kernel $\Sigma_{B}^{\pm,a}(\Omega,\xi)$ and a newly generated
effective vertex $\mathcal{B}_{\pm,11'}^a(\Omega,\xi)$. The lowest order 
diagrams are shown in Fig.~\ref{fig:discrete_LG}.
Using the diagrammatic rules together with (\ref{contraction_decomposition})
and the convention (\ref{E_definition}), we obtain for the first two
diagrams
\begin{equation}
\begin{split}
&\gamma_1^s\,({1\over 2}\gamma_{1'}^s+p'\gamma_{1'}^a)
\,G^{pp}_{11'}\,{1\over \Omega_{11'}\,+\,\bar{\omega}_{11'}\,-\,L_S^{(0)}}
\,\mathcal{B}_\pm\\
&\qquad\qquad\times
{1\over \xi_{11'}\,+\,\bar{\omega}_{11'}\,-\,L_S^{(0)}}
\,G^{p'p'}_{\bar{1}'\bar{1}},
\end{split}
\end{equation}
and for the third one (including the interchange $1\leftrightarrow 1'$)
\begin{equation}
\begin{split}
&p'\,\gamma^s_2\,G^{pp}_{12}\,
{1\over \Omega_{12}\,+\,\bar{\omega}_{12}\,-\,L_S^{(0)}}\,\mathcal{B}_\pm\\
&\qquad\qquad\times
{1\over \xi_{12}\,+\,\bar{\omega}_{12}\,-\,L_S^{(0)}}\, 
G^{p'p'}_{\bar{2}1'}\,-\,(1\leftrightarrow 1').
\end{split}
\end{equation}
We use here the original perturbation series (\ref{value_kernelB}) so that the
unperturbed Liouvillian $L_S^{(0)}$ occurs in the resolvents. Performing the
frequency integrations and assuming the band-width to be large, we obtain
\begin{eqnarray}
  \Sigma_{B}^{\pm,a}(\Omega,\xi)&=&\mathcal{B}_\pm
  -\frac{\pi^2}{32}\,\bar{G}_{11'}\,\mathcal{B}_\pm\,\bar{G}_{\bar{1}'\bar{1}}
  \nonumber\\*
  & &+\ii\frac{\pi}{4}\bar{G}_{11'}\,\mathcal{B}_\pm\,
  \tilde{G}_{\bar{1}'\bar{1}}+\mathcal{O}(G^3,\tfrac{1}{D}),\quad
  \label{eq:Bkernela}\\
  \mathcal{B}_{\pm,11'}^a(\Omega,\xi)&=&\mathcal{O}(G^3,\tfrac{1}{D}),
  \label{eq:B12a}
\end{eqnarray}
where we have performed the sum over the Keldysh indices and used
\eqref{eq:barGdef}. We stress at this point that the kernel $\Sigma_{B}^{\pm}$
and the renormalized vertex $\mathcal{B}_\pm(\Omega,\xi)$ without external
lines are identical,
\begin{equation}
  \Sigma_{B}^{\pm}(\Omega,\xi)=\mathcal{B}_\pm(\Omega,\xi).
\label{eq:SigmaB=B}
\end{equation}
Nevertheless we will retain the distinction between the kernel and the vertex
in the following, as the former appears in the final formulas for the
correlation functions \eqref{eq:barCresult}, whereas the latter appears in the
diagrammatic expressions for the r.h.s. of the RG equations. We note that the
frequency dependence in \eqref{eq:SigmaB=B} is generated during the flow as
shown below.

After integrating out the symmetric part of the Fermi function in this way, we
obtain a new diagrammatic series for the kernel analog to
(\ref{value_kernelB}). The Liouvillian and the vertices have to be replaced by
the effective ones and the contractions between the effective vertices contain
only the antisymmetric part $\gamma^a$. Due to \eqref{eq:B12a} there occur no
diagrams including the new vertex $\mathcal{B}^{a}_{\pm,11'}(\Omega,\xi)$.
Furthermore, since the effective quantities have become energy dependent (also
the effective vertex $\bar{G}^a$ becomes energy dependent in higher order
perturbation theory), one has to replace
\begin{equation}
{1\over z+X_i-L_S^{(0)}}\,G\,\rightarrow\,
{1\over z+X_i-L_S^{a}(z+X_i)}\,\bar{G}^a(z+X_i)
\end{equation}
(with $z=\Omega,\xi$) in (\ref{value_kernelB}). Since the antisymmetric part of
the contraction (\ref{contraction_decomposition}) does not depend on the
Keldysh indices, only the effective vertex $\bar{G}^a$ averaged over the
Keldysh indices occurs in the new perturbative series.

\subsection{Continuous RG equations}
In the second continuous RG procedure we deal with the remaining antisymmetric
part of the Fermi distribution function, where in each infinitesimal step a
small energy shell is integrated out. Instead of integrating out the energies
on the real axis, it has turned out to be more efficient to integrate out the
Matsubara poles of the Fermi distribution function on the imaginary axis
\cite{Jakobs-07,Schoeller09}. This is achieved by introducing a formal cutoff
dependence into the antisymmetric part of the Fermi distribution by
\begin{equation}
\label{Fermi_cutoff}
f^\Lambda_\alpha(\omega)\,=\,
-\,T_\alpha\,\sum_n\,{1\over \omega-\ii\omega^\alpha_n}
\,\,\theta_{T_\alpha}(\Lambda-|\omega_n^\alpha|),
\end{equation}
where $\omega_n^\alpha=(2n+1)\pi T_\alpha$ are the Matsubara frequencies
corresponding to the temperature of reservoir $\alpha$, and
\begin{equation}
\label{theta_T}
\theta_T(\omega)\,=\,
\left\{
\begin{array}{cl}
\theta(\omega) &\quad\mbox{for }|\omega|\,>\,\pi T \\
{1\over 2}+{\omega \over 2\pi T} &\quad\mbox{for }|\omega|\,<\,\pi T
\end{array}
\right.
\end{equation}
is a theta function smeared by temperature. For $\Lambda=\infty$,
(\ref{Fermi_cutoff}) yields the full antisymmetric part
$f_\alpha(\omega)-{1\over 2}$ of the Fermi distribution. In each RG step, one
reduces the cutoff $\Lambda$ by $d\Lambda$, and integrates out the
infinitesimal part
$f_\alpha^\Lambda-f_\alpha^{\Lambda-d\Lambda}=d\Lambda{df^\Lambda_\alpha\over
  d\Lambda}$ of the Fermi distribution. For example, the new effective
Liouvillian at scale $\Lambda-d\Lambda$
\begin{equation}
\label{L_new}
L_S^{\Lambda-d\Lambda}(z)\,=\,L_S^{\Lambda}(z)-dL_S^{\Lambda}(z)
\end{equation}
and similarly the new effective vertices $\bar{G}^{\Lambda-d\Lambda}_{11'}(z)$
and $\mathcal{B}^{\Lambda-d\Lambda}_{\pm,11'}(\Omega,\xi)$ as well as the
kernel $\Sigma^{\pm,\Lambda-d\Lambda}_B(\Omega,\xi)$ can be calculated
technically in the same way as for the first discrete RG step. The only
difference is that an infinitesimal small part is integrated out, so that the
RG diagrams contain only one contraction involving
$d\Lambda{df^\Lambda_\alpha\over d\Lambda}$. Furthermore, since the diagrams
have to be irreducible with respect to this part, this contraction must
connect the first with the last vertex of the diagram. Using this procedure
the RG equations for the dot Liouvillian $L_S^\Lambda(z)$ and the vertex
$\bar{G}^{\Lambda}_{11'}(z)$ have been derived in
Ref.~\onlinecite{SchoellerReininghaus09}. Here we will use this technique to
obtain the RG equations for $\mathcal{B}^\Lambda_{\pm,11'}(\Omega,\xi)$ as
well as $\Sigma^{\pm,\Lambda}_B(\Omega,\xi)$.

\begin{figure}[t]
  \psfrag{Bvertex}{$\mathcal{B}_\pm$}
  \includegraphics[scale=0.28]{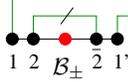}
  \caption{(color online) RG diagram for the renormalization of the vertex
    $\mathcal{B}^{\Lambda}_{\pm,11'}(\Omega,z)$ in $O(G^2)$. The
    slash indicates the contraction where the Fermi function has to be
    replaced by $-d\Lambda{df^\Lambda_\alpha\over d\Lambda}$.}
\label{fig:RGB12}
\end{figure}
\begin{figure}[t]
  \psfrag{Bvertex}{$\mathcal{B}_\pm$}
  \psfrag{B12vertex}{$\mathcal{B}_{\pm,12}$}
  \psfrag{B21vertex}{$\mathcal{B}_{\pm,\bar{2}\bar{1}}$}
  \includegraphics[scale=0.28]{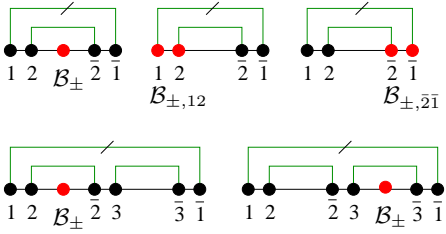}
  \caption{(color online) RG diagrams for the renormalization of the kernel
    $\Sigma^{\pm,\Lambda}_B(\Omega,z)$ up to $O(G^3)$.}
\label{fig:RGsigmaB}
\end{figure}
The diagrams contributing to the RG equations for
$\mathcal{B}_{\pm,11'}(\Omega,\xi)\equiv
\mathcal{B}^{\Lambda}_{\pm,11'}(\Omega,\xi)$ and
$\Sigma^{\pm}_B(\Omega,\xi)\equiv
\Sigma^{\pm,\Lambda}_B(\Omega,\xi)$ are shown in Fig.~\ref{fig:RGB12}
and Fig.~\ref{fig:RGsigmaB}, respectively.  Using the definition
\begin{equation}
\label{gamma_cutoff}
\gamma^\Lambda_1\,=
\,\rho(\bar{\omega})\,f_\alpha^\Lambda(\bar{\omega}),
\end{equation}
together with the convention
\begin{equation}
\label{pi_energy}
\Pi_{1\dots n}(z)\,=\,{1\over z_{1\dots n}+\bar{\omega}_{1\dots n}
-L_S(z_{1\dots n}+\bar{\omega}_{1\dots n})},
\end{equation}
we obtain the following RG equations:
\begin{widetext}
\begin{equation}
\frac{d}{d\Lambda}\mathcal{B}_{\pm,11'}(\Omega,\xi)=
-\left[
\frac{d\gamma^\Lambda_{2}}{d\Lambda}\,\bar{G}_{12}(\Omega)\,\Pi_{12}(\Omega)\,
\mathcal{B}_\pm(\Omega_{12}+\bar{\omega}_{12},\xi_{12}+\bar{\omega}_{12})\,
\Pi_{12}(\xi)\,\bar{G}_{\bar{2}1'}(\xi_{12}+\bar{\omega}_{12})
\,-\,(1\leftrightarrow 1')\right],
\label{eq:BvertexcontRG}
\end{equation}
and
\begin{eqnarray}
\frac{d}{d\Lambda}\Sigma^{\pm}_B(\Omega,\xi)&=&
-\frac{d\gamma^\Lambda_{1}}{d\Lambda}\,\gamma_2^\Lambda\,
\bar{G}_{12}(\Omega)\,\Pi_{12}(\Omega)\,
\mathcal{B}_\pm(\Omega_{12}+\bar{\omega}_{12},\xi_{12}+\bar{\omega}_{12})\,
\Pi_{12}(\xi)\,\bar{G}_{\bar{2}\bar{1}}(\xi_{12}+\bar{\omega}_{12})\nonumber\\
& &-\frac{d\gamma^\Lambda_{1}}{d\Lambda}\,\gamma_2^\Lambda
\biggl[\mathcal{B}_{\pm,12}(\Omega,\xi)\,\Pi_{12}(\xi)\,
\bar{G}_{\bar{2}\bar{1}}(\xi_{12}+\bar{\omega}_{12})
+\bar{G}_{12}(\Omega)\,\Pi_{12}(\Omega)\,
\mathcal{B}_{\pm,\bar{2}\bar{1}}(\Omega_{12}+\bar{\omega}_{12},\xi)\biggr]
\nonumber\\
& &-\frac{d\gamma^\Lambda_{1}}{d\Lambda}\,\gamma_2^\Lambda\,\gamma_3^\Lambda\,
\bar{G}_{12}(\Omega)\,\Pi_{12}(\Omega)
\Bigl[
\mathcal{B}_\pm(\Omega_{12}+\bar{\omega}_{12},\xi_{12}+\bar{\omega}_{12})\,
\Pi_{12}(\xi)\,
\bar{G}_{\bar{2}3}(\xi_{12}+\bar{\omega}_{12})\nonumber\\*
& &\hspace{20mm}
+\bar{G}_{\bar{2}3}(\Omega_{12}+\bar{\omega}_{12})\,\Pi_{13}(\Omega)\,
\mathcal{B}_\pm(\Omega_{13}+\bar{\omega}_{13},\xi_{13}+\bar{\omega}_{13})\,
\Bigr]\Pi_{13}(\xi)\,
\bar{G}_{\bar{3}\bar{1}}(\xi_{13}+\bar{\omega}_{13}).
\label{eq:BkernelcontRG}
\end{eqnarray}
\end{widetext}
We recall here that the kernel $\Sigma^{\pm}_B$ and the vertex without
external lines $\mathcal{B}_\pm$ equal each other, see \eqref{eq:SigmaB=B},
which yields a closed set of RG equations. We will further show in
App.~\ref{sec:B12} that the two-loop diagrams for $\mathcal{B}_{\pm,11'}$ as
well as the one-loop diagrams containing $\mathcal{B}_{\pm,11'}$ itself do not
contribute at second order in the coupling constant and hence can be neglected
on the r.h.s. of \eqref{eq:BvertexcontRG}.

The initial conditions of the RG equations are given by (\ref{eq:Bkernela})
and (\ref{eq:B12a}).  Since $\gamma_1^{\Lambda=0}=0$, the solution at
$\Lambda=0$ provides the result for the kernel
\begin{equation}
  \Sigma_B^{\pm}(\Omega,\xi)=\Sigma_B^{\pm}(\Omega,\xi)|_{\Lambda=0},
\end{equation}
from which the correlation functions can be calculated via
\eqref{eq:barCresult}.

As the resolvents and the vertices on the r.h.s. of the RG equations are
analytic functions in all frequencies $\bar{\omega}_i$ in the upper half of
the complex plane, all frequency integrations can be calculated analytically.
The only poles contributing are the ones of the contractions and their
derivatives given by \eqref{gamma_cutoff} with \eqref{Fermi_cutoff} as well as
\begin{equation}
{d\gamma_1^\Lambda \over d\Lambda}=-\rho(\bar{\omega})\,{1\over 2\pi}\,
\left({1\over \bar{\omega}-\ii\Lambda_{T_\alpha}}\,
+\,{1\over \omega+\ii\Lambda_{T_\alpha}}\right).
\end{equation}
Here $\Lambda_{T_\alpha}$ denotes the Matsubara frequency $\omega_n^\alpha$
which lies closest to the cutoff $\Lambda$.  After performing the integration
we find\cite{Schoeller09} that, due to the presence of the cutoff function
$\rho(\bar{\omega})={D^2\over D^2+\bar{\omega}^2}$, the r.h.s. of the RG
equations gives a negligible contribution for $\Lambda\gg D$. Therefore, we
can start the RG at $\Lambda_0\sim D$ and omit the cutoff function
$\rho(\bar{\omega})$ (the precise ratio between $\Lambda_0$ and $D$ is
determined such that no linear terms in $D$ in the effective Liouvillian are
generated~\cite{Schoeller09,SchoellerReininghaus09}). As a consequence, only
the Matsubara poles of the Fermi function in the upper half plane contribute
and all real frequencies are simply replaced by Matsubara frequencies. From
now on, we write the frequency dependence explicitly and define the analytic
continuation of the Liouvillian and the vertices in imaginary frequency space
by
\begin{eqnarray}
\label{G_matsubara}
\bar{G}_{11'}(E,\omega;\omega_1,\omega_{1'})&=&
\bar{G}_{11'}(E+\ii\omega)|_{\bar{\omega}_i\rightarrow \ii\omega_i},\\
\label{L_matsubara}
L_S(E,\omega)&=&L_S(E+\ii\omega),\\
\label{B12_matsubara}
\mathcal{B}_{\pm,11'}(\Omega,\delta,\xi,\xi';
\omega_1,\omega_{1'})&=&\nonumber\\*
& &\hspace{-15mm}
\mathcal{B}_{\pm,11'}(\Omega+\ii\delta,\xi+\ii \xi')
|_{\bar{\omega}_i\rightarrow \ii\omega_i},\\
\Sigma_B^{\pm}(\Omega,\delta,\xi,\xi')&=&
\Sigma_B^{\pm}(\Omega+\ii\delta,\xi+\ii \xi'),
\end{eqnarray}
where we keep the real and imaginary parts of the Laplace variable
$z=E+\ii\omega$ and the external frequencies $\Omega\equiv\Omega+\ii\delta$
and $\xi\equiv\xi+\ii\xi'$ separated from now on. Furthermore, $\omega\equiv
\omega_n^\alpha$, $\omega_i\equiv\omega_{n_i}^{\alpha_i}$ correspond to
Matsubara frequencies and the compact indices $1$ and $2$ on the l.h.s. do no
longer contain the frequencies $\omega_i$.  With the definition
\begin{equation}
\label{pi_matsubara}
\Pi(E,\omega)\,=\,{1\over E+\ii\omega-L_S(E,\omega)}\quad,
\end{equation}
the RG equations \eqref{eq:BvertexcontRG} and \eqref{eq:BkernelcontRG} in
Matsubara space can be written as
\begin{widetext}
\begin{equation}
\begin{split}
\frac{d}{d\Lambda}\mathcal{B}_{\pm,11'}(\Omega,\delta,\xi,\xi';
\omega_1,\omega_{1'})=&
\ii\,\bar{G}_{12}(\Omega,\delta;\omega_1,\Lambda_{T_{\alpha_2}})\,
\Pi(\Omega_{12},\delta+\omega_1+\Lambda_{T_{\alpha_2}})\,
\mathcal{B}_\pm(\Omega_{12},\delta+\omega_1+\Lambda_{T_{\alpha_2}},
\xi_{12},\xi'+\omega_1+\Lambda_{T_{\alpha_2}})\\
&\times\,\Pi(\xi_{12},\xi'+\omega_1+\Lambda_{T_{\alpha_2}})\,
\bar{G}_{\bar{2}1'}(\xi_{12},\xi'+\omega_1+\Lambda_{T_{\alpha_2}};
-\Lambda_{T_{\alpha_2}},\omega_{1'})
\,-\,(1\leftrightarrow 1'),
\label{eq:Bvertexmatsubara}
\end{split}
\end{equation}
and
\begin{eqnarray}
\frac{d}{d\Lambda}\Sigma^{\pm}_B(\Omega,\delta,\xi,\xi')&=&
\bar{G}_{12}(\Omega,\delta;\Lambda_{T_{\alpha_1}},\omega_2)\,
\Pi(\Omega_{12},\delta+\Lambda_{T_{\alpha_1}}\!\!+\omega_2)\,
\mathcal{B}_\pm(\Omega_{12},\delta+\Lambda_{T_{\alpha_1}}\!\!+\omega_2,
\xi_{12},\xi'+\Lambda_{T_{\alpha_1}}\!\!+\omega_2)
\nonumber\\*
& &\qquad\qquad\qquad\qquad\qquad\qquad
\times\,\Pi(\xi_{12},\xi'+\Lambda_{T_{\alpha_1}}\!\!+\omega_2)\,
\bar{G}_{\bar{2}\bar{1}}(\xi_{12},\xi'+\Lambda_{T_{\alpha_1}}\!\!+\omega_2;
-\omega_2,-\Lambda_{T_{\alpha_1}})\nonumber\\
& &\hspace{-25mm}+\mathcal{B}_{\pm,12}(\Omega,\delta,\xi,\xi';
\Lambda_{T_{\alpha_1}},\omega_2)\,
\Pi(\xi_{12},\xi'+\Lambda_{T_{\alpha_1}}\!\!+\omega_2)\,
\bar{G}_{\bar{2}\bar{1}}(\xi_{12},\xi'+\Lambda_{T_{\alpha_1}}\!\!+\omega_2;
-\omega_2,-\Lambda_{T_{\alpha_1}})\nonumber\\
& &\hspace{-25mm}+\bar{G}_{12}(\Omega,\delta;\Lambda_{T_{\alpha_1}},\omega_2)\,
\Pi(\Omega_{12},\delta+\Lambda_{T_{\alpha_1}}\!\!+\omega_2)\,
\mathcal{B}_{\pm,\bar{2}\bar{1}}
(\Omega_{12},\delta+\Lambda_{T_{\alpha_1}}\!\!+\omega_2,
\xi,\xi';-\omega_2,-\Lambda_{T_{\alpha_1}})\nonumber\\
& &\hspace{-25mm}
-\ii\,\bar{G}_{12}(\Omega,\delta;\Lambda_{T_{\alpha_1}},\omega_2)\,
\Pi(\Omega_{12},\delta+\Lambda_{T_{\alpha_1}}\!\!+\omega_2)\nonumber\\*
& &\hspace{-25mm}\qquad\times\Biggl[\mathcal{B}_\pm(\Omega_{12},
\delta+\Lambda_{T_{\alpha_1}}\!\!+\omega_2,
\xi_{12},\xi'+\Lambda_{T_{\alpha_1}}\!\!+\omega_2)\,
\Pi(\xi_{12},\xi'+\Lambda_{T_{\alpha_1}}\!\!+\omega_2)\,
\bar{G}_{\bar{2}3}(\xi_{12},\xi'+\Lambda_{T_{\alpha_1}}\!\!+\omega_2;
-\omega_2,\omega_3)\nonumber\\*
& &\hspace{-25mm}\qquad\qquad+
\bar{G}_{\bar{2}3}(\Omega_{12},\delta+\Lambda_{T_{\alpha_1}}\!\!+\omega_2;
-\omega_2,\omega_3)\,
\Pi(\Omega_{13},\delta+\Lambda_{T_{\alpha_1}}\!\!+\omega_3)\,
\mathcal{B}_\pm(\Omega_{13},\delta+\Lambda_{T_{\alpha_1}}\!\!+\omega_3,
\xi_{13},\xi'+\Lambda_{T_{\alpha_1}}\!\!+\omega_3)
\Biggr]\nonumber\\*
& &\hspace{-25mm}\qquad\times
\,\Pi(\xi_{13},\xi'+\Lambda_{T_{\alpha_1}}\!\!+\omega_3)\,
\bar{G}_{\bar{3}\bar{1}}(\xi_{13},\xi'+\Lambda_{T_{\alpha_1}}\!\!+\omega_3;
-\omega_3,-\Lambda_{T_{\alpha_1}}).
\label{eq:Bkernelmatsubara}
\end{eqnarray}
\end{widetext}
In these equations we implicitly sum over all indices and Matsubara
frequencies on the r.h.s. of the RG equations which do not occur on the l.h.s.
Only positive Matsubara frequencies smaller than the cutoff $\Lambda$ are
allowed and each sum has to be written as
\begin{equation}
\label{matsubara_sum}
2\pi T_\alpha\,\sum_{n}\,\theta_{T_\alpha}(\Lambda-\omega_n^\alpha)\,
\theta(\omega_n^\alpha)
\end{equation}
which reduces to an integral $\int_0^\Lambda d\omega$ for zero temperature.

In the next two subsections we will solve the RG equations
(\ref{eq:Bvertexmatsubara}) and (\ref{eq:Bkernelmatsubara}) analytically in
the weak coupling regime up to $O(G^2)$. Weak coupling is defined by the
condition that the renormalized vertices
$\bar{G}_{12}(E,\omega,\omega_1,\omega_2)$ stay small compared to one
throughout the RG flow, so that the expansion in powers of $G$ on the r.h.s.
of the RG equations is well defined. This condition is fulfilled if the
various cutoff scales occurring in the resolvents are much larger than the
Kondo temperature $T_K$ at which the vertices would diverge in the absence of
any cutoff scales.

\subsection{Weak coupling analysis above $\bs{\Lambda_c}$}\label{sec:WKabove}
As is discussed in detail in
Refs.~\onlinecite{Schoeller09,SchoellerReininghaus09} there exists a
characteristic energy scale
\begin{equation}
  \Lambda_c=\max\{|E|,|\mu_\alpha|,\tilde{h}\},
  \label{eq:Lambdac1}
\end{equation}
where $\tilde{h}\approx h_0$ is the renormalized magnetic field. For
$\Lambda>\Lambda_c$ the cutoff scales $|E|\equiv|\Omega|$, $|\mu_\alpha|$ and
$\tilde{h}$ can be neglected in the RG equation for the vertex $\bar{G}$ (see
below). This leads to a reference solution $\bar{G}^{(1)}$ which serves as the
starting point for a systematic expansion in powers of the coupling constant
$J^\Lambda$, where $\bar{G}^{(1)}_{12}\propto J^\Lambda$ (see
(\ref{G_vertex_liouville}) together with \eqref{g_kondo}). This yields a
perturbative solution of the RG equations in the regime $\Lambda>\Lambda_c$.
These results serve as initial values for the flow in the second regime
$0<\Lambda<\Lambda_c$. Here the renormalization of the vertex $\bar{G}_{12}$
is at least of order $J_c^2$, where $J_c\equiv J^{\Lambda=\Lambda_c}$.
Provided the weak-coupling condition $J_c \ll 1$ is satisfied all quantities
can be calculated perturbatively. This fact crucially relies on the appearance
of some relaxation/dephasing rate in the resolvents \eqref{eq:resprojectors},
which is guaranteed by \eqref{eq:proj0G} (as then in all resolvents standing
left to a vertex $\bar{G}$ the zero eigenvalue of the Liouvillian cannot
contribute). This analysis has been performed for the Liouvillian and the
current kernel in the anisotropic Kondo model in
Ref.~\onlinecite{SchoellerReininghaus09}. We note that the zero eigenvalue may
appear in the resolvent left to the vertex $\mathcal{B}_\pm$ in
\eqref{value_kernelB}. We will show below that this does not lead to any
problems in the calculation of the spin-spin correlation functions in the
Kondo model up to order $J_c^2$.

Regarding the appearance of the external frequency $|\Omega|$ as one of the
cut-off parameters in \eqref{eq:Lambdac1} we see from the perturbative
expansion \eqref{value_kernelB} that $\Omega$ does not appear in the
resolvents and as a cut-off parameter for all vertices right to
$\mathcal{B}_\pm$.  This fact, however, will only affect the results in the
regime $\Omega\gg V,\tilde{h}$. We will therefore use \eqref{eq:Lambdac1} as
unique cut-off for all vertices appearing in the derivation of the kernel
$\Sigma_B^\pm$. In Sec.~\ref{sec:expl} we will show that for the spin
operator in the Kondo model the difference yields a correction $\propto
1/\Omega$ and can thus be neglected.  Nevertheless we stress that all vertices
appearing in the stationary reduced density matrix $\rho_S^{st}$ in
\eqref{eq:pstdef} do not possess $\Omega$ as cut-off parameter.  We thus
deduce that in order to stay in the perturbative regime we cannot rely on the
external frequency $\Omega$ but have to require $\max\{V,\tilde{h}\}\gg T_K$.

We finally note that temperature serves as a unique cutoff for all terms on
the r.h.s. of the RG equations as for $\Lambda<2\pi T_\alpha$ the Matsubara
sums are reduced to one term and the cutoff $\Lambda_{T_\alpha}=\pi T_\alpha$
becomes independent of $\Lambda$. This trivial cutoff is set to zero in the
following, i.e. we will set $T_\alpha=0$.

After these preliminary remarks let us turn to the evaluation of the RG
equations. The one-loop RG equation for the vertex $\bar{G}$ is at zero
temperature given by~\cite{Schoeller09,SchoellerReininghaus09}
\begin{equation}
  \begin{split}
  &\frac{d}{d\Lambda}\bar{G}_{11'}(E,\omega;\omega_1,\omega_{1'})=
  \ii\,\bar{G}_{12}(E,\omega;\omega_{1},\Lambda)\\
  &\;\times
  \Pi(E_{12},\omega+\Lambda+\omega_2)\,
  \bar{G}_{\bar{2}1'}(E,\omega;-\Lambda,\omega_{1'})
  -(1\leftrightarrow 1').
  \end{split}
  \label{eq:1loopG}
\end{equation}
The RG equation for the reference solution $\bar{G}^{(1)}_{11'}$ is obtained
by assuming $\Lambda$ to be much larger than any other term appearing in the
resolvent, which gives
\begin{equation}
  \frac{d}{d\Lambda}\bar{G}_{11'}^{(1)}=
  \frac{1}{\Lambda}\Bigl[\bar{G}_{12}^{(1)}\bar{G}_{\bar{2}1'}^{(1)}-
  \bar{G}_{1'2}^{(1)}\bar{G}_{\bar{2}1}^{(1)}\Bigr].
  \label{eq:poorman}
\end{equation}
The initial condition for $\bar{G}_{11'}^{(1)}$ at $\Lambda=\Lambda_0\sim D$
is the bare vertex $\bar{G}_{11'}$ defined in \eqref{G_vertex_liouville}.  The
leading order solution is proportional to the coupling constant
$\bar{G}^{(1)}\propto J(\Lambda)$. We stress that the term on the r.h.s.,
which is $\sim J^2/\Lambda$, contributes to the change of the vertex at order
$J$. This is a general feature of the RG above $\Lambda_c$. In order to
calculate the change of a quantity at order $J^n$ one has to analyze those
terms $\sim J^n/\Delta$, where $\Delta\sim\Omega,\mu_\alpha,\tilde{h}$ is some
energy scale, and $\sim J^{n+1}/\Lambda$ on the r.h.s. of the corresponding RG
equation. In contrast, terms $\sim J^{n+1}(\Delta/\Lambda)^k/\Lambda$ $(k\ge
1)$ do not contribute to the change at order $J^n$. The RG equation for the
vertex $\tilde{G}_{11'}^{(1)}$ is given by \eqref{eq:poorman} with the
replacement~\cite{Schoeller09}
$\bar{G}_{\bar{2}1'}^{(1)},\bar{G}_{\bar{2}1}^{(1)}\rightarrow
\tilde{G}_{\bar{2}1'}^{(1)},\tilde{G}_{\bar{2}1}^{(1)}$.

Using this leading order solution we can formally expand all quantities in
powers of $J$, i.e.
\begin{eqnarray}
  & &\!\!\!\!\!\!\!\!\!\!\!\!\!\!\!\!\!\!
  \bar{G}_{11'}(E,\omega;\omega_1,\omega_{1'})=\nonumber\\*
  & &\bar{G}_{11'}^{(1)}+
  \bar{G}_{11'}^{(2)}(E,\omega;\omega_1,\omega_{1'})+\ldots
  \label{eq:expansionG}\\
  & &\!\!\!\!\!\!\!\!\!\!\!\!\!\!\!\!\!\!
  L_S(E,\omega)=\nonumber\\*
  & &L_S^{(0)}+L_S^{(1)}(E,\omega)+L_S^{(2)}(E,\omega)+\ldots,
  \label{eq:expansionL}\\
  & &\!\!\!\!\!\!\!\!\!\!\!\!\!\!\!\!\!\!
  \mathcal{B}_{\pm,11'}(\Omega,\delta,\xi,\xi';\omega_1,\omega_2)=\nonumber\\*
  & &\mathcal{B}_{\pm,11'}^{(2)}
  (\Omega,\delta,\xi,\xi';\omega_1,\omega_2)+\ldots
  \label{eq:expansionB12}\\
  & &\!\!\!\!\!\!\!\!\!\!\!\!\!\!\!\!\!\!
  \Sigma^\pm_B(\Omega,\delta,\xi,\xi')=\nonumber\\*
  & &\Sigma_B^{\pm,(0)}+\Sigma_B^{\pm,(1)}+
  \Sigma_B^{\pm,(2)}(\Omega,\delta,\xi,\xi')+\ldots\label{eq:expansionsigmaB}
\end{eqnarray}
Here $L_S^{(0)}=\comm{H_S}{.}_-$ is the bare dot Liouvillian. We recall
\eqref{eq:SigmaB=B}, which implies
$\Sigma^{\pm,(n)}_B(\Omega,\delta,\xi,\xi')=
\mathcal{B}_\pm^{(n)}(\Omega,\delta,\xi,\xi')$ in all orders in $J$.
Furthermore, we have already indicated which terms will depend on the Matsubara
frequencies, external frequencies $\Omega+\ii\delta$ and $\xi+\ii\xi'$ and the
Laplace variable $E+\ii\omega$. 

The vertex $\bar{G}$ and the Liouvillian were calculated in
Ref.~\onlinecite{SchoellerReininghaus09}. We will here state those results
needed for the calculation of \eqref{eq:expansionB12} and
\eqref{eq:expansionsigmaB}. The second order vertex $\bar{G}^{(2)}$ is further
decomposed as
\begin{equation}
  \begin{split}
  \bar{G}_{11'}^{(2)}(E,\omega;\omega_1,\omega_{1'})=&
  \ii\,\bar{G}_{11'}^{(2a_1)}+\bar{G}_{11'}^{(2a_2)}\\
  &+\bar{G}_{11'}^{(2b)}(E,\omega;\omega_1,\omega_{1'}).
  \end{split}
\end{equation}
Here the vertex $\bar{G}^{(2a_1)}$ is given by~\cite{SchoellerReininghaus09}
\begin{equation}
\bar{G}_{11'}^{(2a_1)}=
-\frac{\pi}{2}\Bigl[\bar{G}_{12}^{(1)}\tilde{G}_{\bar{2}1'}^{(1)}
-\bar{G}_{1'2}^{(1)}\tilde{G}_{\bar{2}1}^{(1)}\Bigr].
\end{equation}
For the Kondo model the vertex $\bar{G}^{(2a_2)}$ turns out to have the same
matrix structure as the leading order solution $\bar{G}^{(1)}$. This implies
that both can be put together by redefining
$\bar{G}^{(1)}\equiv\bar{G}^{(1)}+\bar{G}^{(2a_2)}$, which amounts to a
two-loop renormalization of the Kondo
temperature~\cite{Andrei-83,DoyonAndrei06,SchoellerReininghaus09}.
Furthermore, the vertex $\bar{G}^{(2b)}$ is generically given by
\begin{equation}
  \begin{split}
  &\bar{G}_{11'}^{(2b)}(E,\omega;\omega_1,\omega_{1'})=\\
  &\quad\bar{G}_{12}^{(1)}\,
  \ln\frac{\Lambda+\omega+\omega_1-\ii E_{12}+\ii L_S^{(0)}}{\Lambda}\,
  \bar{G}_{\bar{2}1'}^{(1)}-(1\leftrightarrow 1').
  \end{split}
  \label{eq:G2b}
\end{equation}
The zeroth-order Liouvillian is given by the initial condition
$L_S^{(0)}=\comm{H_S}{.}_-$ with \eqref{H_S_kondo} while the first-order
Liouvillian is further decomposed as 
\begin{equation}
  L_S^{(1)}(E,\omega)=L_S^{(1)}-(E+\ii\omega) Z^{(1)},
\end{equation}
where $L_S^{(1)}$ and $Z^{(1)}$ do not depend on the Laplace variable. The
second-order Liouvillian was calculated in
Ref.~\onlinecite{SchoellerReininghaus09}, however, we will not need it for the
solution of (\ref{eq:Bvertexmatsubara}) and (\ref{eq:Bkernelmatsubara}) in the
regime $\Lambda>\Lambda_c$.

Let us now turn to the calculation of \eqref{eq:expansionB12} and
\eqref{eq:expansionsigmaB}. The zeroth-order term of the kernel is just given
by the initial condition \eqref{eq:Bvertexinitial}, i.e.
\begin{equation}
  \Sigma_B^{\pm,(0)}\equiv\mathcal{B}_\pm^{(0)}=\ii\,\comm{B}{.}_\pm.
\end{equation}

For the derivation of an RG equation for $\Sigma_B^{\pm,(1)}$ we have to keep
all terms $\propto J^2/\Lambda$ on the r.h.s. of (\ref{eq:Bkernelmatsubara}).
Taking the zero-temperature limit and keeping only the order $J^0$ in the
resolvents we obtain
\begin{eqnarray}
  & &\frac{d}{d\Lambda}\Sigma_B^{\pm,(1)}=\int_0^\Lambda d\omega_2\,
  \bar{G}_{12}^{(1)}\,
  \frac{1}{\Omega_{12}+\ii\delta+\ii\Lambda+\ii\omega_2-L_S^{(0)}}\quad
  \nonumber\\*
  & &\qquad\times\mathcal{B}_\pm^{(0)}\,
  \frac{1}{\xi_{12}+\ii\xi'+\ii\Lambda+\ii\omega_2-L_S^{(0)}}\,
  \bar{G}_{\bar{2}\bar{1}}^{(1)}.
  \label{eq:RGSigmaB1}
\end{eqnarray}
We evaluate this integral by assuming
\begin{equation}
  \comm{L_S^{(0)}}{\mathcal{B}_\pm^{(0)}}_-=\kappa\,\mathcal{B}_\pm^{(0)},
  \label{eq:defkappa}
\end{equation}
which has to be checked for the specific model at hand. In the Kondo model we
will find $\kappa=\pm h_0$ for $B=S^\pm$ and $\kappa=0$ for $B=S^z$ (see
Sec.~\ref{sec:expl} below). \eqref{eq:defkappa} can be used to shift
$\mathcal{B}_\pm^{(0)}$ to the right and evaluate the remaining integral by a
partial fraction expansion
\begin{eqnarray}
  \frac{d}{d\Lambda}\Sigma_B^{\pm,(1)}&=&
  \frac{\ii}{\Omega-\xi-\kappa+\ii(\delta-\xi')}\nonumber\\*
  & &\times\bar{G}_{12}^{(1)}\Bigl[
  \mathcal{K}_\Lambda(\Omega_{12}+\ii\delta-L_S^{(0)})\,\mathcal{B}_\pm^{(0)}
  \nonumber\\*
  & &\;-
  \mathcal{B}_\pm^{(0)}\,\mathcal{K}_\Lambda(\xi_{12}+\ii\xi'-L_S^{(0)})
  \Bigr]\bar{G}_{\bar{2}\bar{1}}^{(1)},\quad
  \label{eq:RGB1}
\end{eqnarray}
where we have defined
\begin{equation}
  \mathcal{K}_\Lambda(z)=\ln\frac{2\Lambda-\ii z}{\Lambda-\ii z}.
\end{equation}
The leading term in \eqref{eq:RGB1} is extracted by treating the terms $\sim
z/\Lambda$ separately,
\begin{equation}
  \mathcal{K}_\Lambda(z)=\tilde{\mathcal{K}}_\Lambda(z)+\frac{\ii
  z}{2\Lambda},
\label{eq:tildeK}
\end{equation}
where $\tilde{\mathcal{K}}_\Lambda(z)$ can be integrated by
$\tilde{\mathcal{K}}_\Lambda(z)=\frac{d}{d\Lambda}\tilde{F}_\Lambda(z)$ with 
\begin{equation}
  \tilde{F}_\Lambda(z)=\Lambda\,\ln\frac{2\Lambda-\ii z}{\Lambda-\ii z}
  -\frac{\ii z}{2}\left(
    \ln\frac{\Lambda(2\Lambda-\ii z)}{2(\Lambda-\ii z)^2}+1\right),
  \label{eq:tildeF}
\end{equation}
which has the asymptotic behavior $\tilde{F}_\Lambda(z)=\Lambda(\ln 2
+O(z^2/\Lambda^2))$ as $\Lambda\rightarrow\infty$.
Using \eqref{eq:tildeK} in \eqref{eq:RGB1} we obtain 
the frequency independent result
\begin{equation}
  \frac{d}{d\Lambda}\Sigma_B^{\pm,(1)}=-\frac{1}{2\Lambda}\,
  \bar{G}_{12}^{(1)}\,\mathcal{B}_\pm^{(0)}\,\bar{G}_{\bar{2}\bar{1}}^{(1)}.
  \label{eq:RGB1ende}
\end{equation}
The initial condition is given by \eqref{eq:Bkernela}, i.e.
$\Sigma_B^{\pm,(1)}|_{\Lambda=\Lambda_0}=0$. As mentioned above we identify
this with the vertex in first order,
$\mathcal{B}_\pm^{(1)}\equiv\Sigma_B^{\pm,(1)}$.

Using the result for $\mathcal{B}_\pm^{(1)}$ we derive in App.~\ref{sec:B12}
the vertex \eqref{eq:expansionB12}, which is given by
\begin{widetext}
\begin{eqnarray}
    \mathcal{B}_{\pm,11'}^{(2)}(\Omega,\delta,\xi,\xi';\omega_1,\omega_2)&=&
    -\frac{\ii}{\Omega-\xi-\tilde{\kappa}+\ii(\delta-\xi')}
    \Bigl[\bar{G}_{12}^{(1)}\,
    \ln\frac{\Lambda+\omega_1-\ii(\Omega_{12}+\ii\delta-L_S^{(0)})}{\Lambda}\,
    \mathcal{B}_\pm^{(0)}\,\bar{G}_{\bar{2}1'}^{(1)}-(1\leftrightarrow 1')
    \nonumber\\*
    & &\qquad\qquad-\bar{G}_{12}^{(1)}\,\mathcal{B}_\pm^{(0)}\,
    \ln\frac{\Lambda+\omega_1-\ii(\xi_{12}+\ii\xi'-L_S^{(0)})}{\Lambda}\,
    \bar{G}_{\bar{2}1'}^{(1)}+(1\leftrightarrow 1')
    \Bigr],
  \label{eq:B12result}
\end{eqnarray}
where $\tilde{\kappa}=\kappa+O(J)$ (see Sec.~\ref{sec:expl} below). We note
that $\mathcal{B}_{\pm,11'}^{(2)}\propto J^2/\Lambda$.  For large $\Lambda$
and using the result
\eqref{eq:B12result} we can derive the RG equation for the kernel in
second order (see App.~\ref{sec:Bkernel})
\begin{eqnarray}
  \frac{d}{d\Lambda}\Sigma_B^{\pm,(2)}(\Omega,\delta,\xi,\xi')&=&
  \frac{\ii}{\Omega-\xi-\kappa+\ii(\delta-\xi')}\,
  \frac{d}{d\Lambda}\bar{G}_{12}^{(1)}\Bigl[
  \tilde{F}'_\Lambda(\Omega_{12}+\ii\delta-L_S^{(0)})\,\mathcal{B}_\pm^{(0)}-
  \mathcal{B}_\pm^{(0)}\,\tilde{F}'_\Lambda(\xi_{12}+\ii\xi'-L_S^{(0)})
  \Bigr]\bar{G}_{\bar{2}\bar{1}}^{(1)}\nonumber\\*
  & &\hspace{-20mm}-\frac{\ii}{2\Lambda}\Bigl[
  \bar{G}_{12}^{(2a_1)}\,\mathcal{B}_\pm^{(0)}\,\bar{G}_{\bar{2}\bar{1}}^{(1)}
  +\bar{G}_{12}^{(1)}\,\mathcal{B}_\pm^{(0)}\,\bar{G}_{\bar{2}\bar{1}}^{(2a_1)}
  \Bigr]-\frac{1}{2\Lambda}\,\bar{G}_{12}^{(1)}\,\mathcal{B}_\pm^{(1)}\,
    \bar{G}_{\bar{2}\bar{1}}^{(1)}
    +\frac{1}{2\Lambda}\,\bar{G}_{12}^{(1)}\,
    \comm{Z^{(1)}}{\mathcal{B}_\pm^{(0)}}_+\,
    \bar{G}_{\bar{2}\bar{1}}^{(1)},
  \label{eq:B2kernelRG}
\end{eqnarray}
\end{widetext}
where $\tilde{F}'_\Lambda(z)=\tilde{F}_\Lambda(z)-\ii\frac{z}{2}\ln 2$. The
initial condition is given by \eqref{eq:Bkernela}. At this point it turns out
to be useful to decompose the second-order kernel as
\begin{equation}
  \begin{split}
  &\Sigma_B^{\pm,(2)}(\Omega,\delta,\xi,\xi')=\\
  &\qquad\Sigma_B^{\pm,(2a)}(\Omega,\delta,\xi,\xi')+
  \Sigma_B^{\pm,(2b)}+\Sigma_B^{\pm,(2c)}.
  \end{split}
  \label{eq:sigmaB2decomp}
\end{equation}
Here $\Sigma_B^{\pm,(2a)}(\Omega,\delta,\xi,\xi')$ is given by
\begin{eqnarray}
  \Sigma_B^{\pm,(2a)}(\Omega,\delta,\xi,\xi')&=&
  \frac{\ii}{\Omega-\xi-\kappa+\ii(\delta-\xi')}\nonumber\\*
  & &\hspace{-20mm}\times\bar{G}_{12}^{(1)}\Bigl[
  \tilde{F}'_\Lambda(\Omega_{12}+\ii\delta-L_S^{(0)})\,\mathcal{B}_\pm^{(0)}
  \nonumber\\*
  & &\hspace{-20mm}\;
  \quad-\mathcal{B}_\pm^{(0)}\,\tilde{F}'_\Lambda(\xi_{12}+\ii\xi'-L_S^{(0)})
  \Bigr]\bar{G}_{\bar{2}\bar{1}}^{(1)},\label{eq:SigmaB2aaboveLc}
\end{eqnarray}
which satisfies the initial condition 
\begin{equation}
  \Sigma_B^{\pm,(2a)}(\Omega,\delta,\xi,\xi')\Big|_{\Lambda=\Lambda_0}=
  \frac{\ln 2}{2}\,\bar{G}_{12}\,\mathcal{B}_\pm^{(0)}\,
  \bar{G}_{\bar{2}\bar{1}}.
  \label{eq:sigmaB2a_initial}
\end{equation}
The remaining terms in \eqref{eq:sigmaB2decomp} satisfy
\begin{eqnarray}
  & &\frac{d}{d\Lambda}\Sigma_B^{\pm,(2b)}=\nonumber\\*
  & &\quad-\frac{\ii}{2\Lambda}\Bigl[
  \bar{G}_{12}^{(2a_1)}\,\mathcal{B}_\pm^{(0)}\,\bar{G}_{\bar{2}\bar{1}}^{(1)}
  +\bar{G}_{12}^{(1)}\,\mathcal{B}_\pm^{(0)}\,\bar{G}_{\bar{2}\bar{1}}^{(2a_1)}
  \Bigr],\quad\label{eq:SigmaB2baboveLc}\\
  & &\frac{d}{d\Lambda}\Sigma_B^{\pm,(2c)}=
  -\frac{1}{2\Lambda}\,\bar{G}_{12}^{(1)}\,\mathcal{B}_\pm^{(1)}\,
    \bar{G}_{\bar{2}\bar{1}}^{(1)}\nonumber\\*
  & &\quad+\frac{1}{2\Lambda}\,\bar{G}_{12}^{(1)}\,
    \comm{Z^{(1)}}{\mathcal{B}_\pm^{(0)}}_+\,
    \bar{G}_{\bar{2}\bar{1}}^{(1)},\label{eq:SigmaB2caboveLc}
\end{eqnarray}
with initial condition (according to (\ref{eq:Bkernela}) and 
(\ref{eq:sigmaB2a_initial}))
\begin{eqnarray}
  & &\Sigma_B^{\pm,(2b)}\Big|_{\Lambda=\Lambda_0}+
  \Sigma_B^{\pm,(2c)}\Big|_{\Lambda=\Lambda_0}=
  -\frac{\pi^2}{32}\,\bar{G}_{12}\,\mathcal{B}_\pm^{(0)}\,
  \bar{G}_{\bar{2}\bar{1}}\nonumber\\*
  & &\quad
  -\frac{\ln 2}{2}\,\bar{G}_{12}\,\mathcal{B}_\pm^{(0)}\,
  \bar{G}_{\bar{2}\bar{1}}
  +\ii\frac{\pi}{4}\,\bar{G}_{12}\,\mathcal{B}_\pm^{(0)}\,
  \tilde{G}_{\bar{2}\bar{1}}.
\end{eqnarray}
The RG equations in the regime $\Lambda>\Lambda_c$ derived in this section for
a model describing spin/orbital fluctuations will be specialized and
solved for the isotropic Kondo model in Sec.~\ref{sec:KondolargeL} below. In
the next section we will first study the effect of the RG flow in the regime
$0<\Lambda<\Lambda_c$. 

\subsection{Weak coupling analysis below $\bs{\Lambda_c}$}
\label{sec:genericRGbelowLambdac}
As explained in Ref.~\onlinecite{SchoellerReininghaus09} the RG above
$\Lambda_c$ has resummed all leading and subleading logarithmic contributions
in $\ln{D\over\Lambda_c}$ into the renormalized vertices. At
$\Lambda=\Lambda_c$, the bare coupling constant is replaced by a renormalized
one $J_c$, all logarithmic contributions are eliminated, and a simple power
series in $J_c$ remains. Thus the RG equations can be solved perturbatively
provided $J_c\ll 1$. In addition, the Liouvillian in the resolvents is
replaced by the full effective Liouvillian $L_S^{eff}(z)$.

When calculating diagrams containing the resolvent \eqref{eq:resprojectors} an
obvious problem is that the frequency dependence of the effective Liouvillian
$L_S^{eff}(z)$ is not known explicitly. To circumnavigate this complication we
use the following approximation for the resolvents,
\begin{equation}
  \sum_i \frac{1}{z-\lambda_i(z)}\,P_i(z)\approx
  \sum_i \frac{a_i}{z-z_i}\,P_i(z_i).
  \label{eq:resolventapproximation}
\end{equation}
Here the poles $z_i$ of the resolvent follow from the self-consistency
equation
\begin{equation}
  z_i=\lambda_i(z_i).
  \label{eq:defzi}
\end{equation}
The residue satisfy~\cite{SchoellerReininghaus09} $a_i=1+O(J)$ and hence can
be set to one in the following.

Starting with the one-loop RG equation \eqref{eq:1loopG} for the vertex
$\bar{G}$ we observe that the terms on the r.h.s. are already of $O(J_c^2)$.
Therefore, the renormalization of the leading-order vertex $\bar{G}^{(1)}$
stops at $\Lambda=\Lambda_c$ and we have to use its value
$\bar{G}^{(1)c}\propto J_c$ in all calculations from now on. Note that we
indicate the use of the coupling constant at $\Lambda_c$ by the additional
superscript $c$. Furthermore, as we are eventually interested in the kernel
$\Sigma_B^{\pm}$ up to $O(J_c^2)$ we deduce that the higher-order vertices
$\bar{G}^{(2a_1)}$ and $\bar{G}^{(2b)}$ will not be needed below $\Lambda_c$.

Using the replacement $\bar{G}^{(1)}\rightarrow\bar{G}^{(1)c}\propto J_c$ in
\eqref{eq:Bkernelmatsubara} we obtain up to $O(J_c^2)$
\begin{equation}
  \Sigma_B^{\pm}(\Omega,\delta,\xi,\xi')=
  \Sigma_B^{\pm,(0)}+\Sigma_B^{\pm,(1)c}+
  \Sigma_B^{\pm,(2)}(\Omega,\delta,\xi,\xi'),
\end{equation}
where we have already used that the flow of the kernel in order $J$ also stops
at $\Lambda_c$. The second-order kernel satisfies
\begin{widetext} 
\begin{eqnarray}
  \frac{d}{d\Lambda}\Sigma_B^{\pm,(2)}(\Omega,\delta,\xi,\xi')&=&
  \int_0^\Lambda d\omega_2\,\bar{G}^{(1)c}_{12}\,
  \Pi(\Omega_{12},\delta+\Lambda+\omega_2)\,\mathcal{B}_\pm^{(0)}\,
  \Pi(\xi_{12},\xi'+\Lambda+\omega_2)\,\bar{G}^{(1)c}_{\bar{2}\bar{1}}
  \nonumber\\
  &=&\ii\,\sum_{i,j}\frac{\mathcal{K}_\Lambda(\Omega_{12}+\ii\delta-z_i)
  -\mathcal{K}_\Lambda(\xi_{12}+\ii\xi'-z_j)}
  {\Omega-\xi-(z_i-z_j)+\ii(\delta-\xi')}\,
  \bar{G}^{(1)c}_{12}\,P_i(z_i)\,\mathcal{B}_\pm^{(0)}\,P_j(z_j)\,
  \bar{G}^{(1)c}_{\bar{2}\bar{1}},\label{eq:SigmaB2aflow}
\end{eqnarray}
\end{widetext}
where we have used the approximation \eqref{eq:resolventapproximation} and
only kept terms $\propto J_c^2$. The initial condition is given by the
solution at $\Lambda=\Lambda_c$. Using the decomposition
\eqref{eq:sigmaB2decomp} we obtain
\begin{equation}
  \begin{split}
  &\Sigma_B^{\pm,(2)}(\Omega,\delta,\xi,\xi')=\\
  &\qquad\Sigma_B^{\pm,(2a)}(\Omega,\delta,\xi,\xi')+
  \Sigma_B^{\pm,(2b)c}+\Sigma_B^{\pm,(2c)c},
  \end{split}
\end{equation}
where the flow of $\Sigma_B^{\pm,(2a)}$ below $\Lambda_c$ is governed by
\eqref{eq:SigmaB2aflow}. The initial value at $\Lambda=\Lambda_c$ is given by
\eqref{eq:SigmaB2aaboveLc}, which we can rewrite as
\begin{eqnarray}
  \Sigma_B^{\pm,(2a)}(\Omega,\delta,\xi,\xi')\Big|_{\Lambda=\Lambda_c}&=&
  \frac{\ii}{\Omega-\xi-\kappa+\ii(\delta-\xi')}\nonumber\\*
  & &\hspace{-30mm}\times\sum_{i,j}\Bigl[
  \tilde{F}'_{\Lambda_c}(\Omega_{12}+\ii\delta-z_i)-
  \tilde{F}'_{\Lambda_c}(\xi_{12}+\ii\xi'-z_j)
  \Bigr]\nonumber\\*
  & &\hspace{-30mm}\times
  \bar{G}^{(1)c}_{12}\,P_i(z_i)\,\mathcal{B}_\pm^{(0)}\,P_j(z_j)\,
  \bar{G}^{(1)c}_{\bar{2}\bar{1}}.
  \label{eq:SigmaB2ainitialLambdac}
\end{eqnarray}
In doing so we have assumed
\begin{eqnarray}
  L_S^{(0)}&=&\sum_i z_i\,P_i(z_i)+O(J_c),\label{eq:assumption1}\\
  \mathbf{1}&=&\sum_i P_i(z_i)+O(J_c),\label{eq:assumption2}
\end{eqnarray}
and neglected terms of order $J_c^3$. We will show in App.~\ref{sec:ass} that
these assumptions are fulfilled for the Kondo model in a magnetic field. We
will further show that $\kappa=z_i-z_j+O(J_c)$ for all pairs $(i,j)$ for which
the last line in \eqref{eq:SigmaB2ainitialLambdac} is non-zero. When applying
the results of this section to other models of spin/orbital fluctuations one
has to ensure the validity of the assumptions made above. The initial value
problem \eqref{eq:SigmaB2aflow} with \eqref{eq:SigmaB2ainitialLambdac} is
readily solved using $\mathcal{K}_\Lambda(z)=\frac{d}{d\Lambda} F_\Lambda(z)$
with
\begin{equation}
  F_\Lambda(z)=\tilde{F}_{\Lambda}'(z)+
  \frac{\ii z}{2}\,\left(\ln\frac{\ii\Lambda}{2z}+1\right).
\end{equation}
The result for the kernel at $\Lambda=0$ is then obtained using
$F_{\Lambda=0}(z)=-\ii\frac{z}{2}\ln 2$: 
\begin{widetext} 
\begin{eqnarray}
  \Sigma_B^{\pm,(2a)}(\Omega,\delta,\xi,\xi')\Big|_{\Lambda=0}&=&
  \frac{1}{2}\sum_{i,j}\frac{1}{\Omega-\xi-(z_i-z_j)+\ii(\delta-\xi')}\,
  \bar{G}^{(1)c}_{12}\,P_i(z_i)\,\mathcal{B}_\pm^{(0)}\,P_j(z_j)\,
  \bar{G}^{(1)c}_{\bar{2}\bar{1}}\nonumber\\*
  & &\hspace{-20mm}\times\Biggl[(\Omega_{12}+\ii\delta-z_i)
    \left(\ln\frac{\ii\Lambda_c}{\Omega_{12}+\ii\delta-z_i}+1\right)-
  (\xi_{12}+\ii\xi'-z_j)
    \left(\ln\frac{\ii\Lambda_c}{\xi_{12}+\ii\xi'-z_j}+1\right)\Biggr].
  \label{eq:SigmaB2aresult}
\end{eqnarray}
\end{widetext}
Together with $\Sigma_B^{\pm,(2b)c}$ and $\Sigma_B^{\pm,(2c)c}$ determined in
the RG procedure above $\Lambda_c$ this yields the final result for the kernel
in second order in $J_c$. It is applicable to any operator $B$ which does not
couple the dot and reservoir degrees of freedom, i.e. whose initial value
satisfies $n=0$ in \eqref{eq:ABdef}. Furthermore the calculations were done
for a generic model describing spin or orbital fluctuations as the initial
vertex $G$ was assumed to have two external legs.  The only assumptions we
have made regarding the model specifics are the commutation relations
\eqref{eq:defkappa} and \eqref{eq:commapp}, Eqs.~\eqref{eq:assumption1}
and~\eqref{eq:assumption2} as well as the specific relation between the
parameter $\kappa$ and the poles $z_i$ and $z_j$, which have to be determined
from the self-consistency equation \eqref{eq:defzi}.

In the next section we will apply the results derived above to the spin-spin
correlation functions in the isotropic Kondo model. In particular, we will
show that the assumptions discussed above are justified in this model. Finally
we note that similar results for the effective Liouvillian have been derived
in Refs.~\onlinecite{Schoeller09,SchoellerReininghaus09}.

\section{Explicit RG equations for the Kondo model}\label{sec:expl}
In this section we will specialize the generic results derived above to the
case of the spin-spin correlation functions in the isotropic,
antiferromagnetic Kondo model in a magnetic field. The Hamiltonian was
presented in Sec.~\ref{sec:Kondo}; in particular, the dot Hamiltonian and the
coupling to the leads as given in \eqref{H_S_kondo} and \eqref{g_kondo}.

\subsection{RG flow above $\bs{\Lambda_c}$}
\label{sec:KondolargeL}
The first step is to represent the initial vertex and Hamiltonian in Liouville
space,
\begin{eqnarray}
G^{pp}_{11'}\!\!&=&\!\!{1\over 2}\,\left\{
\begin{array}{cl}
(J^i_{\alpha\alpha'})_0\,L^{pi}\,\sigma^i_{\sigma\sigma'}\, 
&\mbox{for }\eta=-\eta'=+ \\
-(J^i_{\alpha'\alpha})_0\,L^{pi}\,\sigma^i_{\sigma'\sigma}\, 
&\mbox{for }\eta=-\eta'=-
\end{array}\right.\!\!,\qquad\label{G_initial_kondo}\\
\label{L_initial_kondo}
L_S^{(0)}\!\!&=&\!\!\comm{H_S}{.}_-=h_0\,(L^{+z}+L^{-z})=h_0\,L^h,
\end{eqnarray}
where the spin superoperators $\underline{L}^p=(L^{px},L^{py},L^{pz})$ are
defined by their action on an arbitrary operator $A$ on the dot Hilbert space
via
\begin{equation}
\label{eq:defLpLm}
\underline{L}^+A\,=\,\underline{S}A,\quad
\underline{L}^-A\,=\,-A\underline{S}.
\end{equation}
An explicit matrix representation for the superoperators $\underline{L}^p$ is
given in App.~\ref{sec:appL}, where also further superoperators are defined.

The leading-order vertex $\bar{G}^{(1)}$ was derived in
Refs.~\onlinecite{Schoeller09,SchoellerReininghaus09}. It can be parametrized
for $\eta=-\eta'=+$ as
\begin{equation}
  \bar{G}_{11'}^{(1)}=-J_{\alpha\alpha'}\,
  \underline{L}^2\cdot\underline{\sigma}_{\sigma\sigma'},
  \label{eq:G1parametrization}
\end{equation}
where $\underline{\sigma}$ is the vector formed by the Pauli matrices. The
vertex for $\eta=-\eta'=-$ is obtained using the antisymmetry
$\bar{G}_{11'}=-\bar{G}_{1'1}$. Inserting \eqref{eq:G1parametrization} into
the RG equation \eqref{eq:poorman} and using the antisymmetry
$\bar{G}^{(1)}_{12}=-\bar{G}^{(1)}_{21}$,
\eqref{eq:AppEsigma1}--\eqref{eq:AppE1}, and
$J_{\alpha\beta}=J_{\beta\alpha}$ we obtain
\begin{equation}
  \frac{d}{d\Lambda}J_{\alpha\alpha'}=-\frac{1}{\Lambda}\,
  J_{\alpha\beta}\,J_{\beta\alpha'}.
  \label{eq:poormanJ}
\end{equation}
If we assume the form \eqref{J_form}, i.e. $J_{\alpha\alpha'}=2\sqrt{x_\alpha
  x_{\alpha'}}\,\bar{J}$ with $\sum_\alpha\,x_\alpha=1$, we obtain the usual
poor-man scaling equation
\begin{equation}
  \frac{d}{d\Lambda}\bar{J}(\Lambda)=-\frac{2}{\Lambda}\,\bar{J}(\Lambda)^2,
  \quad \bar{J}(\Lambda_0)=\bar{J}_0,
  \label{eq:poormanJbar}
\end{equation}
with the solution
\begin{equation}
  \bar{J}(\Lambda)=\frac{1}{2\,\ln\frac{\Lambda}{T_K}},\quad
  T_K=\Lambda_0\,e^{-1/2\bar{J}_0}.
\end{equation}
Eq.~\eqref{eq:poormanJ} explicitly shows that the term $\sim J^2/\Lambda$ on
the r.h.s. contributes to the renormalization in order $J$. Similarly, the
renormalization of a quantity in order $J^n$ is determined by the terms $\sim
J^n/\Delta$ as well as $\sim J^{n+1}/\Lambda$. In contrast, a term $\sim
J^{n+1} (\Delta/\Lambda)^k/\Lambda$ with $k\ge 1$ does not contribute at order
$J^n$, as can be seen from
\begin{eqnarray}
  & &\int^\Lambda d\Lambda'\,\frac{\Delta^k}{\Lambda'^k}
    \frac{\bar{J}^{n+1}}{\Lambda'}=
    -\frac{\Delta^k}{k}\int^\Lambda d\Lambda'\,
    \left(\frac{d}{d\Lambda'}\frac{1}{\Lambda'^k}\right)
    \bar{J}^{n+1}\nonumber\\*
  & &\quad=-\frac{\Delta^k}{\Lambda^k}\frac{\bar{J}^{n+1}}{k}-
  2\frac{n+1}{k}\int^\Lambda\!\!d\Lambda'\,
  \frac{\Delta^k}{\Lambda'^k}\frac{\bar{J}^{n+2}}{\Lambda}.\qquad
\end{eqnarray}
The vertex $\tilde{G}_{11'}^{(1)}$ is given by
\begin{equation}
  \tilde{G}_{11'}^{(1)}=\frac{1}{2}\,J_{\alpha\alpha'}\,
  \bigl(\underline{L}^1+\underline{L}^3\bigr)
  \cdot\underline{\sigma}_{\sigma\sigma'},
\end{equation}

Beside the leading-order vertex \eqref{eq:G1parametrization} and the
zero-order Liouvillian \eqref{L_initial_kondo} we also need explicit
expressions for the vertex $\bar{G}^{(2a_1)}$ as well as the Liouvillian in
first order. These are given by~\cite{Schoeller09,SchoellerReininghaus09}
\begin{eqnarray}
  \bar{G}^{(2a_1)}_{11'}&=&\frac{\pi}{2}\,J_{\alpha\beta}\,
  J_{\beta\alpha'}\,
  \underline{L}^3\cdot\underline{\sigma}_{\sigma\sigma'},\\
  L_S^{(1)}&=&\frac{1}{2}\,\text{tr}\,J\,h_0\,L^h,\\
  Z^{(1)}&=&\text{tr}\,J\,L^a,\label{eq:defZ1}
\end{eqnarray}
where $J=(J_{\alpha\alpha'})$ is the coupling of the leading-order vertex
$\bar{G}^{(1)}$ and the trace tr is taken in the reservoir indices,
$\text{tr}\,J=J_{RR}+J_{LL}$. Furthermore we have taken the scaling limit
$J_0\rightarrow 0$, $\Lambda_0\approx D\rightarrow\infty$ such that the Kondo
temperature $T_K$ remains constant.

As the next step it is straightforward to derive the following results for the
initial vertex $\mathcal{B}_\pm^{(0)}$:
\begin{eqnarray}
  \mathcal{B}_+^{(0)}&=&\ii\,\bigl(L^{1}_s+\ii\,L^{3}_s\bigr),
  \label{eq:initB+}\\
  \mathcal{B}_-^{(0)}&=&-2\ii\,L^{2}_s,\label{eq:initB-}
\end{eqnarray}
where $s$ takes the values $s=z,\pm$ for $B=S^z,S^\pm$ and we have set
$L^{j}_z\equiv L^{jz}$, $j=1,2,3$.  We stress that in this way the spin
operators are directly represented by their matrices in Liouville space.
Hence, we do not have to use a pseudo-fermion representation of the Kondo
spin. The vertex $\mathcal{A}$ defined in \eqref{eq:Avertexinitial} is just
given by $\mathcal{A}=\frac{1}{2}\,\mathcal{B}_+^{(0)}$. Furthermore, we
recall that the kernels in zeroth order are just given by
$\Sigma_A(\Omega)=\mathcal{A}$ and
$\Sigma_B^\pm(\Omega,\xi)=\mathcal{B}_\pm^{(0)}$, respectively.  Now the RG
equation \eqref{eq:RGB1ende} for $\Sigma_B^{\pm,(1)}$ reads
\begin{equation}
  \frac{d}{d\Lambda}\Sigma_B^{\pm,(1)}=
  \left\{\begin{array}{ll}0,& \text{for }\Sigma_B^{+,(1)},\\
  -\frac{1}{2\Lambda}\,\text{tr}\,J^2\,\mathcal{B}_-^{(0)},& 
  \text{for }\Sigma_B^{-,(1)},\end{array}\right.
  \label{eq:RGB1ende2}
\end{equation}
where we have used \eqref{eq:AppEsigma1} and \eqref{eq:AppE2}. The additional
factor of two is due to the implicit summation over $\eta$. Hence the solution
in the scaling limit is given by
\begin{eqnarray}
  \Sigma_B^{+,(1)}=\mathcal{B}_+^{(1)}&=&0,\label{eq:Sigma+B1result}\\
  \Sigma_B^{-,(1)}=\mathcal{B}_-^{(1)}&=&
  \frac{1}{2}\,\text{tr}\,J\,\mathcal{B}_-^{(0)}=
  -\ii\,\text{tr}\,J\,L^2_s.\label{eq:Sigma-B1result}
\end{eqnarray}
In second order we will here determine only the terms $\Sigma_B^{\pm,(2b)}$
and $\Sigma_B^{\pm,(2c)}$ as their flow is cut-off at $\Lambda=\Lambda_c$. The
remaining term $\Sigma_B^{\pm,(2a)}$ will be derived in the next section.
Hence we have to solve the RG equation \eqref{eq:SigmaB2baboveLc}, which using
\eqref{eq:AppE3} and \eqref{eq:AppE4} reads
\begin{equation}
  \frac{d}{d\Lambda}\Sigma_B^{\pm,(2b)}=
  \pm\frac{\pi}{\Lambda}\,\text{tr}\,J^3\,
  \left\{\begin{array}{ll}L^2_s,& \text{for }\Sigma_B^{+,(2b)},\\
  L^{3}_s,&\text{for }\Sigma_B^{-,(2b)}.\end{array}\right.
  \label{eq:RGB2bende}
\end{equation}
Analogously we obtain for \eqref{eq:SigmaB2caboveLc} using \eqref{eq:defZ1},
\eqref{eq:AppE2} and \eqref{eq:AppE5}
\begin{equation}
  \frac{d}{d\Lambda}\Sigma_B^{\pm,(2c)}=-\ii\frac{3}{2\Lambda}\,
  \text{tr}\,J^2\,\text{tr}\,J\,
  \left\{\begin{array}{ll}0,&\text{for }\Sigma_B^{+,(2c)},\\
      L^2_s,&\text{for }\Sigma_B^{-,(2c)}.\end{array}\right.
\end{equation}
The solutions read
\begin{eqnarray}
  \Sigma_B^{+,(2b)}&=&-\frac{\pi}{2}\,\text{tr}\,J^2\,L^2_s,
  \label{eq:Sigma+B2bresult}\\
  \Sigma_B^{-,(2b)}&=&\frac{\pi}{2}\,\text{tr}\,J^2\,L^3_s,
  \label{eq:Sigma-B2bresult}\\
  \Sigma_B^{+,(2c)}&=&0,\label{eq:Sigma+B2cresult}\\
  \Sigma_B^{-,(2c)}&=&\ii\frac{3}{4}\,\text{tr}\,J^2\,L^2_s,
  \label{eq:Sigma-B2cresult}
\end{eqnarray}
where we have already taken the scaling limit so that the contributions from
the initial condition are negligible.

\subsection{RG flow below $\bs{\Lambda_c}$}\label{sec:resultsbelowLc}
As we have already explained above the RG flow of the leading order solution
$\bar{G}^{(1)}$ stops at the scale $\Lambda_c$ which is given by the maximal
one of the external parameters, 
\begin{equation}
  \Lambda_c=\max\{|\Omega|,V,h_0\}.
  \label{eq:defLambdac}
\end{equation}
The value of the vertex at $\Lambda_c$ is given by
\eqref{eq:G1parametrization} with 
\begin{equation}
  J_c=J(\Lambda_c)\equiv\left(\begin{array}{cc}J_R&J_{nd}\\
      J_{nd}&J_L\end{array}\right),
\end{equation}
where due to \eqref{J_form} the couplings satisfy 
\begin{equation}
  J_R=2x_R\bar{J}_c,\;J_L=2x_L\bar{J}_c,\;J_{nd}=\sqrt{J_RJ_L},
\end{equation}
with $x_R+x_L=1$ and
\begin{equation}
  \bar{J}_c=\frac{1}{2\,\ln\frac{\Lambda_c}{T_K}}.
  \label{eq:defbarJc}
\end{equation}
The asymmetry ratio is defined as $r=J_L/J_R$. As already mentioned the flow
of all vertices right to $\mathcal{B}_\pm$ does not stop at $\Lambda_c$ as
defined in \eqref{eq:defLambdac} but rather at $\max\{V,h_0\}$. This affects
the result for the kernel $\Sigma_B^\pm$ only in the regime $\Omega\gg V,h_0$.
We will discuss the changes in this case separately at the end of this
section. We stress, however, that the flow of all vertices $\bar{G}$ needed
for the derivation of the stationary reduced density matrix $\rho_S^{st}$ is
cut off by $\max\{V,h_0\}$. Hence in order to stay in the weak-coupling regime
we need
\begin{equation}
  \max\{V,h_0\}\gg T_K\;\leftrightarrow\;J_c\ll 1.
\end{equation}
As the definition of the scale $\Lambda_c$ is to some extent arbitrary as long
as it remains of the order of the external energy scales in the problem, it is
necessary to study the effect of a redefinition
$\Lambda_c\rightarrow\Lambda_c'$ with $\Lambda_c'/\Lambda_c\sim 1$. This will
induce a redefinition of the coupling as
\begin{equation}
  \bar{J}_c'=\frac{1}{2\,\ln\frac{\Lambda_c'}{T_K}}
  =\bar{J}_c-2\bar{J}_c^2\ln\frac{\Lambda_c'}{\Lambda_c}+\mathcal{O}(J_c^3).
\end{equation}
As we will show below, the redefinition $\Lambda_c\rightarrow\Lambda_c'$ does
not change the final results for the Liouvillian or the kernel $\Sigma_B^\pm$
up to order $J_c^2$.

The Liouvillian up to $O(J_c^2)$ can be parametrized
as~\cite{SchoellerReininghaus09}
\begin{equation}
  \begin{split}
    L_S(E,\omega)=&h(E,\omega)\,L^h-\ii\,\Gamma^a(E,\omega)\,L^a\\
    &-\ii\,\Gamma^{c}(E,\omega)\,L^c-\ii\,\Gamma^{3z}(E,\omega)\,L^{3z}.
  \end{split}
  \label{eq:Lparametrization}
\end{equation}
The Liouvillian can be diagonalized (see \eqref{eq:Ldiagonalization})
using the eigenvalues ($z=E+\ii\omega$)
\begin{eqnarray}
  \lambda_0(E,\omega)&=&0,\label{eq:lambda0}\\
  \lambda_1(E,\omega)&=&-\ii\,\Gamma^a(E,\omega),\label{eq:lambda1}\\
  \lambda_\pm(E,\omega)&=&\pm h(E,\omega)-\ii\,\Gamma^a(E,\omega)
  -\ii\,\Gamma^c(E,\omega),\qquad\label{eq:lambdapm}
\end{eqnarray}
and the projectors
\begin{eqnarray}
  P_0(E,\omega)&=&L^b-
  \frac{\Gamma^{3z}(E,\omega)}{\Gamma^a(E,\omega)}\,L^{3z},\label{eq:proj0}\\
  P_1(E,\omega)&=&L^a-L^c+
  \frac{\Gamma^{3z}(E,\omega)}{\Gamma^a(E,\omega)}\,L^{3z},\label{eq:proj1}\\
  P_\pm&=&\frac{1}{2}(L^c\pm L^h).\label{eq:projpm}
\end{eqnarray}
The eigenvalues \eqref{eq:lambda0}--\eqref{eq:lambdapm} can now be used to
determine the poles $z_i$ of the resolvent defined in \eqref{eq:defzi}.
Solving the self-consistency equation one finds $z_0=0$,
$z_1=-\ii\,\tilde{\Gamma}_1=-\ii\,\Gamma^a(0,0)$, and
$z_\pm=\pm\tilde{h}-\ii\,\tilde{\Gamma}_2$, where the spin relaxation and
dephasing rates and the renormalized magnetic field are given up to $O(J_c^2)$
by
\begin{widetext}
\begin{eqnarray}
  \tilde{\Gamma}_1&=&(J_R^2+J_L^2)\,Im\,\mathcal{H}_2(\tilde{h})
  +J_RJ_L\left(Im\,\mathcal{H}_2(V+\tilde{h})+
  Im\,\mathcal{H}_2(V-\tilde{h})\right),\label{eq:tildegamma1}\\
  \tilde{\Gamma}_2&=&\frac{J_R^2+J_L^2}{2}\,
  Im\,\mathcal{H}_1(\tilde{h})+\frac{J_RJ_L}{2}\left(2\,Im\,\mathcal{H}_1(V)+
  Im\,\mathcal{H}_2(V+\tilde{h})+
  Im\,\mathcal{H}_2(V-\tilde{h})\right),\label{eq:tildegamma2}\\
  \tilde{h}&=&\left(1-\frac{J_R+J_L}{2}+\frac{(J_R+J_L)^2}{2}\right)h_0
  -\frac{J_R^2+J_L^2}{2}\,Re\,\mathcal{H}_2(\tilde{h})
  -\frac{J_RJ_L}{2}\left(Re\,\mathcal{H}_2(V+\tilde{h})
  -Re\,\mathcal{H}_2(V-\tilde{h})\right).\label{eq:tildeh}
\end{eqnarray}
\end{widetext}
The higher order terms $\sim J_c^3 \ln\ldots$ for the rates were obtained in
Ref.~\onlinecite{SchoellerReininghaus09}. The voltage was defined in
\eqref{eq:voltagedef} and we always assume $V,h_0>0$.  As these rates are
obtained from the Liouvillian at $z=0$ and $z=\pm\tilde{h}$, respectively, the
external frequency $\Omega$ does not appear as a cut-off in the definition of
$\Lambda_c$. Furthermore, we have defined the auxiliary functions
\begin{equation}
  \mathcal{H}_i(E)=E\left(\ln\frac{\Lambda_c}{\sqrt{E^2+\tilde{\Gamma}_i^2}}
    +1\right)+\ii\,E\arctan\frac{E}{\tilde{\Gamma}_i},
  \label{eq:defHi}
\end{equation}
which arises from terms like
\begin{equation}
  (z-z_j)\left(\ln\frac{\ii\Lambda_c}{z-z_j}+1\right)
\end{equation}
by neglecting the imaginary part of $z_j$, which is proportional to
$\tilde{\Gamma}_{1/2}\propto J_c^2$, in the prefactor and takeing the real and
imaginary parts. We note that as $z_\pm=\pm\tilde{h}-\ii\,\tilde{\Gamma}_2$
the renormalized magnetic field automatically appears in the logarithm. We have
therefore also kept $\tilde{h}$ in the linear prefactor. The deviation of
$Im\,\mathcal{H}_i(E)$ from $\frac{\pi}{2}|E|$ is only important for
$|E|<\tilde{\Gamma}_i$ and will be neglected otherwise. Furthermore, we have
omitted the imaginary part of the Laplace variable, $\omega$, since for the
correlation functions calculated below we only need the Liouvillian on the
real axis.  For $\omega=0$ and using \eqref{eq:defHi}, the functions in the
parametrization \eqref{eq:Lparametrization} are given up to order $J_c^2$
by~\cite{SchoellerReininghaus09}
\begin{widetext}
\begin{eqnarray}
    h(E)&=&\left(1+\frac{J_R+J_L}{2}-\frac{3}{8}(J_R+J_L)^2\right)h_0
    +\frac{J_R^2+J_L^2}{4}
    \left(\mathcal{H}_2(E+\tilde{h})-\mathcal{H}_2(E-\tilde{h})\right)
    \nonumber\\*
    & &+\frac{J_RJ_L}{4}\left(\mathcal{H}_2(E+V+\tilde{h})+
      \mathcal{H}_2(E-V+\tilde{h})-\mathcal{H}_2(E+V-\tilde{h})
      -\mathcal{H}_2(E-V-\tilde{h})\right),\label{eq:Liouvillianh}\\
    \Gamma^a(E)&=&
    -\ii\,(J_R+J_L)\left(1-\frac{J_R+J_L}{2}\right)E
    -\ii\,\frac{J_R^2+J_L^2}{2}\left(\mathcal{H}_2(E+\tilde{h})
    +\mathcal{H}_2(E-\tilde{h})\right)\nonumber\\*
    & &-\ii\,\frac{J_RJ_L}{2}\left(\mathcal{H}_2(E+V+\tilde{h})+
      \mathcal{H}_2(E-V+\tilde{h})+\mathcal{H}_2(E+V-\tilde{h})
      +\mathcal{H}_2(E-V-\tilde{h})\right),\label{eq:LiouvillianGammaa}\\
    \Gamma^c(E)&=&-\ii\,\frac{J_R^2+J_L^2}{2}
    \left(\mathcal{H}_1(E)-\frac{1}{2}\mathcal{H}_2(E+\tilde{h})
    -\frac{1}{2}\mathcal{H}_2(E-\tilde{h})\right)
  -\ii\,\frac{J_RJ_L}{2}\biggl(\mathcal{H}_1(E+V)+\mathcal{H}_1(E-V)
  \nonumber\\*
  & &-\frac{1}{2}\mathcal{H}_2(E+V+\tilde{h})
    -\frac{1}{2}\mathcal{H}_2(E-V+\tilde{h})
    -\frac{1}{2}\mathcal{H}_2(E+V-\tilde{h})
    -\frac{1}{2}\mathcal{H}_2(E-V-\tilde{h})\biggr),
    \label{eq:LiouvillianGammac}\\
    \Gamma^{3z}(E)&=&\frac{\pi}{2}\,(J_R+J_L)^2\,h_0
    =\frac{\pi}{2}\,(J_R+J_L)^2\,\tilde{h}.\label{eq:LiouvillianGamma3z}
\end{eqnarray}
\end{widetext}
In the last line we have replaced the bare magnetic field by the renormalized
field, such that the latter appears consistently in all functions
\eqref{eq:Liouvillianh}--\eqref{eq:LiouvillianGamma3z}. The change is of
$O(J_c^3)$. We further note that a redefinition
$\Lambda_c\rightarrow\Lambda_c'$ yields the same result for the Liouvillian
with the replacement $J_c\rightarrow J_c'$. Naively, the linear terms in
$h(E)$ and $\Gamma^a(E)$ yield additional contributions in order $J_c'^2$.
These are, however, exactly cancelled by terms appearing from the logarithms
using $\mathcal{H}_i(E)=\mathcal{H}_i'(E)-E\ln\frac{\Lambda_c'}{\Lambda_c}$,
where $\mathcal{H}_i'(E)$ is given by \eqref{eq:defHi} with
$\Lambda_c\rightarrow\Lambda_c'$.  The facts presented above allow us to
justify the assumptions \eqref{eq:assumption1} and \eqref{eq:assumption2} made
in Sec.~\ref{sec:genericRGbelowLambdac} above for the specific case of the
Kondo model (see App.~\ref{sec:ass}).

After the recall of the Liouvillian in second order we will finally evaluate
the kernel $\Sigma_B^{\pm}$ in the Kondo model. The flow of the kernel also
stops at $\Lambda_c$ except for one term, namely $\Sigma_B^{\pm,(2a)}$. Hence
we find 
\begin{eqnarray}
  & &\Sigma_B^{\pm}(\Omega,\delta,\xi,\xi')=
  \Sigma_B^{\pm,(0)}+\Sigma_B^{\pm,(1)c}\nonumber\\*
  & &\quad+\Sigma_B^{\pm,(2a)}(\Omega,\delta,\xi,\xi')+\Sigma_B^{\pm,(2b)c}+
  \Sigma_B^{\pm,(2c)c},\quad
  \label{eq:SigmaBkernelcompleteresult}
\end{eqnarray}
where the first term is given by \eqref{eq:initB+} or \eqref{eq:initB-}
depending on the operator studied, and the second, fourth, and fifth is
obtained from \eqref{eq:Sigma+B1result}, \eqref{eq:Sigma-B1result} as well as
\eqref{eq:Sigma+B2bresult}--\eqref{eq:Sigma-B2cresult} using the replacement
$J\rightarrow J_c$. For the evaluation of the remaining contribution
$\Sigma_B^{\pm,(2a)}|_{\Lambda=0}$ from \eqref{eq:SigmaB2aresult}
let us first consider the operator $B=S^z$.
Using
\begin{equation}
  \sum_{k=x,y,z} L^{2k}\,P_i(z_i)\,\mathcal{B}_+^{(0)}\,P_j(z_j)\,L^{2k}
  \propto L^c
  \label{eq:GPB+PG}
\end{equation}
for $j=1$ and $i=0,1$ (in all other cases the l.h.s. vanishes) we find
\begin{equation}
  \Sigma_{S^z}^{+,(2a)}(\Omega,\delta,\xi,\xi')\propto L^c.
  \label{eq:GPB2a+PG}
\end{equation}
We note that in \eqref{eq:GPB+PG} the zero eigenvalue of the effective
Liouvillian appears in the resolvent left to $\mathcal{B}_+^{(0)}$. As we will
show in the next section the resulting term \eqref{eq:GPB2a+PG} does not
contribute to the correlation functions. For the evaluation of the kernel for
the calculation of the susceptibility we use
\begin{widetext} 
\begin{equation}
  \sum_{k=x,y,z} L^{2k}\,P_i(z_i)\,\mathcal{B}_-^{(0)}\,P_j(z_j)\,L^{2k}
  =\left\{\begin{array}{ll}
  \pm\frac{\ii}{4}\left(L^a-\frac{1}{2}(L^c\mp L^h)\right),& i,j=\pm,\\[2mm]
  0,& \text{otherwise}, \end{array}\right.
  \label{eq:GPB-PG}
\end{equation}
which results in
\begin{eqnarray}
  \Sigma_{S^z}^{-,(2a)}(\Omega,\delta,\xi,\xi')\Big|_{\Lambda=0}&=&
  -\frac{\ii}{2}\frac{J_R^2+J_L^2}{\Omega-\xi+\ii(\delta-\xi')}
  \biggl[\left(\mathcal{H}_2(\Omega+\tilde{h})-
    \mathcal{H}_2(\Omega-\tilde{h})-\mathcal{H}_2(\xi+\tilde{h})
    +\mathcal{H}_2(\xi-\tilde{h})\right)\left(L^a-\tfrac{1}{2}\,L^c\right)
    \nonumber\\*
    & &\hspace{32mm}-\frac{1}{2}\left(\mathcal{H}_2(\Omega+\tilde{h})+
    \mathcal{H}_2(\Omega-\tilde{h})-\mathcal{H}_2(\xi+\tilde{h})
    -\mathcal{H}_2(\xi-\tilde{h})\right)L^h\biggr]\nonumber\\*
  & &\hspace{-30mm}
  -\frac{\ii}{2}\frac{J_RJ_L}{\Omega-\xi+\ii(\delta-\xi')}
  \biggl[\left(\mathcal{H}_2(\Omega+V+\tilde{h})
    +\mathcal{H}_2(\Omega-V+\tilde{h})-\mathcal{H}_2(\Omega+V-\tilde{h})
    -\mathcal{H}_2(\Omega-V-\tilde{h})\right.\nonumber\\*
    & &\left.\quad
      -\mathcal{H}_2(\xi+V+\tilde{h})-\mathcal{H}_2(\xi-V+\tilde{h})
    +\mathcal{H}_2(\xi+V-\tilde{h})+\mathcal{H}_2(\xi-V-\tilde{h})\right)
  \left(L^a-\tfrac{1}{2}\,L^c\right)\nonumber\\*
  & &-\frac{1}{2}\left(\mathcal{H}_2(\Omega+V+\tilde{h})
    +\mathcal{H}_2(\Omega-V+\tilde{h})+\mathcal{H}_2(\Omega+V-\tilde{h})
    +\mathcal{H}_2(\Omega-V-\tilde{h})\right.\nonumber\\*
    & &\left.\quad
      -\mathcal{H}_2(\xi+V+\tilde{h})-\mathcal{H}_2(\xi-V+\tilde{h})
    -\mathcal{H}_2(\xi+V-\tilde{h})-\mathcal{H}_2(\xi-V-\tilde{h})\right)L^h
  \biggr].
  \label{eq:SigmaB2aresult2}
\end{eqnarray}
\end{widetext}

Next we consider the case $B=S^{\pm}$ and start with the evalution of
the kernel $\Sigma_{S^\pm}^{+,(2a)}$ from \eqref{eq:SigmaB2aresult} by
using
\begin{eqnarray}
  & &\sum_{k=x,y,z} L^{2k}\,P_i(z_i)\,\mathcal{B}_+^{(0)}\,P_j(z_j)\,L^{2k}
  \nonumber\\*
  & &\qquad=\left\{\begin{array}{ll}
  \pm\frac{\ii}{4}\frac{\Gamma^{3z}(z_0)}{\Gamma^a(z_0)}L^5_\pm,& 
  i=0,j=-,\\[2mm]
  \mp\frac{\ii}{4}\frac{\Gamma^{3z}(z_1)}{\Gamma^a(z_1)}L^5_\pm,& 
  i=1,j=-,\\[2mm]
  0,& \text{otherwise}. \end{array}\right.
\end{eqnarray}
Hence the double sum in \eqref{eq:SigmaB2aresult} reduces to a sum over
$i=0,1$, where the two terms have opposite signs and otherwise equal each
other up to the appearance of the rate $\tilde{\Gamma}_1$ in the second
term. As can be easily shown this sum vanishes in second order, i.e.
\begin{equation}
  \Sigma_{S^\pm}^{+,(2a)}(\Omega,\delta,\xi,\xi')=O(J_c^3).
\end{equation}
Similarly, the kernel $\Sigma_{S^\pm}^{-,(2a)}$ is evaluated using
\begin{equation}
\sum_{k=x,y,z}\!\!\! L^{2k}\,P_i(z_i)\,\mathcal{B}_-^{(0)}\,P_j(z_j)\,L^{2k}
\!=\!\left\{\begin{array}{ll}
  \!\!-\frac{\ii}{4}L^5_\pm,& i=0,j=\mp,\\[2mm]
  \!\!-\frac{\ii}{4}L^4_\pm,& i=\pm ,j=1,\\[2mm]
  0,& \text{otherwise}, \end{array}\right.
\end{equation}
which results in
\begin{widetext}
\begin{eqnarray}
  \Sigma_{S^\pm}^{-,(2a)}(\Omega,\delta,\xi,\xi')\Big|_{\Lambda=0}&=&
  -\frac{\ii}{2}\frac{J_R^2+J_L^2}{\Omega-\xi-\tilde{h}+\ii(\delta-\xi')}
  \biggl[\left(\mathcal{H}_1(\Omega)-\mathcal{H}_2(\xi+\tilde{h})\right)
  L^5_\pm+\left(\mathcal{H}_2(\Omega-\tilde{h})-\mathcal{H}_1(\xi)\right)
  L^4_\pm\biggr]\nonumber\\*
  & &\hspace{-30mm}
  -\frac{\ii}{2}\frac{J_RJ_L}{\Omega-\xi-\tilde{h}+\ii(\delta-\xi')}
  \biggl[\left(\mathcal{H}_1(\Omega+V)+\mathcal{H}_1(\Omega-V)
    -\mathcal{H}_2(\xi+V+\tilde{h})-\mathcal{H}_2(\xi-V+\tilde{h})\right)
  L^5_\pm\nonumber\\*
  & &+\left(\mathcal{H}_2(\Omega+V-\tilde{h})+\mathcal{H}_2(\Omega-V-\tilde{h})
    -\mathcal{H}_1(\xi+V)-\mathcal{H}_1(\xi-V)\right)
  L^4_\pm\biggr].
  \label{eq:SigmaB2aresult2pm}
\end{eqnarray}
\end{widetext}
A redefinition $\Lambda_c\rightarrow\Lambda_c'$ yields the same result for the
kernels $\Sigma_B^\pm$ with the replacement $J_c\rightarrow J_c'$ as can be
easily shown in the same way as for the Liouvillian. 

Finally, let us consider the kernel in first order in the regime $\Omega\gg
V,h_0$. Starting from \eqref{eq:RGB1} we obtain using
$\mathcal{K}_\Lambda(z)=\ii\,\Lambda/z+\ldots$ ($|z|\gg\Lambda$) 
\begin{eqnarray}
  \frac{d}{d\Lambda}\Sigma_B^{\pm,(1)}&=&\frac{1}{\Omega}\frac{1}{2\Lambda}\,
  \bar{G}_{12}^{(1)}\Big|_{\Lambda_c}\,\mathcal{B}_\pm^{(0)}\nonumber\\*
  & &\hspace{-10mm}\times
  (\xi_{12}+\ii\xi'-L_S^{(0)})\,
  \bar{G}_{\bar{2}\bar{1}}^{(1)}+O\left(\frac{\Lambda}{\Omega^2}\right).
\end{eqnarray}
Here the second vertex still depends on $\Lambda$.  When integrating this from
$\max\{V,h_0\}$ to $\Lambda_c=|\Omega|$ we obtain a contribution $\propto
1/\Omega$. Hence we deduce
\begin{equation}
  \Sigma_B^{\pm,(1)c}-\Sigma_B^{\pm,(1)}\Big|_{\max\{V,h_0\}}
  \propto\frac{1}{\Omega}\quad\text{for}\;\Omega\gg V,h_0.
\end{equation}
A similar analysis shows that the difference between
$\Sigma_B^{\pm,(2)}|_{\max\{V,h_0\}}$ and $\Sigma_B^{\pm,(2)c}$ is at least
$\propto 1/\Omega$.

\section{Longitudinal Spin-spin correlation functions}\label{sec:CF}
In this section we will use the results for the Liouvillian and the kernel to
calculate the correlation functions \eqref{eq:defPhi} and \eqref{eq:defchi}.
We first calculate the auxiliary correlation functions \eqref{eq:defbarCOmega}
using \eqref{eq:barCresult}. We start with the parametrization
\eqref{eq:Lparametrization} of the Liouvillian, which implies the form
\eqref{eq:resprojectors} for the resolvent in $C_{AB}^\pm(\Omega)$.  We
further deduce from the previous section that the kernels
$\Sigma_{S^z}^{\pm}(\Omega,\ii 0+)$ admit the parametrizations
\begin{eqnarray}
  \Sigma_{S^z}^+(\Omega,\ii 0+)&=&\ii\,L^{1z}+\ii\,L^{3z}\nonumber\\*
  & &+h_{S^z}^+(\Omega)\,L^h+\Gamma_{S^z}^{+c}(\Omega)\,L^c,
  \label{eq:S+}\\
  \Sigma_{S^z}^-(\Omega,\ii 0+)&=&h_{S^z}^-(\Omega)\,L^h
  +\Gamma_{S^z}^{-,3z}(\Omega)\,L^{3z}\nonumber\\*
  & &+\Gamma_{S^z}^{-,a}(\Omega)\,L^a+\Gamma_{S^z}^{-,c}(\Omega)\,L^c,
  \label{eq:S-}
\end{eqnarray}
where for example $\Gamma_{S^z}^{-,3z}(\Omega)=\frac{\pi}{2}\mathrm{tr}J_c^2=
\frac{\pi}{2}(J_R+J_L)^2$.  We stress that in contrast to the parametrization
of the Liouvillian we have not introduced additional factors of $\ii$ here.
The stationary reduced density matrix has the form
\begin{equation}
  \rho_S^{st}=\left(\begin{array}{cc}\rho_{\up\up}&0\\
      0&\rho_{\dw\dw}\end{array}\right),
\end{equation}
with $\rho_{\up\up}+\rho_{\dw\dw}=1$.

\subsection{Longitudinal correlation functions without magnetic field}
\begin{figure}[t]
  \includegraphics[scale=0.3,clip=true]{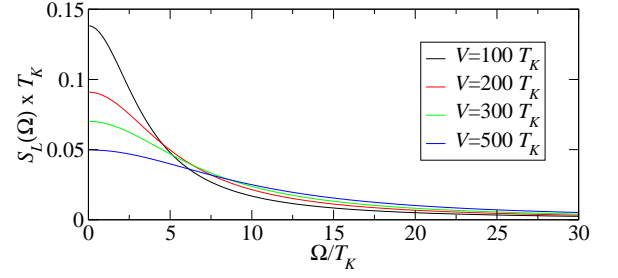}
  \caption{(color online) Longitudinal correlation function
    $S_L(\Omega)\equiv S_{S^zS^z}(\Omega)$ in the symmetric Kondo model ($r=1$)
    for various values of the applied voltage $V$.}
\label{fig:plot1}
\end{figure}

\begin{figure}[t]
  \includegraphics[scale=0.3,clip=true]{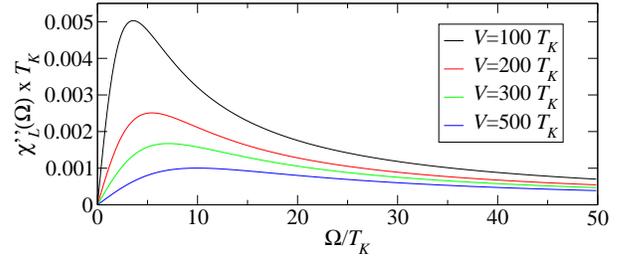}
  \caption{(color online) Imaginary part of the longitudinal susceptibility 
    $\chi_L''(\Omega)\equiv\chi_{S^zS^z}''(\Omega)$ in the symmetric Kondo
    model ($r=1$) for various values of the applied voltage $V$.}
\label{fig:plot2}
\end{figure}

\begin{figure}[t]
  \includegraphics[scale=0.3,clip=true]{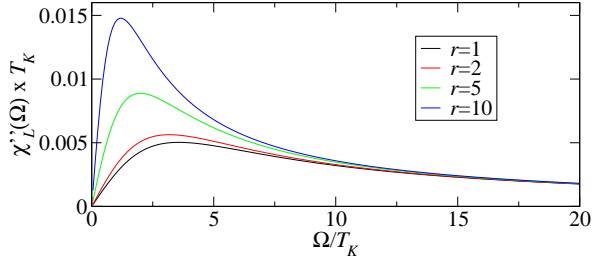}
  \caption{(color online) Imaginary part of the longitudinal susceptibility 
    $\chi_{S^zS^z}''(\Omega)$ for $V=100\,T_K$ and various values of the
    asymmetry ratio $r=J_L/J_R$.}
\label{fig:plot3}
\end{figure}

The stationary reduced density matrix can be determined using
\eqref{eq:pstdef}. Without magnetic field one simply finds
$\rho_{\up\up}=\rho_{\dw\dw}=1/2$. Furthermore, the rates
\eqref{eq:tildegamma1} and \eqref{eq:tildegamma2} are given by
\begin{equation}
  \tilde{\Gamma}_1=\tilde{\Gamma}_2=\pi\,J_RJ_L\,V.
\end{equation}
We now rewrite the resolvent using the projectors \eqref{eq:resprojectors} and
use
\begin{eqnarray}
  \mbox{Tr}_S\Bigl[\Sigma_{S^z}(\Omega)\,P_i(\Omega)\,
  \Sigma_{S^z}^+(\Omega,\ii 0+)\,\rho_S^{st}\Bigr]&=&-\frac{1}{2}\,\delta_{i1},
  \quad\label{eq:bla1}\\
  \mbox{Tr}_S\Bigl[\Sigma_{S^z}(\Omega)\,P_i(\Omega)\,
  \Sigma_{S^z}^-(\Omega,\ii 0+)\,\rho_S^{st}\Bigr]&=&\nonumber\\*
  & &\hspace{-20mm}\ii\frac{\pi}{4}(J_R+J_L)^2\,\delta_{i1},
\end{eqnarray}
where we have applied \eqref{eq:resultSigmaA} as well as
\eqref{eq:Avertexinitial}. We note in particular that the term
$\Gamma_{S^z}^{+c}(\Omega)\,L^c$ does not contribute to \eqref{eq:bla1}.  This
yields with \eqref{eq:barCresult} and
$\lambda_1(\Omega)=-\ii\,\Gamma^a(\Omega)$:
\begin{eqnarray}
  C^+_{S^zS^z}(\Omega)&=&\frac{\ii}{2}
  \frac{\Omega-Im\,\Gamma^a(\Omega)-\ii\,Re\,\Gamma^a(\Omega)}
  {(\Omega-Im\,\Gamma^a(\Omega))^2+Re\,\Gamma^a(\Omega)^2},\qquad
  \label{eq:resC+h0}\\
  C_{S^zS^z}^-(\Omega)&=&-\ii\frac{\pi}{2}\,(J_R+J_L)^2\,
  C^+_{S^zS^z}(\Omega).
  \label{eq:resC-h0}
\end{eqnarray}
Since
\begin{equation}
|Im\,\Gamma^a(\Omega)|~\sim~\Omega\,J_c~\ll~\Omega
  \label{eq:expansiondenominator}
\end{equation}
we can neglect $Im\,\Gamma^a(\Omega)$ in \eqref{eq:resC+h0}.  On the other
hand, the real part of $\Gamma^a$ in the denominator has to be kept as it
becomes large compared to $\Omega$ in the small-frequency limit.  Hence we
arrive at
\begin{eqnarray}
  S_{S^zS^z}(\Omega)&=&\frac{1}{2}\,\frac{Re\,\Gamma^a(\Omega)}
  {\Omega^2+Re\,\Gamma^a(\Omega)^2},\label{eq:resphih0}\\
  \chi_{S^zS^z}(\Omega)&=&\frac{\pi}{4}\,(J_R+J_L)^2\,
  \frac{Re\,\Gamma^a(\Omega)+\ii\,\Omega}{\Omega^2+Re\,\Gamma^a(\Omega)^2}.
  \label{eq:reschih0}
\end{eqnarray}
We note that the leading term of the susceptibility is of order $J_c^2$. The
correlation functions are plotted in Figs.~\ref{fig:plot1}--\ref{fig:plot3}.
We observe very good agreement with the results obtained by Fritsch and
Kehrein using the flow-equation method~\cite{FritschKehrein09ap,Kehreinbook}.

Let us further study the behavior of the correlation functions analytically.
For small values of the frequency $S_{S^zS^z}(\Omega)$ has the Lorentzian form
\begin{equation}
  S_{S^zS^z}(\Omega)=\frac{1}{2}
  \frac{\tilde{\Gamma}_1}{\Omega^2+\tilde{\Gamma}_1^2}
  \stackrel{\Omega\rightarrow 0}{\longrightarrow}
  \frac{1}{2\tilde{\Gamma}_1}=\frac{1}{2\pi\,J_RJ_L\,V},
\end{equation}
where we have used $Re\,\Gamma^a(0)=\tilde{\Gamma}_1$. This result for the
small-frequency regime agrees with conclusions drawn from a mapping of the
spin correlators to the one-particle Green's function of Majorana
fermions~\cite{Mao-03,ShnirmanMakhlin03}. On the other hand, in the limit of
large frequencies we find
\begin{equation}
  S_{S^zS^z}(\Omega)=\frac{\pi}{4}\frac{(J_R+J_L)^2}{\Omega}
  \propto\frac{1}{\Omega}\frac{1}{\ln^2\frac{\Omega}{T_K}}
  \label{eq:corrh0largeomega}
\end{equation}
in agreement with the flow-equation
method~\cite{FritschKehrein09ap,Kehreinbook}.  We note that the $J's$
appearing in the correlation function \eqref{eq:resphih0} have their origin in
the resolvent $1/(\Omega-L_S^{eff}(\Omega))$, hence the external frequency
$\Omega$ serves as a cut-off parameter in $\Lambda_c$. This results in the
logarithmic corrections at large frequencies.

The susceptibility in the limiting regimes reads
\begin{eqnarray}
  \chi_{S^zS^z}''(\Omega)&=&\frac{(1+r)^2}{4\pi\,r\,J_RJ_L\,V^2}\,\Omega,
  \quad\Omega\rightarrow 0,\quad\\
  \chi_{S^zS^z}''(\Omega)&=&S_{S^zS^z}(\Omega),
  \quad\Omega\rightarrow\infty.\label{eq:chipph0largew}
\end{eqnarray}
The first result shows a dependence of the gradient at small $\Omega$ on the
asymmetry ratio $r=J_L/J_R$, while the second result indicates the revival of
the fluctuation-dissipation theorem \eqref{eq:FDT} for $\Omega\gg V$. We note
that the derivation of \eqref{eq:chipph0largew} relies on the fact that the
coupling constants $J_c$ appearing in the kernel $\Sigma_{S^z}^-$ are
cut-off by the external frequency $\Omega$, as it was discussed at the end of
Sec.~\ref{sec:expl}.  Furthermore, the susceptibility
$\chi_{S^zS^z}''(\Omega)$ has a maximum at $\Omega\approx\tilde{\Gamma}_1$,
where it takes the value
\begin{equation}
  \chi_{S^zS^z}''(\Omega\approx\tilde{\Gamma}_1)\approx\frac{(1+r)^2}{8rV}.
\end{equation}
This behavior was also deduced using the flow-equation
method~\cite{Kehreinbook}.

\begin{figure}[t]
  \includegraphics[scale=0.3,clip=true]{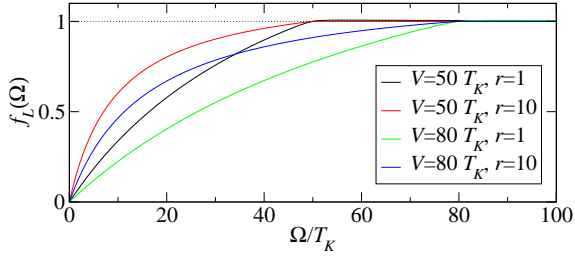}
  \caption{(color online) Longitudinal fluctuation-dissipation ratio
    $f_L(\Omega)$ for various values of the asymmetry $r$ and applied voltage
    $V$. In order to get a smooth behavior at $\Omega\approx V$ we have kept
    the $\arctan$ in the definition of $\mathcal{H}_i$ in this region. The
    dotted line is a guide to the eye.}
\label{fig:plot4}
\end{figure}
In order to investigate the revival of the fluctuation-dissipation theorem we
introduce the longitudinal fluctuation-dissipation
ratio~\cite{Mao-03,ShnirmanMakhlin03,MitraMillis05}
\begin{equation}
  f_L(\Omega)=\frac{\chi_{S^zS^z}''(\Omega)}{S_{S^zS^z}(\Omega)},
  \label{eq:defratio}
\end{equation}
which is in equilibrium simply given by
$f_L(\Omega)=\tanh\frac{\Omega}{2T}\rightarrow\sgn{\Omega}$ ($T\rightarrow
0$). Using our results \eqref{eq:resphih0} and \eqref{eq:reschih0} we obtain
\begin{equation}
  f_L(\Omega)=\frac{\pi}{2}
  \frac{(J_R+J_L)^2}{Re\,\Gamma^a(\Omega)}\,\Omega,
\end{equation}
which is plotted in Fig.~\ref{fig:plot4}. We find $f_L(\Omega>V)=1$, i.e. the
equilibrium result, whereas for small frequencies we get
\begin{equation}
  f_L(\Omega\ll V)=\frac{(1+r)^2}{2r}\,\frac{\Omega}{V},
\end{equation}
in agreement with Refs.~\onlinecite{Mao-03,ShnirmanMakhlin03}. We note that
$f_L(\Omega<V)$ increases with increasing asymmetry $r$ as the coupling of
the voltage to the dot becomes less effective.

\subsection{Longitudinal correlation functions in a weak magnetic field 
  ($\bs{V>\tilde{h}}$)}
\begin{figure}[t]
  \includegraphics[scale=0.3,clip=true]{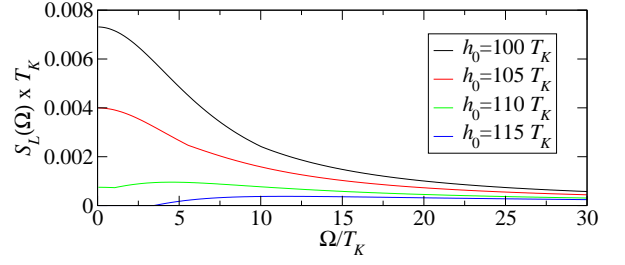}
  \caption{(color online) Longitudinal correlation function
    $S_{S^zS^z}(\Omega)$ in the symmetric Kondo model ($r=1$) for $V=100\,T_K$
    and various values of the applied magnetic field $h_0$.  For
    $h_0=115\,T_K$ we already have $V<\tilde{h}$, which implies
    $S_{S^zS^z}(\Omega<\tilde{h}-V)=0$ in order $J_c^2$ (see
    Sec.~\ref{sec:longlargeh}).}
\label{fig:plot5}
\end{figure}
In the presence of an external magnetic field the stationary reduced density
matrix is given by $\rho_{\up\up}=1+M$, $\rho_{\dw\dw}=1-M$,
$\rho_{\up\dw}=\rho_{\dw\up}=0$, with the magnetization in leading
order~\cite{SchoellerReininghaus09} (see also
Refs.~\onlinecite{ParcolletHooley02,Rosch-03prl,Paaske-04prb1,Rosch-05})
\begin{equation}
  M=-\frac{1}{2}\frac{\Gamma^{3z}(0)}{\Gamma^a(0)}
  =-\frac{1}{2}\frac{(1+r)^2\tilde{h}}{(1+r^2)\tilde{h}+2rV}.
\end{equation}
To evaluate the correlation function we now use
\begin{equation}
  \begin{split}
  &\mbox{Tr}_S\Bigl[\Sigma_{S^z}(\Omega)\,P_i(\Omega)\,
  \Sigma_{S^z}^+(\Omega,\ii 0+)\,\rho_S^{st}\Bigr]\\
  &\qquad\qquad=
  \left\{\begin{array}{ll}M\frac{\Gamma^{3z}(\Omega)}{\Gamma^a(\Omega)},&i=0,\\
      -\frac{1}{2}-M\frac{\Gamma^{3z}(\Omega)}{\Gamma^a(\Omega)},&i=1,\\
      0,& i=\pm,
    \end{array}\right.
  \end{split}
\end{equation}
where the term $\Gamma_{S^z}^{+c}(\Omega)\,L^c$ again does not contribute.
From this a straightforward calculation using \eqref{eq:expansiondenominator}
yields the correlation function up to $O(J_c^2)$
\begin{equation}
  S_{S^zS^z}(\Omega)=\frac{1}{2}
  \frac{Re\,\Gamma^a(\Omega)+2M\,\Gamma^{3z}(\Omega)}
  {\Omega^2+Re\,\Gamma^a(\Omega)^2},
  \label{eq:resultphifiniteh}
\end{equation}
where the zero-frequency $\delta$-peak does not appear because of our
definition \eqref{eq:defPhi}. The suppression of the correlation function by
the finite magnetic field is shown in Fig.~\ref{fig:plot5}, which agrees very
well with similar plots obtained using the flow-equation
method~\cite{FritschKehrein09}. In the zero-frequency limit we find
\begin{equation}
  S_{S^zS^z}(\Omega\rightarrow 0)=\frac{4r^2}{\pi J_RJ_L}
  \frac{(1+r+r^2)\tilde{h}+rV}{((1+r^2)\tilde{h}+2rV)^3}(V-\tilde{h}),
\end{equation}
while the leading term $\propto 1/\Omega$ in the large frequency regime is
given by \eqref{eq:corrh0largeomega} (including the logarithmic corrections in
the coupling constants). Furthermore we observe a weak feature at
$\Omega=V-\tilde{h}$ which has for $\tilde{\Gamma}_2\ll
V-\tilde{h}\ll\tilde{h}$ the line shape
\begin{eqnarray}
  S_{S^zS^z}(\Omega)\!&\approx&\!
  \frac{\pi J_RJ_L}{8\Omega^2}\biggl[
    \bigl(2+r+2r^2+4M(1+r)^2\bigr)\frac{\tilde{h}}{r}+3V
  \nonumber\\*
  & &\hspace{-8mm}
  +\Omega+\frac{2}{\pi}(\Omega-V+\tilde{h})
    \arctan\frac{\Omega-V+\tilde{h}}{\tilde{\Gamma}_2}\biggr].\quad
  \label{eq:featureV-h}
\end{eqnarray}
Similar features appear at $\Omega=\tilde{h},V+\tilde{h}$.

\begin{figure}[t]
  \includegraphics[scale=0.3,clip=true]{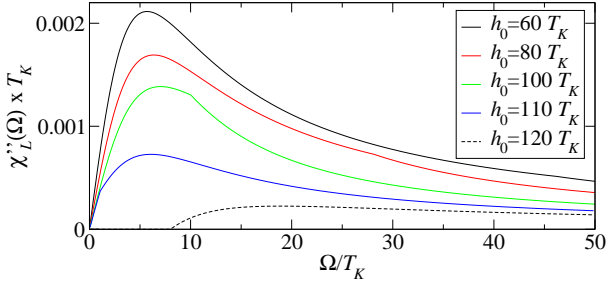}
  \caption{(color online) Imaginary part of the longitudinal susceptibility
    $\chi_{S^zS^z}''(\Omega)$ in the symmetric Kondo model ($r=1$) for
    $V=100\,T_K$ and various values of the applied magnetic field $h_0$. For
    $h_0=120\,T_K$ we have $V<\tilde{h}$, which implies
    $\chi_{S^zS^z}(\Omega<\tilde{h}-V)=0$ in order $J_c^2$ (see
    Sec.~\ref{sec:longlargeh}).}
\label{fig:plot6}
\end{figure}
For the calculation of the susceptibility we need
\begin{equation}
  \begin{split}
  &\mbox{Tr}_S\Bigl[\Sigma_{S^z}(\Omega)\,P_i(\Omega)\,
  \Sigma_{S^z}^-(\Omega,\ii 0+)\,\rho_S^{st}\Bigr]\\
  &\qquad\qquad=\left[\frac{\ii}{2}\,\Gamma_{S^z}^{-,3z}(\Omega)
    +\ii\,M\,\Gamma_{S^z}^{-,a}(\Omega)\right]\delta_{i1},
  \end{split}
\end{equation}
which directly yields
\begin{eqnarray}
  \chi_{S^zS^z}'(\Omega)&=&
  \frac{1}{\Omega^2+Re\,\Gamma^a(\Omega)^2}
  \biggl[-M\,\Omega\,Im\,\Gamma_{S^z}^{-,a}(\Omega)\nonumber\\*
  & &\hspace{-20mm}+\left(\frac{\pi}{4}(J_R+J_L)^2+
    M\,Re\,\Gamma_{S^z}^{-,a}(\Omega)
    \right)Re\,\Gamma^a(\Omega)\biggr],
    \quad\label{eq:chipresultfiniteh}\\
  \chi_{S^zS^z}''(\Omega)&=&
  \left[\frac{\pi}{4}(J_R+J_L)^2+M\,Re\,\Gamma_{S^z}^{-,a}(\Omega)\right]
  \nonumber\\*
  & &\qquad\times\frac{\Omega}{\Omega^2+Re\,\Gamma^a(\Omega)^2}.
  \label{eq:chippresultfiniteh}
\end{eqnarray}
In \eqref{eq:chipresultfiniteh} we have kept the terms in the second line,
which are of $O(J_c^4)$, as due to $\Omega\,Im\,\Gamma_{S^z}^{-,a}(\Omega)
\rightarrow 0$ ($\Omega\rightarrow 0$) they become dominant in the
small-frequency limit.  For larger frequencies these terms have to be
neglected. Thus the static susceptibility \eqref{eq:staticchi} is given in
leading order by
\begin{equation}
  \chi_{S^zS^z}=-\frac{r(1+r)^2V}{((1+r^2)\tilde{h}+2rV)^2}
\end{equation}
in agreement with the
literature~\cite{ParcolletHooley02,Rosch-03prl,Paaske-04prb1,Rosch-05,SchoellerReininghaus09}.
For larger frequencies the real part of the susceptibility possesses
logarithmic features at $\Omega=\tilde{h},V\pm\tilde{h}$ due to the term
$Im\,\Gamma_{S^z}^{-,a}(\Omega)$. For example,
\begin{equation}
  \begin{split}
  &\chi_{S^zS^z}'(\Omega\approx\tilde{h})\approx\\
  &\quad-M(J_R+J_L)^2\frac{\Omega-\tilde{h}}{2\Omega^2}
  \ln\frac{\Lambda_c}{\sqrt{(\Omega-\tilde{h})^2+\tilde{\Gamma}_2^2}}
  +\ldots,
  \end{split}
  \label{eq:logs}
\end{equation}
where the terms represented by the dots do not contain any logarithmic
features at $\Omega=\tilde{h}$. The imaginary part of the susceptibility is
plotted in Fig.~\ref{fig:plot6}. It has a finite gradient at $\Omega=0$ given
by
\begin{equation}
  \chi_{S^zS^z}''(\Omega\rightarrow 0)=\frac{2}{\pi}
  \frac{r^2(1+r)^2V}{((1+r^2)\tilde{h}+2rV)^3}\frac{\Omega}{J_RJ_L}
\end{equation}
as well as a maximum at $\Omega\approx\tilde{\Gamma}_1$, where it takes the
value
\begin{equation}
  \chi_{S^zS^z}''(\Omega\approx\tilde{\Gamma}_1)\approx
  \frac{r}{2}\frac{(1+r)^2V}{((1+r^2)\tilde{h}+2rV)^2}
  =-\frac{1}{2}\chi_{S^zS^z}.
\end{equation}
In the large-frequency limit $\chi_{S^zS^z}''(\Omega\gg V,\tilde{h})$
coincides with the correlation function \eqref{eq:corrh0largeomega}.
Furthermore, the imaginary part of the susceptibility has features at
$\Omega=\tilde{h},V\pm\tilde{h}$ which have their origin in the function
$Im\,\mathcal{H}_2$ contained in $Re\,\Gamma_{S^z}^{-,a}(\Omega)$ and hence
have a line shape similar to \eqref{eq:featureV-h}.

\begin{figure}[t]
  \includegraphics[scale=0.3,clip=true]{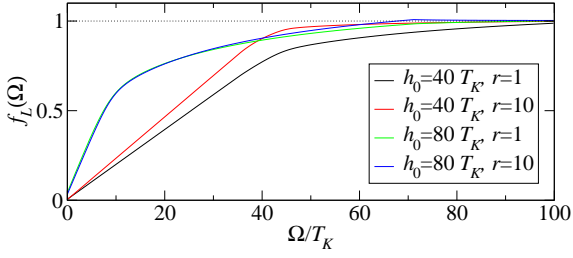}
  \caption{(color online) Fluctuation-dissipation ratio
    $f_L(\Omega)$ for $V=80\,T_K$ and different values of the asymmetry $r$ and
    magnetic field $h_0$. In order to get a smooth behavior at $\Omega\approx
    V-\tilde{h}$ we have kept the $\arctan$ in the definition of
    $\mathcal{H}_i$. For $V<\tilde{h}$ the fluctuation-dissipation ratio is
    independently of $\Omega$ given by $f_L(\Omega)=1$. The dotted line is a
    guide to the eye.}
\label{fig:plot7}
\end{figure}
The fluctuation-dissipation ratio $f_L(\Omega)$ defined in \eqref{eq:defratio}
reads in the presence of a magnetic field
\begin{equation}
  f_L(\Omega)=
  \frac{\frac{\pi}{2}(J_R+J_L)^2\Omega+2M\,Re\,\Gamma_{S^z}^{-,a}(\Omega)\,
    \Omega}
  {Re\,\Gamma^a(\Omega)+\pi(J_R+J_L)^2 M\tilde{h}},
  \label{eq:fLVh}
\end{equation}
which is plotted in Fig.~\ref{fig:plot7}. Larger values of the magnetic field
push the system closer to its equilibrium behavior, as only those lead
electrons in the energy interval $V-\tilde{h}$ can couple to the dot and thus
induce the non-equilibrium behavior. We note, however, that the equilibrium
result $f_L(\Omega)=1$ is only reached for $\Omega>V+\tilde{h}$.  Furthermore,
increasing the asymmetry $r$ drives the system towards the equilibrium
situation as the coupling of the voltage to the dot becomes less effective.
This effect is suppressed by increasing the magnetic field as overall less
electrons couple to the dot. For small frequencies we obtain
\begin{equation}
  f_L(\Omega\rightarrow 0)=\frac{1}{2}\frac{(1+r)^2V}{(1+r+r^2)\tilde{h}+rV}
  \frac{\Omega}{V-\tilde{h}}.
\end{equation}

\subsection{Longitudinal correlation functions in a strong magnetic field 
  ($\bs{V<\tilde{h}}$)}\label{sec:longlargeh}
\begin{figure}[t]
  \includegraphics[scale=0.3,clip=true]{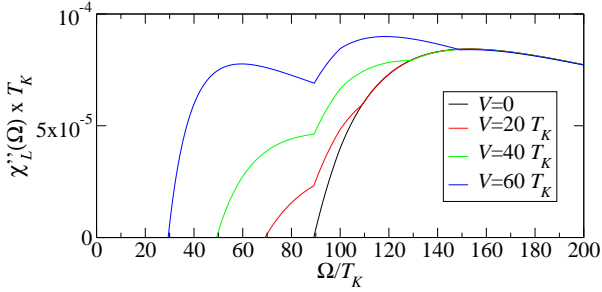}
  \caption{(color online) Imaginary part of the longitudinal susceptibility
    $\chi_{S^zS^z}''(\Omega)$ in the symmetric Kondo model ($r=1$) for
    $h_0=100\,T_K$ and various values of the applied voltage $V$. The
    correlation function is given by
    $S_{S^zS^z}''(\Omega)=\chi_{S^zS^z}(\Omega)$. For $\Omega<\tilde{h}-V$ the
    susceptibility vanishes in order $J_c^2$. The line shape close to
    $\Omega=\tilde{h}$ is given by \eqref{eq:kinkexample}.}
\label{fig:plot8}
\end{figure}
\begin{figure}[t]
  \includegraphics[scale=0.3,clip=true]{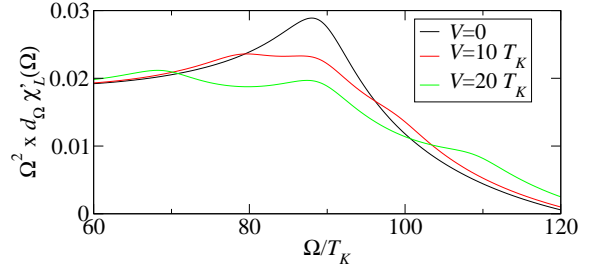}
  \caption{(color online) Derivative of the real part of the longitudinal 
    susceptibility $\Omega^2\,\frac{d}{d\Omega}\chi_{S^zS^z}'(\Omega)$ in the
    symmetric Kondo model ($r=1$) for $h_0=100\,T_K$ and various values of the
    applied voltage $V$. We observe characteristic logarithmic features at
    $\Omega=\tilde{h},\tilde{h}\pm V$.}
\label{fig:plot8a}
\end{figure}
In the case of a strong magnetic field, $V<\tilde{h}$, the correlation
functions up to quadratic order in the coupling are still given by
\eqref{eq:resultphifiniteh}, \eqref{eq:chipresultfiniteh} and
\eqref{eq:chippresultfiniteh}, respectively, where the magnetization is simply
$M=-1/2$. One can easily show using
$Im\,\mathcal{H}_2(\Omega)=\frac{\pi}{2}|\Omega|$ that
$\chi_{S^zS^z}''(\Omega>0)=S_{S^zS^z}(\Omega>0)$, which implies the equilibrium
result $f_L(\Omega>0)=1$.  Furthermore, the correlation function vanishes
identically in order $J_c^2$ for $\Omega<\tilde{h}-V$.  Physically the Kondo
spin is in its ground state $\ket{\dw}$ and the energy difference to the state
$\ket{\up}$ due to the external magnetic field is given by $\tilde{h}$.  Hence
one has to apply at least the frequency $\tilde{h}-V$ to obtain any response
from the spin, where the energy $V$ is provided by the applied voltage.

This has to be contrasted with the result for the susceptibility in the
equilibrium Kondo model derived by Garst et al.~\cite{Garst-05}. They used a
relation between the inelastic electron scattering and the correlation
function to show that the susceptibility in equilibrium has the
small-frequency behavior $\chi_{S^zS^z}''(\Omega)\propto J_c^4\,\Omega$, i.e.
it is non-zero for $\Omega<\tilde{h}$. This linear behavior was also observed
by Costi and Kieffer~\cite{CostiKieffer96} as well as Hewson~\cite{Hewson06}
using a numerical renormalization group calculation. In analogy, we expect the
non-equilibrium correlation functions to be nonzero for $\Omega<\tilde{h}-V$
in higher order in $J_c$. The consistent calculation of terms $\sim J_c^4$ in
the real-time RG procedure applied here would involve, however, 5-loop
diagrams and is hence beyond the scope of this work.

The correlation function in the regime $V<\tilde{h}$ is plotted in
Fig.~\ref{fig:plot8}. We find excellent agreement with numerical results
recently obtained by Fritsch and Kehrein using the flow-equation
method~\cite{FritschKehrein09}. In particular, we observe a splitting of the
sharp edge at $\Omega=\tilde{h}$ due to the applied voltage, which leads to
characteristic features at $\Omega=\tilde{h},\tilde{h}\pm V$. Using our result
\eqref{eq:resultphifiniteh} we can derive analytic expressions for the line
shape close to these frequencies. For example, at $\Omega\approx\tilde{h}$ we
find
\begin{eqnarray}
  \chi_{S^zS^z}''(\Omega)&\approx&
  \frac{\pi J_RJ_L\,V}{4\Omega^2}
  +\frac{\pi}{8}(J_R+J_L)^2\frac{\Omega-\tilde{h}}{\Omega^2}\nonumber\\*
  & &\hspace{-10mm}
  +\frac{1}{4}(J_R^2+J_L^2)\frac{\Omega-\tilde{h}}{\Omega^2}\,
  \arctan\frac{\Omega-\tilde{h}}{\tilde{\Gamma}_2}.
  \label{eq:kinkexample}
\end{eqnarray}
The first term shows that the gradient of $\chi_{S^zS^z}''(\Omega)$ will
become negative for $\Omega<\tilde{h}$ if the applied voltage is large enough,
i.e.  $V>\tilde{h}/2$. In the vicinity of $\Omega=\tilde{h}\pm V$ the
correlation function shows similar kink-like behavior \eqref{eq:kinkexample}.
The physical origin of these kinks lies in the fact that at each of the
energies $\Omega=\tilde{h},\tilde{h}\pm V$ a new process sets in, which
involves a spin-flip on the dot costing the Zeeman energy $\tilde{h}$ as well
as the virtual hopping of an electron on and off the dot gaining or costing
the energy $-V$, $0$, or $V$, respectively. The real part
$\chi_{S^zS^z}'(\Omega)$ of the susceptibility shows logarihmic features
\eqref{eq:logs} at $\Omega=\tilde{h},\tilde{h}\pm V$ as is shown in
Fig.~\ref{fig:plot8a}.\pagebreak We stress that the splitting of the sharp
edge at $\Omega=\tilde{h}$ is a true non-equilibrium effect.

\section{Transverse correlation functions}\label{sec:TCF}
Finally let us discuss the transverse correlation functions in the presence of
a magnetic field. We note that by virtue of \eqref{eq:spectralbarC} we
can restrict ourselves to the $S^-S^+$-correlations. The corresponding kernels
up to second order in $J_c$ were calculated in \eqref{eq:initB+},
\eqref{eq:initB-}, \eqref{eq:Sigma+B1result}, \eqref{eq:Sigma-B1result}, 
\eqref{eq:Sigma+B2bresult}--\eqref{eq:Sigma-B2cresult} as well as
\eqref{eq:SigmaB2aresult2pm}. We will first discuss the susceptibility and
present the results for the correlation function afterwards.

\begin{figure}[t]
  \includegraphics[scale=0.3,clip=true]{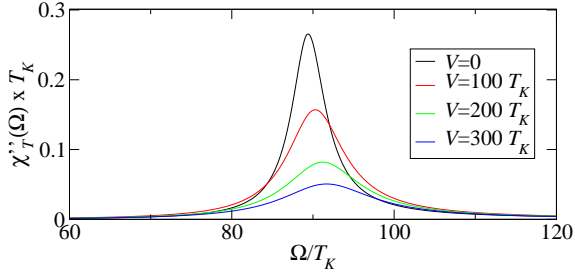}
  \caption{(color online) Imaginary part of the transverse susceptibility
    $\chi_T''(\Omega)\equiv\chi_{S^-S^+}''(\Omega)$ in the symmetric Kondo
    model ($r=1$) for $h_0=100\,T_K$ and various values of the applied voltage
    $V$.}
\label{fig:plot9}
\end{figure}
\begin{figure}[t]
  \includegraphics[scale=0.3,clip=true]{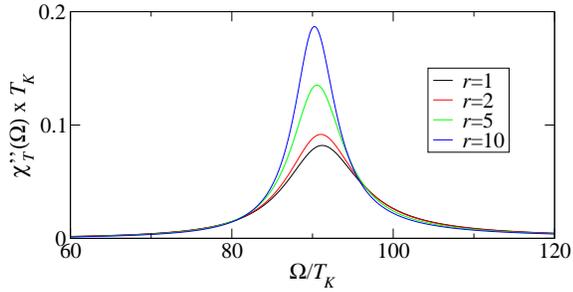}
  \caption{(color online) Imaginary part of the transverse susceptibility 
    $\chi_{S^-S^+}''(\Omega)$ for $V=200\,T_K$, $h_0=100\,T_K$, and various
    values of the asymmetry ratio $r=J_L/J_R$. The result is invariant under
    $r\rightarrow 1/r$.}
\label{fig:plot10}
\end{figure}
In order to derive the susceptibility we start with the parametrization 
\begin{equation}
  \Sigma_{S^+}^-(\Omega,\ii 0+)=\Gamma_{S^+}^{-,2}\,L^2_+
  +\Gamma_{S^+}^{-,3}\,L^3_++\Gamma_{S^+}^{-,5}(\Omega)\,L^5_+,
\end{equation}
where for example
$\Gamma_{S^+}^{-,2}=-2\ii-\ii\,\mathrm{tr}J_c+\ii\frac{3}{4}\mathrm{tr}J_c^2=
-2\ii-\ii(J_R+J_L)+\ii\frac{3}{4}(J_R+J_L)^2$.  Note that we already indicated
that the explicit frequency dependence in order $J_c^2$ appears in
$\Gamma_{S^+}^{-,5}$ exclusively. (There is of course an implicit frequency
dependence of $\Gamma_{S^+}^{-,2}$ and $\Gamma_{S^+}^{-,3}$ through $J_c$.)
Now using
\begin{eqnarray}
  & &\mbox{Tr}_S\Bigl[\Sigma_{S^-}(\Omega)\,P_i(\Omega)\,
  \Sigma_{S^+}^-(\Omega,\ii 0+)\,\rho_S^{st}\Bigr]\nonumber\\*[1mm]
  & &\qquad=\ii\left(\Gamma_{S^+}^{-,3}+\bigl(\Gamma_{S^+}^{-,2}
    +2\Gamma_{S^+}^{-,5}(\Omega)\bigr)M\right)\delta_{i+}\qquad
\end{eqnarray}
we obtain
\begin{equation}
  \chi_{S^-S^+}(\Omega)=\ii\,
  \frac{\Gamma_{S^+}^{-,3}+\bigl(\Gamma_{S^+}^{-,2}
    +2\Gamma_{S^+}^{-,5}(\Omega)\bigr)M}
  {\Omega-h(\Omega)+\ii\Gamma_2(\Omega)},
  \label{eq:transsusc}
\end{equation}
where we have introduced the short-hand notation
$\Gamma_2(\Omega)=\Gamma^a(\Omega)+\Gamma^c(\Omega)$. For $h_0=0$ we find
$\chi_{S^-S^+}(\Omega)=2\chi_{S^zS^z}(\Omega)$. The transverse susceptibility
has a peak at the solution of
\begin{equation}
  \Omega-Re\,h(\Omega)-Im\,\Gamma_2(\Omega)=0,
\end{equation}
which is up to first order solved by
\begin{equation}
  \Omega=\left(1-\frac{1}{2}(J_R+J_L)\right)h_0=\tilde{h}.
\end{equation}
In a finite magnetic field the spin on the dot will be in its ground state
$\ket{\dw}$. The energy difference to the excited state $\ket{\up}$ is given
by $\tilde{h}$, leading to an enhanced response of the system at this
frequency. At the peak the imaginary part of the susceptibility takes the
value
\begin{equation}
  \chi_{S^-S^+}''(\Omega\approx\tilde{h})\approx-\frac{2M}{\tilde{\Gamma}_2},
\end{equation}
as is shown in Figs.~\ref{fig:plot9} and~\ref{fig:plot10}. The peak is
suppressed by increasing the voltage, since this reduces the probability for
the Kondo spin to be in its ground state. On the other hand, for a fixed value
of the voltage the peak increases with increasing asymmetry ratio as the
coupling of the voltage to the dot becomes less effective. The width of the
peak is up to order $J_c^2$ given by
\begin{equation}
  Re\,\Gamma_2(\tilde{h})-Im\,h(\tilde{h})=\tilde{\Gamma}_2
\end{equation}
with the limiting cases
\begin{eqnarray}
  V\ll\tilde{h}:& &\frac{\pi}{4}(J_R+J_L)^2\,h_0,\\
  \tilde{h}\ll V:& &\pi J_RJ_LV.
\end{eqnarray}
In the equilibrium limit, $V=0$, this corresponds to the result obtained in
Ref.~\onlinecite{GoetzeWoelfle71}. The real part of the transverse
susceptibility possesses logarithmic features similar to \eqref{eq:logs} at
$\Omega=\tilde{h},\tilde{h}\pm V$.

Using a pseudo-fermion representation of the Kondo spin together with
non-equilibrium perturbation theory Paaske et al.~\cite{Paaske-04prb2}
previously obtained the transverse susceptibility.  In order to compare these
results to \eqref{eq:transsusc} we first make the approximations
$h(\Omega)\rightarrow h_0$ and $\Gamma_2(\Omega)\rightarrow\tilde{\Gamma}_2$.
In the limit $h_0\rightarrow 0$ we then obtain using
$\Gamma_{S^+}^{-,3}=\frac{\pi}{2}(J_R+J_L)^2$
\begin{equation}
  \chi_{S^-S^+}(\Omega)\approx\ii\frac{\pi}{2}
  \frac{(J_R+J_L)^2}{\Omega+\ii\tilde{\Gamma}_2}.
\end{equation}
In the regime $h_0\gg\tilde{\Gamma}_2$ we have $M=O(J_c^0)$, thus we can
neglect $\Gamma_{S^+}^{-,3}$ as well as $\Gamma_{S^+}^{-,5}$ in the numerator
in \eqref{eq:transsusc}, which results in
\begin{equation}
  \chi_{S^-S^+}(\Omega)\approx\frac{2M}{\Omega-h_0+\ii\tilde{\Gamma}_2}.
\end{equation}
These approximations agree with the results obtained in
Ref.~\onlinecite{Paaske-04prb2} (we have to replace $h_0\rightarrow -B$ due to
a different sign in the definition of the bare dot Hamiltonian $H_S$).
Furthermore, in the regime $\Omega,h_0<V$ we can use
\begin{equation}
  \Gamma_{S^+}^{-,3}+
  \bigl(\Gamma_{S^+}^{-,2}+2\Gamma_{S^+}^{-,5}(\Omega)\bigr)\,M
  \rightarrow-2\frac{M}{h_0}\tilde{\Gamma}_2-2\ii\,M,
\end{equation}
where we have replaced
$Im\,\mathcal{H}_i(\Omega)\rightarrow\frac{\pi}{2}|\Omega|$ in the real part
of $\Gamma_{S^+}^{-,5}$, to obtain
\begin{equation}
  \chi_{S^-S^+}(\Omega)\approx\frac{2M}{h_0}
  \frac{h_0-\ii\tilde{\Gamma}_2}{\Omega-h_0+\ii\tilde{\Gamma}_2}.
  \label{eq:Paaske}
\end{equation}
This confirms a conjecture by Paaske et al.~\cite{Paaske-04prb2}.  We would
like to stress, however, that our result \eqref{eq:transsusc} goes beyond the
approximation \eqref{eq:Paaske}.

\begin{figure}[t]
  \includegraphics[scale=0.3,clip=true]{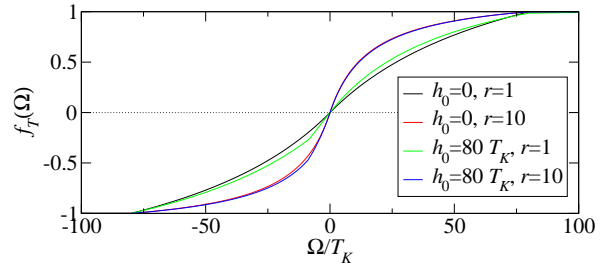}
  \caption{(color online) Transverse fluctuation-dissipation ratio
    $f_T(\Omega)$ for $V=80\,T_K$ and various values of the asymmetry ration
    $r$ and the applied magnetic field $h_0$.  The plot shows good agreement
    with similar results obtained in Ref.~\onlinecite{MitraMillis05}. The
    dotted line is a guide to the eye.}
\label{fig:plot11}
\end{figure}

In analogy to the susceptibility one finds for the correlation function
\begin{equation}
  \begin{split}
    &S_{S^-S^+}(\Omega)=\\
    &\frac{Re\,\Gamma_2(\Omega)-Im\,h(\Omega)
      -\frac{\pi}{2}M(J_R+J_L)^2(\Omega-h_0)}
  {(\Omega\!-\!Re\,h(\Omega)\!-\!Im\,\Gamma_2(\Omega))^2\!+\!
    (Re\,\Gamma_2(\Omega)\!-\!Im\,h(\Omega))^2},
  \end{split}
\end{equation}
where we have neglected all terms of order $J_c^3$ in the numerator. This
allows the calculation of the transverse fluctuation-dissipation
ratio
\begin{equation}
  f_T(\Omega)=\frac{\chi_{S^-S^+}''(\Omega)}{S_{S^-S^+}(\Omega)}
  \label{eq:defratiot}
\end{equation}
which is plotted in Fig.~\ref{fig:plot11}. For negative frequencies
$\Omega<-V$ the fluctuation-dissipation ratio takes the value
$f_T(\Omega)=-1$, whereas for frequencies $\Omega>V$ we find $f_T(\Omega)=1$,
thus recovering the equilibrium situation in these limits.  As for the
longitudinal fluctuation-dissipation ration we observe that increasing the
magnetic field or the asymmetry ratio $r$ drives the system towards the
equilibrium situation. 

\section{Conclusions}
In this article we have generalized the real-time renormalization group method
in frequency space to allow the calculation of dynamical correlation functions
of arbitrary dot operators in systems describing spin and/or orbital
fluctuations. We applied this to the two-lead Kondo model in a magnetic field,
where we calcualted the longitudinal and transverse spin-spin correlation and
response functions up to second order in the exchange coupling. We wish to
stress that within this formalism the Kondo spin is directly represented by
matrices in Liouville space, hence there is no need to apply a pseudo-fermion
representation. Specifically, we derived the two-loop RG equations for the dot
operators and solved them analytically up to order $J_c^2$ in the
weak-coupling regime. Here $J_c$ denotes the effective coupling at the energy
scale $\Lambda_c=\max\{V,h_0\}$ which has to satisfy $\Lambda_c\gg T_K$. Our
results show several features attributed to the non-equilibrium situation,
e.g. the splitting of the edge at $\Omega=\tilde{h}$ of the longitudinal
correlation function in a strong magnetic field or the suppression of the peak
in the transverse susceptibility by a finite applied voltage. Furthermore, we
find very good agreement with results for the longitudinal correlation
function recently obtained by Fritsch and Kehrein using the flow-equation
method~\cite{FritschKehrein09ap,FritschKehrein09}. A particular advantage of
our approach is the possibility to obtain analytic expressions for all
correlation functions in the weak coupling limit.

We have calculated the spin-spin correlation functions for the nonequilibrium
Kondo model in the weak coupling regime $\Lambda_c\gg T_K$. The regime of
strong coupling, $\Lambda_c < T_K$, is still an open problem. In this case,
the exchange couplings $J_c$ become of order $O(1)$ and a controlled
truncation of the RG equations is no longer possible. Within the present
RTRG-FS method it was shown in Ref.~\onlinecite{Schoeller09} that the
relaxation/dephasing rates saturate to the Kondo temperature in the strong
coupling regime. As an effect the coupling constants do not diverge as in poor
man scaling methods but remain finite. However, the numerical solution of the
RG equations in lowest order showed an instability against an exponentionally
small change in the initial condition for the relaxation/dephasing rates.
Although it was possible to find excellent agreement for the temperature
dependence of the linear conductance with NRG calculations, it was necessary
to fine tune the initial condition for the rates. Therefore, up to now, it is
not yet clear whether a controlled solution of the strong coupling regime is
possible by using RTRG-FS.

The nonequilibrium Kondo model describes the spin fluctuation (or Coulomb
blockade) regime of the more general nonequilibrium Anderson impurity model
(for a systematic derivation of the Kondo model from the Anderson model using
a Schrieffer-Wolff transformation, see e.g. Ref.~\onlinecite{Korb-07}). In
this model the single-particle spectral function is of most interest, which
has recently been studied within NRG in a scattering wave basis
\cite{Anders08prl,Anders08}. The present RTRG-FS method can also be applied to
this model but, as explained in detail in Ref.~\onlinecite{Schoeller09}, in
the charge fluctuation regime it is not yet clear whether a well-defined weak
coupling regime exists at zero temperature. At resonance (i.e. when the
renormalized single-particle level is identical to one of the chemical
potentials of the leads) there is no energy scale except the broadening of the
level itself and the expansion parameter is of order $O(1)$. Nevertheless the
results obtained in Ref.~\onlinecite{SchoellerKoenig00} (using a previous
version of the real-time RG method) after an a priori uncontrolled truncation
of the RG equations were in excellent agreement with the Bethe Ansatz
solution~\cite{TsvelikWiegmann83} for the equilibrium occupation of the local
level. These and related topics are of high interest and will be the subject
of forthcoming research.

\section*{Acknowledgments}
We would like to thank Benjamin Doyon, Markus Garst, Stefan Kehrein, Verena
K\"orting, Mikhail Pletyukhov, and Frank Reininghaus for valuable discussions.
This work was supported the DFG-Forschergruppe 723 ``Functional
Renormalization Group in Correlated Fermion Systems'' and the
DFG-Forschergruppe 912 ``Coherence and
relaxation properties of electron spins''.

\appendix
\section{Derivation of (\ref{eq:rel1}) and (\ref{eq:rel2})}\label{sec:Crel}
Obviously the time-dependent correlation function can be written as
\begin{equation}
  S_{AB}(t)=\frac{1}{2}\Bigl\langle
  \comm{A(t)_\mathrm{H}}{B(0)_\mathrm{H}}_+\Bigr\rangle_{st}-
\bigl\langle A\bigr\rangle_{st}\bigl\langle B\bigr\rangle_{st}.
\end{equation}
Now, applying the time-translational invariance
\begin{equation}
  \Bigl\langle A(t_1)_\mathrm{H}\,B(t_2)_\mathrm{H}\Bigr\rangle_{st}
  =\Bigl\langle A(t_1+t)_\mathrm{H}\,B(t_2+t)_\mathrm{H}\Bigr\rangle_{st}
  \label{eq:apphom}
\end{equation}
for all $t,t_1,t_2$ fixed and finite as well as the relation
\begin{equation}
  \Bigl\langle\comm{A(0)_\mathrm{H}}{B(t)_\mathrm{H}}_\pm\Bigr\rangle_{st}^*=
  \Bigl\langle\comm{A(t)_\mathrm{H}}{B(0)_\mathrm{H}}_\pm\Bigr\rangle_{st},
\end{equation}
which can be verified by a straightforward calculation using $B=A^\dagger$, we
obtain \eqref{eq:rel1}. In the same way \eqref{eq:apphom} can be used to
derive the relation $\chi_{AB}(\Omega)=\ii\,C_{AB}^-(\Omega)$.

\section{Derivation of $\bs{\mathcal{B}_{\pm,11'}^{(2)}}$}\label{sec:B12}
In this appendix we will calculate the vertex $\mathcal{B}_{\pm,11'}^{(2)}$.
Starting from (\ref{eq:Bvertexmatsubara}) we first take the zero-temperature
limit.  Furthermore, we can expand the resolvents up to $O(J)$
as\cite{SchoellerReininghaus09}
\begin{eqnarray}
    & &\!\!\!\!\!\!\!\!\!\!\!\!\!\!\!\!\!
    \Pi(E,\omega)=\frac{1}{E+\ii\omega-L_S^{(0)}-L_S^{(1)}
      -(E+\ii\omega)Z^{(1)}}\nonumber\\
    & &\!\!\!\!\!\!\!\!\!\!\!\!\!=\left(1-\frac{Z^{(1)}}{2}\right)
      \frac{1}{E+\ii\omega-L_S^{(0)}-\tilde{L}_S^{(1)}}
      \left(1-\frac{Z^{(1)}}{2}\right)
  \label{eq:restildeL}
\end{eqnarray}
with
\begin{equation}
  \tilde{L}_S^{(1)}=L_S^{(1)}-\frac{1}{2}\comm{Z^{(1)}}{L_S^{(0)}}_+.
\end{equation}
Using this we obtain together with the expansion
$\mathcal{B}_\pm=\mathcal{B}_\pm^{(0)}+\mathcal{B}_\pm^{(1)}$ in
(\ref{eq:Bvertexmatsubara}) 
\begin{widetext}
\begin{eqnarray}
& &\frac{d}{d\Lambda}\mathcal{B}_{\pm,11'}(\Omega,\delta,\xi,\xi';
\omega_1,\omega_{1'})=
\ii\,\bar{G}_{12}\left(1-\frac{Z^{(1)}}{2}\right)
\frac{1}{\Omega_{12}+\ii\delta+\ii\Lambda+\ii\omega_1-
  L_S^{(0)}-\tilde{L}_S^{(1)}}\nonumber\\*
& &\qquad\times\left(\mathcal{B}_\pm^{(0)}+\mathcal{B}_\pm^{(1)}-\frac{1}{2}
  \comm{Z^{(1)}}{\mathcal{B}_\pm^{(0)}}_+\right)
\frac{1}{\xi_{12}+\ii\xi'+\ii\Lambda+\ii\omega_1-L_S^{(0)}-\tilde{L}_S^{(1)}}
\left(1-\frac{Z^{(1)}}{2}\right)\bar{G}_{\bar{2}1'}
-(1\leftrightarrow 1'),
\end{eqnarray}
where we have omitted the arguments of the vertices $\bar{G}$ for simplicity.
Using
\begin{equation}
  \comm{L_S^{(0)}+\tilde{L}_S^{(1)}}{\mathcal{B}_\pm^{(0)}}_-=
  \tilde{\kappa}\,\mathcal{B}_\pm^{(0)},\quad
  \comm{L_S^{(0)}+\tilde{L}_S^{(1)}}{\mathcal{B}_\pm^{(1)}}_-=
  \tilde{\kappa}\,\mathcal{B}_\pm^{(1)},\quad
  \comm{L_S^{(0)}+\tilde{L}_S^{(1)}}{\comm{Z^{(1)}}{\mathcal{B}_\pm^{(0)}}_+}_-
    =\tilde{\kappa}\,\comm{Z^{(1)}}{\mathcal{B}_\pm^{(0)}}_+,
    \label{eq:commapp}
\end{equation}
with $\tilde{\kappa}=\pm h_0\mp\frac{h_0}{2}\,\mathrm{tr}\,J$ for $B=S^\pm$
and $\tilde{\kappa}=0$ for $B=S^z$ (see Sec.~\ref{sec:expl}) we 
obtain after a partial fraction expansion
\begin{eqnarray}
& &\frac{d}{d\Lambda}\mathcal{B}_{\pm,11'}(\Omega,\delta,\xi,\xi';
\omega_1,\omega_{1'})=
-\frac{\ii}{\Omega-\xi-\tilde{\kappa}+\ii(\delta-\xi')}
\bar{G}_{12}\left(1-\frac{Z^{(1)}}{2}\right)
\left(\mathcal{B}_\pm^{(0)}+\mathcal{B}_\pm^{(1)}-\frac{1}{2}
  \comm{Z^{(1)}}{\mathcal{B}_\pm^{(0)}}_+\right)\nonumber\\*
& &\,\times
\left[
\frac{1}{\Omega_{12}-\tilde{\kappa}+\ii\delta+\ii\Lambda+\ii\omega_1
  -L_S^{(0)}-\tilde{L}_S^{(1)}}-
\frac{1}{\xi_{12}+\ii\xi'+\ii\Lambda+\ii\omega_1-L_S^{(0)}-\tilde{L}_S^{(1)}}
\right]\left(1-\frac{Z^{(1)}}{2}\right)\bar{G}_{\bar{2}1'}
+(1\leftrightarrow 1').\qquad
\end{eqnarray}
We see that no terms $\propto 1/\Lambda$ can occur on the r.h.s. and, hence,
in order to determine the RG equation for $\mathcal{B}_{\pm,11'}^{(2)}$ we
have to replace the vertices $\bar{G}$ by the leading order ones
$\bar{G}^{(1)}$ and omit all terms containing $\mathcal{B}_\pm^{(1)}$,
$Z^{(1)}$ and $\tilde{L}_S^{(1)}$. This yields using
$\tilde{\kappa}=\kappa+O(J)$ as well as $\frac{1}{\Lambda-\ii
  z}=\frac{1}{\Lambda}+\frac{d}{d\Lambda}\ln\frac{\Lambda-\ii z}{\Lambda}$:
\begin{eqnarray}
\frac{d}{d\Lambda}\mathcal{B}_{\pm,11'}^{(2)}(\Omega,\delta,\xi,\xi';
\omega_1,\omega_{1'})&=&-\frac{1}{\Omega-\xi-\kappa+\ii(\delta-\xi')}\,
\bar{G}_{12}^{(1)}\biggl[\biggl(\frac{d}{d\Lambda}
\ln\frac{\Lambda+\omega_1-\ii(\Omega_{12}+\ii\delta-L_S^{(0)})}{\Lambda}
\biggr)\,\mathcal{B}_\pm^{(0)}\nonumber\\*
& &\qquad-\mathcal{B}_\pm^{(0)}\,\biggl(\frac{d}{d\Lambda}
\ln\frac{\Lambda+\omega_1-\ii(\xi_{12}+\ii\xi'-L_S^{(0)})}{\Lambda}
\biggr)\biggr]\bar{G}_{\bar{2}1'}^{(1)}+(1\leftrightarrow 1').
\label{eq:B12RGapp}
\end{eqnarray}
\end{widetext}
Using the RG equation for the leading-order vertex \eqref{eq:poorman} we can
integrate \eqref{eq:B12RGapp} up to higher-order corrections and obtain the
solution \eqref{eq:B12result}, which in particular satisfies the initial
condition
\begin{equation}
  \mathcal{B}_{\pm,11'}^{(2)}(\Omega,\delta,\xi,\xi';
\omega_1,\omega_{1'})\Big|_{\Lambda=\Lambda_0}=0
\end{equation}
as given by \eqref{eq:B12a} from the discrete RG step. 

\begin{figure}[t]
  \psfrag{Bvertex}{$\mathcal{B}_\pm$}
  \includegraphics[scale=0.28]{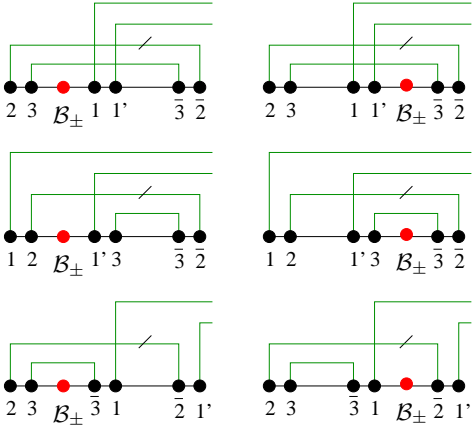}
  \caption{(color online) Two-loop diagrams for $\mathcal{B}_{\pm,11'}$.}
\label{fig:twoloopB12}
\end{figure}
\begin{figure}[t]
  \psfrag{B12vertex}{$\mathcal{B}_{\pm,12}$}
  \psfrag{B21vertex}{$\mathcal{B}_{\pm,\bar{2}1'}$}
  \includegraphics[scale=0.28]{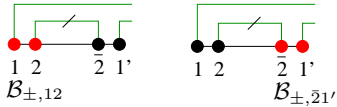}
  \caption{(color online) One-loop RG diagrams for $\mathcal{B}_{\pm,11'}$
  which contain $\mathcal{B}_{\pm,11'}$ itself.}
\label{fig:RGoneloopB12withB12}
\end{figure}
The two-loop diagrams for the vertex $\mathcal{B}_{\pm,11'}$ are shown in
Fig.~\ref{fig:twoloopB12}. As each diagram contains three vertices
$\bar{G}\propto J$ and we are interested in the vertex up to second order, we
have to extract the terms $\sim 1/\Lambda$. This is done by expanding the
resolvents in lowest order
\begin{eqnarray}
  \Pi(E_{23},\omega+\Lambda+\omega_3)&\approx&-\frac{\ii}{\Lambda+\omega_3},\\
  \Pi(E_{12},\omega+\Lambda+\omega_1)&\approx&-\frac{\ii}{\Lambda},\\
  \Pi(E_{11'23},\omega+\Lambda+\omega_1+\omega_{1'}+\omega_3)
  &\approx&-\frac{\ii}{\Lambda+\omega_3},\qquad
\end{eqnarray}
where $E$ ($\omega$) stands for either $\Omega$ ($\delta$) or $\xi$ ($\xi'$),
respectively. Using this we immediately see that the first two diagrams are
proportional to 
\begin{equation}
  \int_0^\Lambda\,\frac{d\omega_3}{(\Lambda+\omega_3)^3}
  \propto\frac{1}{\Lambda^2},
\end{equation}
the fourth and fifth to
\begin{equation}
  \frac{1}{\Lambda}\int_0^\Lambda\,\frac{d\omega_3}{(\Lambda+\omega_3)^2}
  \propto\frac{1}{\Lambda^2},
\end{equation}
and the third and sixth to 
\begin{equation}
  \frac{1}{\Lambda^2}\int_0^\Lambda\,\frac{d\omega_3}{\Lambda+\omega_3}
  \propto\frac{1}{\Lambda^2}.
\end{equation}
Thus the two-loop diagrams behave as $\sim J^3/\Lambda^2$ and hence do not
contribute to the renormalization of the second-order vertex
$\mathcal{B}_{\pm,11'}^{(2)}$.

Finally we have to study the one-loop diagrams which contain
$\mathcal{B}_{\pm,11'}$ itself. These diagrams are shown in
Fig.~\ref{fig:RGoneloopB12withB12}. As we can easily see from
\eqref{eq:B12result}, the leading-order result for $\mathcal{B}_{\pm,11'}$
behaves for large $\Lambda$ as
\begin{equation}
  \mathcal{B}_{\pm,11'}^{(2)}(\Omega,\delta,\xi,\xi';\omega_1,\omega_2)
  \sim \frac{1}{\Lambda}\,\bar{G}^2\sim \frac{J^2}{\Lambda}.
\end{equation}
Now the one-loop diagrams contain an additional vertex $\bar{G}$ as well as a
resolvent $\Pi\sim 1/\Lambda$. Hence we deduce that the diagrams yield terms
proportional to $J^3/\Lambda^2$ which do not contribute to the renormalization
of the second-order vertex $\mathcal{B}_{\pm,11'}^{(2)}$.

\section{Derivation of $\bs{\Sigma_B^{\pm,(2)}}$}\label{sec:Bkernel}
In this appendix we will derive the RG equation \eqref{eq:B2kernelRG} for the
second-order kernel $\Sigma_B^{\pm,(2)}(\Omega,\delta,\xi,\xi')$. For this we
have to evaluate all terms on the r.h.s. of \eqref{eq:Bkernelmatsubara}. We
start with the first line. Using (\ref{eq:restildeL}) and expanding $\bar{G}$
in powers of $J$ yields
\begin{widetext}
\begin{eqnarray}
  & &\int_0^\Lambda d\omega_2\,\bar{G}_{12}^{(1)}\,
  \left(1-\frac{Z^{(1)}}{2}\right)
  \frac{1}{\Omega_{12}+\ii\delta+\ii\Lambda+\ii\omega_2-
    L_S^{(0)}-\tilde{L}_S^{(1)}}
  \left(\mathcal{B}_\pm^{(0)}+\mathcal{B}_\pm^{(1)}
    -\frac{1}{2}\comm{Z^{(1)}}{\mathcal{B}_\pm^{(0)}}_+\right)\nonumber\\*
  & &\qquad\qquad\qquad\qquad\qquad\qquad\qquad\qquad\qquad\times
  \frac{1}{\xi_{12}+\ii\xi'+\ii\Lambda+\ii\omega_2-L_S^{(0)}-\tilde{L}_S^{(1)}}
  \left(1-\frac{Z^{(1)}}{2}\right)\bar{G}_{\bar{2}\bar{1}}^{(1)}
  \label{eq:AppC1}\\
  & &+\int_0^\Lambda d\omega_2\,
  \bigl(\ii\,\bar{G}_{12}^{(2a_1)}+
  \bar{G}_{12}^{(2b)}(\Omega,\delta;\Lambda,\omega_2)\bigr)\,
  \frac{1}{\Omega_{12}+\ii\delta+\ii\Lambda+\ii\omega_2-L_S^{(0)}}\,
  \mathcal{B}_\pm^{(0)}\,
  \frac{1}{\xi_{12}+\ii\xi'+\ii\Lambda+\ii\omega_2-L_S^{(0)}}\,
  \bar{G}_{\bar{2}\bar{1}}^{(1)}\label{eq:AppC2}\\
  & &+\int_0^\Lambda d\omega_2\,\bar{G}_{12}^{(1)}\,
  \frac{1}{\Omega_{12}+\ii\delta+\ii\Lambda+\ii\omega_2-L_S^{(0)}}\,
  \mathcal{B}_\pm^{(0)}\,
  \frac{1}{\xi_{12}+\ii\xi'+\ii\Lambda+\ii\omega_2-L_S^{(0)}}\nonumber\\*
  & &\qquad\qquad\qquad\qquad\qquad\qquad\qquad\qquad\qquad\times
  \bigl(\ii\,\bar{G}_{\bar{2}\bar{1}}^{(2a_1)}+
  \bar{G}_{\bar{2}\bar{1}}^{(2b)}(\xi_{12},\xi'+\Lambda+\omega_2;
  -\omega_2,-\Lambda)\bigr),\label{eq:AppC3}
\end{eqnarray}
where we have already neglected the terms $Z^{(1)}$ and $\tilde{L}_S^{(1)}$ in
the second and third line as they lead only to higher-order corrections.
Using the commutators \eqref{eq:commapp} the first line \eqref{eq:AppC1} can
be treated similarly to $\Sigma_B^{\pm,(1)}$ derived in
Sec.~\ref{sec:WKabove}. In the term $\propto\mathcal{B}_\pm^{(0)}$ we only
keep the term containing $\tilde{\mathcal{K}}_\Lambda(z)$ (the term $\propto
1/\Lambda$ was already used to calculate $\Sigma_B^{\pm,(1)}$), while in
the other two terms we have to extract the term $\propto 1/\Lambda$. Thus we
arrive at
\begin{equation}
  \begin{split}
  &\frac{\ii}{\Omega-\xi-\tilde{\kappa}+\ii(\delta-\xi')}\,
  \bar{G}_{12}^{(1)}\Bigl[
  \tilde{\mathcal{K}}_\Lambda(\Omega_{12}+\ii\delta-L_S^{(0)})\,
  \mathcal{B}_\pm^{(0)}-\mathcal{B}_\pm^{(0)}\,
  \tilde{\mathcal{K}}_\Lambda(\xi_{12}+\ii\xi'-L_S^{(0)})
  \Bigr]\bar{G}_{\bar{2}\bar{1}}^{(1)}\\
  &\qquad-\frac{1}{2\Lambda}\,\bar{G}_{12}^{(1)}\,\mathcal{B}_\pm^{(1)}\,
  \bar{G}_{\bar{2}\bar{1}}^{(1)}
  +\frac{1}{2\Lambda}\,\bar{G}_{12}^{(1)}\,
  \comm{Z^{(1)}}{\mathcal{B}_\pm^{(0)}}_+\,\bar{G}_{\bar{2}\bar{1}}^{(1)}.
  \end{split}
  \label{eq:appBkernelresult1a}
\end{equation}
Using the same steps for the terms containing $\bar{G}^{(2a_1)}$ in
\eqref{eq:AppC2} and \eqref{eq:AppC3} one finds
\begin{equation}
  -\frac{\ii}{2\Lambda}\,\Bigl[\bar{G}_{12}^{(2a_1)}\,\mathcal{B}_\pm^{(0)}\,
  \bar{G}_{\bar{2}\bar{1}}^{(1)}+\bar{G}_{12}^{(1)}\,\mathcal{B}_\pm^{(0)}\,
  \bar{G}_{\bar{2}\bar{1}}^{(2a_1)}\Bigr].
  \label{eq:appBkernelresult1b}
\end{equation}
For the terms of \eqref{eq:AppC2} and \eqref{eq:AppC3} containing
$\bar{G}^{(2b)}$ we use \eqref{eq:G2b} in the form
\begin{eqnarray}
  \bar{G}^{(2b)}_{12}(\Omega,\delta;\Lambda,\omega_2)\!\!\!&=&\!\!\!
  \bar{G}^{(1)}_{13}
  \ln\frac{2\Lambda-\ii(\Omega_{13}+\ii\delta-L_S^{(0)})}{\Lambda}
  \,\bar{G}^{(1)}_{\bar{3}2}-
  \bar{G}^{(1)}_{23}
  \ln\frac{\Lambda+\omega_2-\ii(\Omega_{23}+\ii\delta-L_S^{(0)})}
  {\Lambda}\,\bar{G}^{(1)}_{\bar{3}1},\label{eq:G2bB1}\\
  \!\!\!\!\!\!\!\!\!
  \bar{G}^{(2b)}_{\bar{2}\bar{1}}(\xi_{12},\xi'\!+\!\Lambda\!+\!\omega_2;
  -\omega_2,-\Lambda)\!\!\!&=&\!\!\!
  \bar{G}^{(1)}_{\bar{2}3}
  \ln\frac{2\Lambda-\ii(\xi_{13}+\ii\xi'-L_S^{(0)})}{\Lambda}
  \,\bar{G}^{(1)}_{\bar{3}\bar{1}}-
  \bar{G}^{(1)}_{\bar{1}3}
  \ln\frac{\Lambda+\omega_2-\ii(\xi_{23}+\ii\xi'-L_S^{(0)})}
  {\Lambda}\,\bar{G}^{(1)}_{\bar{3}\bar{2}}.\label{eq:G2bB2}
\end{eqnarray}
When inserted into \eqref{eq:AppC2} and \eqref{eq:AppC3} the first terms do
not depend on the integration variable $\omega_2$. The remaining integral can
be done as usual by a partial fraction expansion. This yields ($\kappa=\pm
h_0$ for $B=S^\pm$ and $\kappa=0$ for $B=S^z$)
\begin{eqnarray}
  & &\hspace{-8mm}
  \frac{\ii}{\Omega\!-\!\xi\!-\!\kappa\!+\!\ii(\delta\!-\!\xi')}\,
  \bar{G}^{(1)}_{13}
  \ln\frac{2\Lambda-\ii(\Omega_{13}+\ii\delta-L_S^{(0)})}{\Lambda}
  \,\bar{G}^{(1)}_{\bar{3}2}\left[
    \mathcal{K}_\Lambda(\Omega_{12}+\ii\delta-L_S^{(0)})\,
    \mathcal{B}_\pm^{(0)}-\mathcal{B}_\pm^{(0)}\,
    \mathcal{K}_\Lambda(\xi_{12}+\ii\xi'-L_S^{(0)})\right]\,
  \bar{G}_{\bar{2}\bar{1}}^{(1)}\nonumber\\*
  & &\hspace{-8mm}+
 \frac{\ii}{\Omega\!-\!\xi\!-\!\kappa\!+\!\ii(\delta\!-\!\xi')}\,
 \bar{G}_{12}^{(1)}\,\left[
    \mathcal{K}_\Lambda(\Omega_{12}+\ii\delta-L_S^{(0)})
    \mathcal{B}_\pm^{(0)}-\mathcal{B}_\pm^{(0)}\,
    \mathcal{K}_\Lambda(\xi_{12}+\ii\xi'-L_S^{(0)})\right]\,
  \bar{G}^{(1)}_{\bar{2}3}
    \ln\frac{2\Lambda-\ii(\xi_{13}+\ii\xi'-L_S^{(0)})}{\Lambda}
  \,\bar{G}^{(1)}_{\bar{3}\bar{1}}.\nonumber
\end{eqnarray}
\end{widetext}
If we now expand $\mathcal{K}_\Lambda$ and the logarithm for large $\Lambda$,
$\mathcal{K}_\Lambda(z)=\ln 2+\ii z/2\Lambda$ and $\ln\frac{2\Lambda-\ii
  z}{\Lambda}=\ln 2-\ii z/2\Lambda$, and keep only the terms proportional to
$\frac{J^3}{\Lambda}$ we arrive at
\begin{equation}
  -\frac{\ln 2}{2\Lambda}\,\bar{G}_{12}^{(1)}
  \Bigl[\bar{G}^{(1)}_{\bar{2}3}\,\mathcal{B}_\pm^{(0)}+
  \mathcal{B}_\pm^{(0)}\,\bar{G}^{(1)}_{\bar{2}3}\Bigr]
  \bar{G}_{\bar{3}\bar{1}}^{(1)}.
  \label{eq:BG2b1stterm}
\end{equation}
In contrast, the second terms of \eqref{eq:G2bB1} and \eqref{eq:G2bB2} do
depend on the integration variable $\omega_2$. The evaluations is, however,
straightforward. We use a partial fraction expansion for the resolvents left
and right to the vertex $\mathcal{B}_\pm^{(0)}$ as well as 
\begin{equation}
  \ln\frac{\Lambda+\omega_2-\ii z}{\Lambda}=
  \ln\frac{\Lambda+\omega_2}{\Lambda}-
  \frac{\ii z}{\Lambda+\omega_2}.
\end{equation}
This leads to integrals of the form
\begin{eqnarray}
  \int_0^\Lambda d\omega_2\,\frac{1}{z+\ii\Lambda+\ii\omega_2}\,
  \ln\frac{\Lambda+\omega_2}{\Lambda}\approx\frac{1-\ln 2}{2\Lambda}\,z,\\
  \int_0^\Lambda d\omega_2\,\frac{1}{z+\ii\Lambda+\ii\omega_2}\,
  \frac{1}{\Lambda+\omega_2}\approx-\frac{\ii}{2\Lambda},
\end{eqnarray}
where we are only interested in the $\sim 1/\Lambda$ terms. Now using the
asymmetry $\bar{G}^{(1)}_{12}=-\bar{G}^{(1)}_{21}$ we find
\begin{equation}
  -\frac{1-\ln 2}{2\Lambda}\,\bar{G}_{12}^{(1)}
  \Bigl[\bar{G}^{(1)}_{\bar{2}3}\,\mathcal{B}_\pm^{(0)}+
  \mathcal{B}_\pm^{(0)}\,\bar{G}^{(1)}_{\bar{2}3}\Bigr]
  \bar{G}_{\bar{3}\bar{1}}^{(1)}.
  \label{eq:BG2b2ndterm}
\end{equation}
which has to be combined with \eqref{eq:BG2b1stterm} for the full result from
the terms containing $\bar{G}^{(2b)}$. 

The second and third line of \eqref{eq:Bkernelmatsubara} containing the vertex
$\mathcal{B}_{\pm,11'}^{(2)}$ can be treated using the same steps as were used
to evaluate the $\bar{G}^{(2b)}$-dependent parts of \eqref{eq:AppC2} and
\eqref{eq:AppC3}. The result reads after some tedious but straightforward
algebra
\begin{equation}
  \frac{1+\ln 2}{2\Lambda}\,\bar{G}_{12}^{(1)}
  \Bigl[\bar{G}^{(1)}_{\bar{2}3}\,\mathcal{B}_\pm^{(0)}+
  \mathcal{B}_\pm^{(0)}\,\bar{G}^{(1)}_{\bar{2}3}\Bigr]
  \bar{G}_{\bar{3}\bar{1}}^{(1)}.
\label{eq:appBkernelresult2}
\end{equation}

Finally, to extract the leading term $\sim J^3/\Lambda$ of the fourth line of
\eqref{eq:Bkernelmatsubara} one can simply replace the resolvents
$\Pi(z,z'+\Lambda+\omega_i)$ by $1/(\Lambda+\omega_i)$. This yields
\begin{equation}
  \frac{\ln 2}{2\Lambda}\,\bar{G}_{12}^{(1)}
  \Bigl[\bar{G}^{(1)}_{\bar{2}3}\,\mathcal{B}_\pm^{(0)}+
  \mathcal{B}_\pm^{(0)}\,\bar{G}^{(1)}_{\bar{2}3}\Bigr]
  \bar{G}_{\bar{3}\bar{1}}^{(1)}.
  \label{eq:appBkernelresult3}
\end{equation}

Hence, the result for the RG equation of the kernel
$\Sigma_B^{\pm,(2)}(\Omega,\delta,\xi,\xi')$ is obtained by summing
\eqref{eq:appBkernelresult1a}, \eqref{eq:appBkernelresult1b},
\eqref{eq:BG2b1stterm}, \eqref{eq:BG2b2ndterm}, \eqref{eq:appBkernelresult2}
and \eqref{eq:appBkernelresult3}, using
$\tilde{\mathcal{K}}_\Lambda(z)=\frac{d}{d\Lambda}\tilde{F}_\Lambda(z)$ with
\eqref{eq:tildeF} in \eqref{eq:appBkernelresult1a} as well as
\begin{equation}
  \frac{2}{\Lambda}\,\bar{G}_{12}^{(1)}
  \Bigl[\bar{G}^{(1)}_{\bar{2}3}\,\mathcal{B}_\pm^{(0)}+
  \mathcal{B}_\pm^{(0)}\,\bar{G}^{(1)}_{\bar{2}3}\Bigr]
  \bar{G}_{\bar{3}\bar{1}}^{(1)}=
  \frac{d}{d\Lambda}\,\bar{G}_{12}^{(1)}\,\mathcal{B}_\pm^{(0)}\,
  \bar{G}_{\bar{2}\bar{1}}^{(1)}.
\end{equation}
This yields \eqref{eq:B2kernelRG}.

\section{Algebra in Liouville space}\label{sec:appL}
Consider an operator $A$ acting on the dot Hilbert space having matrix
elements $A_{ab}$ with respect to the basis $\{\ket{\up},\ket{\dw}\}$. If $K$
denotes the superoperator acting on dot operators via $O.=\comm{A}{.}_\pm$
then for an arbitrary dot operator $B$ we have
\begin{equation}
  (OB)_{ab}=O_{ab,cd}\,B_{cd},\quad 
  O_{ab,cd}=A_{ac}\,\delta_{bd}\pm \delta_{ac}\,A_{db}.
\end{equation}
Furthermore, we represent superoperators in the matrix representation 
\begin{equation}
  O=(O_{ab,cd})=\left(\begin{array}{cc|cc}
      O_{\up\up,\up\up}&O_{\up\up,\dw\dw}&O_{\up\up,\up\dw}&O_{\up\up,\dw\up}\\
      O_{\dw\dw,\up\up}&O_{\dw\dw,\dw\dw}&O_{\dw\dw,\up\dw}&O_{\dw\dw,\dw\up}\\
      \hline
      O_{\up\dw,\up\up}&O_{\up\dw,\dw\dw}&O_{\up\dw,\up\dw}&O_{\up\dw,\dw\up}\\
      O_{\dw\up,\up\up}&O_{\dw\up,\dw\dw}&O_{\dw\up,\up\dw}&O_{\dw\up,\dw\up}
      \end{array}\right).
    \label{eq:supermatrices}
\end{equation}
If $O=PQ$ is the product of two superoperators, then
$O_{ab,cd}=P_{ab,ef}\,Q_{ef,cd}$ and the matrix \eqref{eq:supermatrices} of
$O$ is simply given by the matrix product of the matrices of $P$ and $Q$.

A basis for the operators in the Liouville space of the Kondo dot can be built
up by the spin superoperators $\underline{L}^+$ and $\underline{L}^-$ defined
in \eqref{eq:defLpLm}. An explicit representation in the basis
\eqref{eq:supermatrices} is provided by
\begin{displaymath}
\begin{split}
&L^{+x}=\left(\begin{array}{cc|cc}
      0&0&0&\tfrac{1}{2}\\ 0&0&\tfrac{1}{2}&0\\ \hline
      0&\tfrac{1}{2}&0&0 \\ \tfrac{1}{2}&0&0&0
      \end{array}\right)\!\!,\;
 L^{+y}=\left(\begin{array}{cc|cc}
      0&0&0&-\tfrac{\ii}{2}\\ 0&0&\tfrac{\ii}{2}&0\\ \hline
      0&-\tfrac{\ii}{2}&0&0 \\ \tfrac{\ii}{2}&0&0&0
      \end{array}\right)\!\!,
\end{split}
\end{displaymath}
\begin{displaymath}
\begin{split}
&L^{+z}=\left(\begin{array}{cc|cc}
      \tfrac{1}{2}&0&0&0\\ 0&-\tfrac{1}{2}&0&0\\ \hline
      0&0&\tfrac{1}{2}&0 \\ 0&0&0&-\tfrac{1}{2}
      \end{array}\right)\!\!,\;
 L^{-x}=\left(\begin{array}{cc|cc}
      0&0&-\tfrac{1}{2}&0\\ 0&0&0&-\tfrac{1}{2}\\ \hline
      -\tfrac{1}{2}&0&0&0 \\ 0&-\tfrac{1}{2}&0&0
      \end{array}\right)\!\!,
\end{split}
\end{displaymath}
\begin{displaymath}
\begin{split}
&L^{-y}=\left(\begin{array}{cc|cc}
      0&0&-\tfrac{\ii}{2}&0\\ 0&0&0&\tfrac{\ii}{2}\\ \hline
      \tfrac{\ii}{2}&0&0&0 \\ 0&-\tfrac{\ii}{2}&0&0
      \end{array}\right)\!\!,\;
 L^{-z}=\left(\begin{array}{cc|cc}
      -\tfrac{1}{2}&0&0&0\\ 0&\tfrac{1}{2}&0&0\\ \hline
      0&0&\tfrac{1}{2}&0 \\ 0&0&0&-\tfrac{1}{2}
      \end{array}\right)\!\!.
\end{split}
\end{displaymath}
Furthermore we define the operators
\begin{eqnarray}
  L^a&=&\frac{3}{4}\,\mathbf{1}+\underline{L}^+\cdot\underline{L}^-,
  \label{eq:defLa}\\
  L^b&=&\frac{1}{4}\,\mathbf{1}-\underline{L}^+\cdot\underline{L}^-,\\
  L^c&=&\frac{1}{2}\,\mathbf{1}+2\,L^{+z}\,L^{-z},\\
  L^h&=&L^{+z}+L^{-z},\\
  \underline{L}^1&=&\frac{1}{2}\bigl(\underline{L}^+-\underline{L}^-\bigr)
  -\ii\,\underline{L}^+\times\underline{L}^-,\\
  \underline{L}^2&=&-\frac{1}{2}\bigl(\underline{L}^++\underline{L}^-\bigr),\\
  \underline{L}^3&=&\frac{1}{2}\bigl(\underline{L}^+-\underline{L}^-\bigr)
  +\ii\,\underline{L}^+\times\underline{L}^-,\label{eq:defL3}
\end{eqnarray}
as well as 
\begin{eqnarray}
  L^a_\pm&=&L^{ax}\pm\ii L^{ay},\quad a=+,-,1,2,3,\\
  L^4_\pm&=&L^2_\pm\pm\bigl(L^+_\pm\,L^{-z}+L^{+z}\,L^-_\pm\bigr),\\
  L^5_\pm&=&L^2_\pm\mp\bigl(L^+_\pm\,L^{-z}+L^{+z}\,L^-_\pm\bigr).
\end{eqnarray}
We note that $L^h=-2L^{2z}$ as well as $L^4_\pm+L^5_\pm=2L^2_\pm$.

In the spin sector we will use frequently
\begin{equation}
  \sigma^a_{\sigma_1\sigma_3}\sigma^a_{\sigma_3\sigma_2}=
  \sigma^a_{\sigma_3\sigma_2}\sigma^a_{\sigma_1\sigma_3}=
  \delta_{\sigma_1\sigma_2}\quad(\text{no sum over }a)
  \label{eq:AppEsigma1}
\end{equation}
as well as
\begin{equation}
  \sigma^a_{\sigma_1\sigma_3}\sigma^b_{\sigma_3\sigma_2}=
  -\sigma^a_{\sigma_3\sigma_2}\sigma^b_{\sigma_1\sigma_3}=
  \ii\sum_c\epsilon_{abc}\,\sigma^c_{\sigma_1\sigma_2}
  \quad(\text{for }a\neq b).
  \label{eq:AppEsigma2}
\end{equation}
The Liouville operators satisfy (the sums are over $i,j=x,y,z$ while the index
$p$ takes the values $p=\pm$)
\begin{eqnarray}
  \ii\sum_{i,j}\epsilon_{ijk}\,L^{2i}\,L^{2j}&=&\frac{1}{2}\,L^{2k},
  \label{eq:AppE1}\\
  \sum_i L^{2i}\,\mathcal{B}_p^{(0)}\,L^{2i}&=&\delta_{p-}\,
  \frac{1}{4}\,\mathcal{B}_-^{(0)},\label{eq:AppE2}\\
  \sum_i L^{2i}\,\mathcal{B}_p^{(0)}\,L^{3i}&=&\delta_{p-}\,\ii\,
  L^{3z},\label{eq:AppE3}\\
  \sum_i L^{3i}\,\mathcal{B}_p^{(0)}\,L^{2i}&=&\delta_{p+}\,\frac{\ii}{2}\,
  L^h,\label{eq:AppE4}\\
  \sum_i L^{2i}\,\comm{Z^{(1)}}{\mathcal{B}_p^{(0)}}_+\,L^{2i}&=&
  \delta_{p-}\,\frac{\ii}{2}\,\text{tr}\,J\,L^h.\label{eq:AppE5}
\end{eqnarray}\\[2mm]

\section{Proof of (\ref{eq:assumption1}) and (\ref{eq:assumption2}) for the
  isotropic Kondo model}\label{sec:ass}
The results for the Liouvillian presented in Sec.~\ref{sec:resultsbelowLc}
allow us to obtain
\begin{equation}
  L_S^{(0)}=h_0\,L^h=h_0(P_+-P_-)=\sum_{i=0,1,\pm} z_i\,P_i(z_i)+O(J_c),
\end{equation}
where we have used $z_0=0$, $z_1=O(J_c^2)$ and $z_\pm=\pm
h_0+O(J_c)$. Furthermore, we can expand the zero-eigenvalue projector as
\begin{eqnarray}
  P_0(z_1)&=&P_0(0)-\ii\frac{\Gamma^{3z}(0)}{\Gamma^a(0)}\,
  \frac{d}{dz}\Gamma^a(z)\Big|_{z=0}\,L^{3z}+\ldots\nonumber\\
  &=&P_0(z_0)+O(J_c),
\end{eqnarray}
where $\frac{d}{dz}\Gamma^a(z)\Big|_{z=0}=-\ii (J_R+J_L)+\ldots$. This
directly yields \eqref{eq:assumption2}. To prove the final statement,
$\kappa=z_i-z_j+O(J_c)$ for all pairs $(i,j)$ for which 
\begin{equation}
  \bar{G}^{(1)c}_{12}\,P_i(z_i)\,\mathcal{B}_\pm^{(0)}\,P_j(z_j)\,
  \bar{G}^{(1)c}_{\bar{2}\bar{1}}
\end{equation}
is non-vanishing, we note that for $B=S^z$ these pairs are given by (see
\eqref{eq:GPB+PG} and \eqref{eq:GPB-PG}) $(i,j)=(0,1),(1,1)$ for
$\mathcal{B}_+^{(0)}$ and $(i,j)=(\pm,\pm)$ for $\mathcal{B}_-^{(0)}$,
respectively. On the other hand we have $\kappa=0$ in both cases, which
implies $z_i-z_j=\kappa+O(J_c^2)$ for $B=S^z$. The same analysis can be
performed for $B=S^\pm$ where the relevant pairs are given by
$(i,j)=(0,\mp),(1,\mp)$ for $\mathcal{B}_+^{(0)}$ and $(i,j)=(1,\mp),(\pm,1)$
for $\mathcal{B}_-^{(0)}$.

\end{document}